\documentclass{emulateapj} 

\usepackage {lscape, graphicx}

\gdef\1054{MS\,1054--03}

\def\simgeq{{\raise.0ex\hbox{$\mathchar"013E$}\mkern-14mu\lower1.2ex\hbox{$\mathchar"0218$}}} 

\begin {document}

\title {Recovering Stellar Population Properties and Redshifts from Broad-Band Photometry of Simulated Galaxies: Lessons for SED Modeling}

\author{Stijn Wuyts\altaffilmark{1,2,3}, Marijn Franx\altaffilmark{2}, Thomas J. Cox\altaffilmark{1,3}, Lars Hernquist\altaffilmark{1}, Philip F. Hopkins\altaffilmark{4}, Brant E. Robertson\altaffilmark{5,6,7}, Pieter G. van Dokkum\altaffilmark{8}}
\altaffiltext{}{Electronic address: swuyts@cfa.harvard.edu}
\altaffiltext{1}{Harvard-Smithsonian Center for Astrophysics, 60 Garden Street, Cambridge, MA 02138}
\altaffiltext{2}{Leiden University, Leiden Observatory, P.O. Box 9513, NL-2300 RA, Leiden, The Netherlands.}
\altaffiltext{3}{W. M. Keck Postdoctoral Fellow}
\altaffiltext{4}{Department of Astronomy, University of California Berkeley, Berkeley, CA 94720}
\altaffiltext{5}{Kavli Institute for Cosmological Physics, and Department of Astronomy and Astrophysics, University of Chicago, Chicago, IL 60637}
\altaffiltext{6}{Enrico Fermi Institute, Chicago, IL 60637}
\altaffiltext{7}{Spitzer Fellow}
\altaffiltext{8}{Department of Astronomy, Yale University, New Haven, CT 06520-8101}

\begin{abstract}
We present a detailed analysis of our ability to determine stellar
masses, ages, reddening and extinction values, and star formation
rates of high-redshift galaxies by modeling broad-band SEDs with
stellar population synthesis.  In order to do so, we computed
synthetic optical-to-NIR SEDs for model galaxies taken from
hydrodynamical merger simulations placed at redshifts $1.5 \leq z \leq
2.9$.  Viewed under different angles and during different evolutionary
phases, the simulations represent a wide variety of galaxy types
(disks, mergers, spheroids).  We show that simulated galaxies span a
wide range in SEDs and color, comparable to these of observed
galaxies.  In all star-forming phases, dust attenuation has a large
effect on colors, SEDs, and fluxes.  The broad-band SEDs were then fed
to a standard SED modeling procedure and resulting stellar population
parameters were compared to their true values.  Disk galaxies
generally show a decent median correspondence between the true and
estimated mass and age, but suffer from large uncertainties ($\Delta
\log M = -0.06^{+0.06}_{-0.13}$, $\Delta \log age_w =
+0.03^{+0.19}_{-0.42}$).  During the merger itself, we find larger
offsets: $\Delta \log M = -0.13^{+0.10}_{-0.14}$ and $\Delta \log
age_w = -0.12^{+0.40}_{-0.26}$.  $E(B-V)$ values are generally
recovered well, but the estimated total visual absorption $A_V$ is
consistently too low, increasingly so for larger optical depths
($\Delta A_V = -0.54^{+0.40}_{-0.46}$ in the merger regime).  Since
the largest optical depths occur during the phases of most intense
star formation, it is for the highest SFRs that we find the largest
underestimates ($\Delta \log SFR = -0.44^{+0.32}_{-0.31}$ in the
merger regime).  The masses, ages, $E(B-V)$, $A_V$, and SFR of
merger remnants (spheroids) are very well reproduced.

We discuss possible biases in SED modeling results caused by mismatch
between the true and template star formation history, dust
distribution, metallicity variations and AGN contribution.  Mismatch
between the real and template star formation history, as is the case
during the merging event, drives the age, and consequently mass
estimate, down with respect to the true age and mass.  However, the
larger optical depth toward young stars during this phase reduces the
effect considerably.  Finally, we tested the photometric redshift code
EAZY on the simulated galaxies placed at high redshift.  We find a
small scatter in $\Delta z / (1+z)$ of 0.031.
\end{abstract}


\section {Introduction}
\label{intro.sec}
Understanding the growth and aging of galaxies over cosmic time
requires reliable estimates of their mass, formation epoch and star
formation history.  With the current generation of telescopes, stellar
velocity dispersion measurements can probe the gravitational potential
in which the baryonic galaxy content resides out to $z \sim 1.3$ (van
Dokkum \& Stanford 2003; Holden et al. 2005).  Beyond this redshift,
gas velocity dispersions can be measured from emission lines, but do
not always trace the potential due to outflows (Franx et al. 1997;
Pettini et al. 1998, 2001; Shapley et al. 2003), and would lead to
biased samples missing quiescent galaxies lacking emission lines in
their spectra (Kriek et al. 2006).  For these reasons, most studies of
high-redshift galaxies have used stellar mass estimates derived by
modeling of the broad-band stellar energy distribution (SED) to characterize
the mass.

Since age estimates from $H\alpha$ equivalent widths (van Dokkum et
al. 2004; Erb et al. 2006c) or Balmer/4000\AA\ break strengths (Kriek
et al. 2006) are very demanding in terms of telescope time and only
attainable for the brightest galaxies, stellar ages as well are
commonly derived from broad-band photometry.

Over the past few years, SED modeling has been proven extremely
valuable in characterizing the galaxy population in the early universe
(e.g., Papovich et al. 2001; Shapley et al. 2001, 2005; F\"{o}rster
Schreiber et al. 2004).  Nevertheless, a number of assumptions are
required for the limited number of data points (11 passbands in our
case, but often less) to lead to a single solution in terms of
physical properties such as stellar mass, stellar age, dust
extinction, star formation rate (SFR), and often redshift.

First, the star formation history (SFH) is generally modelled by a
simple functional form: a single burst, constant star formation, or an
exponentially declining model.  In reality, high-redshift galaxies
show evidence of more complex SFHs, often with brief recurrent
episodes of star formation (e.g. Papovich et al. 2001; Ferguson et
al. 2002; Papovich et al. 2005).  Second, we use the approximation of
a single foreground screen of dust in accounting for the attenuation,
even though in reality the dust will be distributed in between the
stars.  Third, we fit solar metallicity models.  This is a common
approach in modeling of high-redshift galaxy SEDs in order to reduce
the number of degrees of freedom in the fitting procedure by one
(e.g., F\"{o}rster Schreiber et al. 2004; Shapley et al. 2005;
Rigopoulou et al. 2006).  In addition, the tracks and spectral
libraries on which the stellar population synthesis models are based,
have the best empirical calibration for solar metallicity.  Although
consistent with the current metallicity estimates from near-infrared
(NIR) spectroscopy of high-redshift galaxies (van Dokkum et al. 2004;
Erb et al. 2006a; Maiolino et al. 2008), it must be kept in mind that
these measurements are currently limited to the bright end of the
galaxy population.  Fourth, SED modeling generally assumes a purely
stellar origin of the light, while observational evidence for a
substantial fraction of low luminosity AGN at high redshift has been
accumulating (van Dokkum et al. 2004; Reddy et al. 2005; Papovich et
al. 2006; Kriek et al. 2007; Daddi et al. 2007b).  They may contribute
to the optical SEDs.

Finally, one adopts a certain attenuation law, initial mass function
(IMF), and stellar population synthesis code.  Their appropriateness
at low and high redshifts is much debated.

In this paper, we address the impact of the first four assumptions
(related to SFH, dust attenuation, metallicity, and AGN) using
hydrodynamical simulations of merging galaxies (see Robertson et
al. 2006; Cox et al. 2006).  The SPH simulations follow the star
formation on a physical basis, resulting in more complex SFHs than are
allowed in typical SED modeling.  They keep track of the distribution
and metallicity of gas and stellar particles, allowing a determination
of the line-of-sight dependent extinction toward each stellar
particle separately and a knowledge of the stellar metallicity as a
function of time.  Here, we apply the same SED modeling that we use
for observed galaxies to broad-band photometry extracted from the
simulation outputs, and study how well the mass, age, dust content, and SFR
of the simulated galaxies can be recovered.

The reason we use merger simulations for this exercise is threefold.
First, galaxy mergers are believed to play an important role in galaxy
evolution (see, e.g., Holmberg 1941; Zwicky 1956; Toomre \& Toomre
1972; Toomre 1977), increasingly so at high redshift (see, e.g.,
Glazebrook et al. 1995; Driver, Windhorst,\& Griffiths 1995; Abraham
et al. 1996).  Moreover, along their evolutionary path they are
visible as vastly different galaxy types, allowing to test the
recovery of stellar population parameters under a wide range of
conditions: gas-rich star-forming disks, dust-obscured mergers, and
quiescent spheroids.  Finally, in a separate paper we will compare
predictions of the color distribution and mass density of
high-redshift galaxies derived from these simulations with the
observed galaxy population in deep fields.  A good understanding of
what it is we measure with SED modeling is crucial in order to compare
identical mass-limited samples of observed and simulated galaxies.

We start with a description of the simulations in
\S\ref{simulations.sec}.  Next, we explain the methodology of our SED
modeling in \S\ref{SED_modeling.sec}.  \S\ref{results_fixz.sec}
discusses how well we can measure stellar population properties when a
spectroscopic redshift is available.  \S\ref{results_freez.sec}
repeats the analysis, now leaving the redshift as an extra free
parameter (i.e., fitting for the photometric redshift).
Finally, we summarize the results in \S\ref{summary.sec}.

\section {The simulations}
\label{simulations.sec}

\subsection {Main characteristics}
\label{sim_main.sec}
The simulations on which we test our SED modeling were performed by
Robertson et al. (2006).  We refer the reader to that paper for a
detailed description of the simulations.  Briefly, the simulations
were performed with the parallel TreeSPH code GADGET-2 (Springel
2005).  The code uses an entropy-conserving formulation of smoothed
particle hydrodynamics (Springel \& Hernquist 2002), and includes gas
cooling, a multiphase model for the interstellar medium (ISM) to
describe star formation and supernova feedback (Springel \& Hernquist
2003), and a prescription for supermassive black hole growth and
feedback (Springel et al. 2005b).

At the start, each simulation consists of 120000 dark matter
particles, 80000 gas particles, and 80000 stellar particles.  They
represent two stable, coplanar disk galaxies, each embedded in an
extended dark matter halo with Hernquist (1990) profile.  We have
realisations where the disks start with a gas fraction of 40\% and
80\%.  Stellar masses at the start of the simulation varied from $7.0
\times 10^9\ M_{\sun}$ to $2.3 \times 10^{11}\ M_{\sun}$ per disk
galaxy.  In total, we study 6 simulations (2 different gas fractions,
and 3 different masses) for which at 30 timesteps all information was
stored in a snapshot.  Each snapshot was analyzed as seen from 30
different viewing angles and placed at 8 redshifts between $z=1.5$ and
3 (\S\ref{sim_phot.sec}).  We repeated our analysis on a limited
number of snapshots and viewing angles of merger simulations of
non-coplanar disks and conclude that the results presented in this
paper are robust against such variations.

For a given virial velocity, the halo concentration, virial mass and
virial radius were scaled following Robertson et al. (2006) to
approximate the structure of disk galaxies at redshift $z=3$.  In
practice, this means that the mass- and redshift-dependent halo
concentration measured by Bullock et al. (2001) was adopted:
\begin {equation}
C_{vir}(M_{vir}, z) \approx 9 \left( \frac{M_{vir}}{M_{coll,0}} \right)^{-0.13} (1+z)^{-1},
\label {Cvir.eq}
\end {equation}
where $M_{coll,0} \sim 8 \times 10^{12}\ h^{-1}\ M_{\sun}$ is the
linear collapse mass at z=0, and that the following scaling relations were
used for the virial mass and virial radius of the progenitors:
\begin {eqnarray}
M_{vir} = \frac{V_{vir}^3}{10GH(z)} \\
R_{vir} = \frac{V_{vir}}{10H(z)},
\label {Mvir.eq}
\end {eqnarray}
where $V_{vir}$ is the virial velocity and $H(z)$ is the Hubble
parameter.  Disk sizes were initialized according to the Mo et
al. (1998) formalism for dissipational disk galaxy formation assuming
the fraction of the total angular momentum contained in the disk
equals the fraction of the total mass contained in the disk
($\frac{M_{disk}}{M_{vir}} = 0.041$ to match the Milky Way-like model
by Springel et al. 2005a).  The disk scale length is then derived from
the halo concentration $C_{vir}$ (Eq.\ \ref{Cvir.eq}) and the galaxy
spin $\lambda$.  The adopted value of $\lambda=0.033$ is motivated by
cosmological N-body simulations (Vitvitska et al. 2002).

We set the ages of the stars existing at the start of the simulation
such as to represent a constant star formation history prior to the
start of the simulation at a star formation rate (SFR) equal to that
calculated in the first phases of the simulation.  The corresponding
metallicities of stars present at the start of the simulation were
then set according to the closed box model: $Z(t) = -y \ln
[f_{gas}(t)]$, where $Z(t)$ is the metallicity of a stellar particle
formed at time t, the yield y=0.02 and $f_{gas}(t)$ is the gas
fraction of the system at the considered time.  Similarly, the gas at
the start of the simulation was assigned a uniform metallicity
$Z_{gas}(t_S) = -y \ln [f_{gas}(t_S)]$ where $t_S$ represents the
start of the simulation, and $f_{gas}(t_S)=0.4$ or 0.8 respectively
for our 2 gas fraction runs.  The closed box model represents an upper
limit on the allowed enrichment by heavy elements, which in reality
may be reduced by outflows or infall of metal-poor gas (Edmunds 1990).
The fact that we consider 2 gas fractions guarantees a wide range of
progenitor types, with ages of a few 100 Myr and $Z_{gas}=0.004$ at
the start of the simulation for $f_{gas}=0.8$ to typical stellar ages
of a Gyr and nearly solar gas metallicity for $f_{gas}=0.4$.  Starting
from the initial conditions described above, the GADGET-2 code
computed the subsequent evolution of the stellar populations using the
star formation and metal enrichment prescriptions outlined by Springel
\& Hernquist (2003).  The mass-weighted mean stellar metallicity for
the $f_{gas} = 0.4$ runs increases in $\sim 700$ Myr from 0.4
$Z_{\sun}$ to 0.9 $Z_{\sun}$, after which it remains constant.  The
metal enrichment of stars in the $f_{gas} = 0.8$ runs proceeds
rapidly, increasing in a few 100 Myr from 0.1 $Z_{\sun}$ to 0.5
$Z_{\sun}$ and leveling off after $\sim 0.5$ Gyr at values between 0.8
$Z_{\sun}$ and $Z_{\sun}$.

\begin {figure} [htbp]
\centering
\plotone{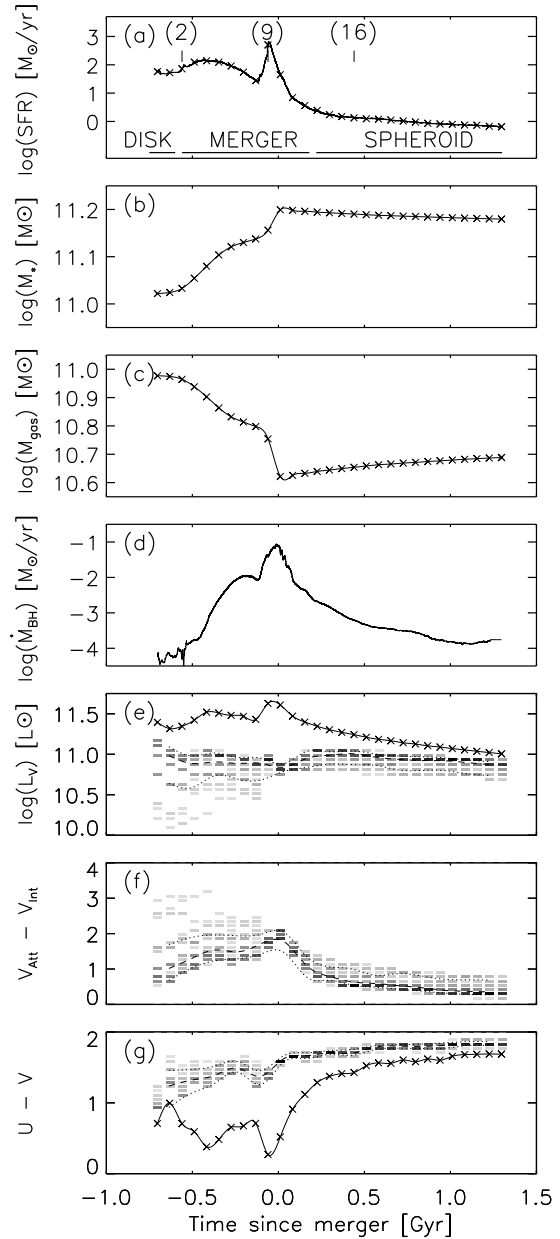} 
\caption{Evolution of a typical merger simulation.  (a) The star
formation history, (b) the mass build-up, (c) the gas exhaustion, (d)
the accretion rate history onto the black hole(s), (e) the evolution
of the intrinsic (i.e., unattenuated) and attenuated V-band
luminosity, (f) the distribution of effective visual extinctions
(attenuated minus intrinsic V-band magnitude) corresponding to
different viewing angles, and (g) the intrinsic and attenuated $U-V$
color.  Cross symbols in panel (e) and (g) present the evolution of
the intrinsic photometry.  For the attenuated photometry in panels (e)
to (g), a binned representation is used where a darker intensity
indicates a larger number of viewing angles.  The dashed curve
represents the median evolution.  The dotted lines indicate the
interval containing the central 68\% of the viewing angles.  The cross
symbols mark the sampling of snapshots when the full physical
information of all SPH particles was stored to disk.  After a first
bump in the star formation rate during the first passage of the
progenitors, a peak in star formation is reached for a brief period
during which several hundreds of solar masses of gas are converted
into stars.  The typical extinction for a random line of sight is
peaking around the same time.  Shortly after, the accretion onto the
supermassive black hole is maximal, coinciding with the merger between
the two progenitor black holes.  The reddest U-V colors are reached
during the merger remnant phase.
\label {sim.fig}
}
\end {figure}

The overall timespan covered by each simulation was 2 Gyr.  Fig.\
\ref{sim.fig}(a) illustrates the star formation history of one of the
merger simulations.  Specifically, we show a simulation starting with
40\% gas fraction and ending with a stellar mass of $1.5 \times
10^{11}\ M_{\sun}$, but the SFH of the other simulations looks
qualitatively similar.  Fig.\ \ref{sim.fig}(b) illustrates the
build-up of stellar mass.  In order to allow a fair test of the SED
modeling with Bruzual \& Charlot (2003) templates (that include mass
loss), we take into account mass loss of each of the stellar particles
for which the time of formation and gas mass from which it was formed
are stored in the simulation snapshots.  This explains the slow
decrease in total stellar mass at late times in Fig.\
\ref{sim.fig}(b).  Fig.\ \ref{sim.fig}(c) shows the exhaustion of gas
from which the stars are formed, and Fig.\ \ref{sim.fig}(d) presents
the accretion history onto the black hole(s).  We draw the time axis
relative to the actual moment of merging, defined as the timestep when
the two black hole particles become one, coinciding with the peak in
the accretion history.  Cross symbols indicate the snapshots,
separated by 70 Myr, when all physical information was stored to disk.

As time progresses, the orderly rotation and star formation in the
disks is disturbed by each others gravitational pull.  The star
formation history shows a first, but rather shallow, bump during the
first passage of the disks.  Next, gravitational torques enable the gas to
loose angular momentum and flow to the centers where it triggers a
starburst (Larson \& Tinsley 1978; Noguchi 1988; Hernquist 1989;
Barnes \& Hernquist 1991, 1996; Mihos \& Hernquist 1994, 1996).
Meanwhile, part of the inflowing gas is fed to the central
supermassive black holes (SMBHs).  Once the SMBHs grow massive enough,
they produce a luminous quasar (Sanders et al. 1988a,b; Hernquist
1989; Sanders \& Mirabel 1996; Genzel et al. 1998) whose feedback
contributes to halting subsequent star formation (Di Matteo et al. 2005; Springel et al
2005a), leaving a red spheroid galaxy as remnant (Robertson et
al. 2006; Cox et al. 2006).

\subsection {Extracting photometry from the simulation output}
\label{sim_phot.sec}
The evolutionary path as outlined in \S\ref{sim_main.sec} is followed
by the GADGET-2 code at a fine time resolution ($\Delta t \sim 10^4$
yr).  At sparser timesteps (70 Myr apart), the positions, masses,
ages, and metallicities of all particles were stored.  It is from
these simulation snapshots that we derive the observed SEDs of the
merger as a function of time.  The applied technique is similar to
that used by Robertson et al. (2007) to translate simulations of the
most massive $z \sim 4-14$ galaxies into observables.

The light a virtual observer would receive from the simulated merger,
is composed of stellar and AGN emission, the latter only contributing
significantly during a brief period of time.  We ignore any
contribution from emission lines produced by the gas content of the
galaxies, possibly contributing on the order of 0.1 mag in the optical
broad-band photometry.  Furthermore, we account for attenuation by
interstellar dust and Lyman forest attenuation by the intervening
medium between the redshifted galaxy and the observer following Madau
(1995).  The combination of these steps, described in this section,
leads to observables that are similar to the real observations that we
model with stellar population synthesis codes.

First, we focus on the computation of intrinsic (i.e., unattenuated)
magnitudes from the stellar component.  Each of the stellar particles
is treated as a single stellar population characterized by its mass,
age, and metallicity.  We choose to use the Salpeter (1955) IMF, as
was done in previous observational work (e.g. F\"{o}rster Schreiber et
al. 2004; Wuyts et al. 2007).  We then interpolate the corresponding
luminosity for each stellar particle from a grid of SSP templates with
different ages and metallicities from the stellar population synthesis
code by Bruzual \& Charlot (2003, hereafter BC03).  Fig.\
\ref{sim.fig}(e) illustrates the evolution of the intrinsic rest-frame
$V$-band luminosity for one of the simulations.

For the AGN emission, we scale a luminosity-dependent template SED by
the bolometric black hole luminosity given by the simulation.  The
template SED was derived from the optically blue (i.e., unreddened)
quasar sample by Richards et al. (2006) with locally attenuated light
being reprocessed as an IR bump longward of $\lambda > 1\ \mu$m.  A
full discussion of the AGN template is presented by Hopkins,
Richards,\& Hernquist (2007).  In most of our analysis, we will
consider the stellar light only.
\S\ref{fixz_AGN.sec} addresses the impact AGN can have on the outcome
of SED modeling during the brief period when its contribution to the
total light is significant.

Galaxies, certainly in their actively star-forming phases, are not
devoid of gas and dust.  It is therefore crucial to account for the
obscuring and reddening effect dust has on the stellar and AGN
emission.  We compute the optical depth along the line of sight toward
each stellar particle.  To do so, we compute the local gas density on
a fine grid derived from the SPH formalism and the particle
distribution (Hopkins et al. 2005a) and integrate out from each
particle along the line of sight to large distance.  The simulations
are based on the GADGET multi-phase ISM model developed by Springel \&
Hernquist (2003).  This model calculates the local mass fraction in
the warm/hot ($T=10^5 - 10^7\ K$, diffuse, partially ionized) and cold
($T=10^3\ K$, in the simulations both associated with molecular clouds
and HI cloud cores) phases of dense gas, assuming pressure equilibrium
between the two phases.  Following Hopkins et al. (2005b), the
attenuation along the line of sight is then derived from the density
of the warm/hot-phase component only.  Hopkins et al. (2005b) found
typically small volume filling factors ($< 0.01$) and cross sections
of the cold-phase ``clumps'', motivating their approach.  The
assumption that most of the lines of sight only pass through the
warm/hot-phase component provides effectively a lower limit on the
optical depths.  We use a gas-to-dust ratio equal to that of the Milky
Way, $(A_B/N_{\rm HI})_{\rm MW} = 8.47 \times 10^{-22}\ {\rm cm}^2$,
with a linear scaling factor accounting for gas metallicities
deviating from solar: $A_B/N_{\rm HI} = (Z/0.02)(A_B/N_{\rm HI})_{\rm
MW}$.  As default, we adopt the Calzetti et al. (2000) attenuation law
for the wavelength dependence of the optical depth.  Changes in the
synthetic photometry when adopting a SMC-like or Milky Way-like
attenuation law from Pei (1992) will be discussed.  The computation of
optical depths was repeated for 30 viewing angles, with directions
uniformly spaced in solid angle $d\cos
\theta d\phi$.  Fig.\ \ref{sim.fig}(f) presents the distribution of
effective visual extinction values (observed minus intrinsic V-band
magnitude) as a function of time since the merger.  The extinction
varies in the following way.  In the early stages typical extinction
values are modest, with the exception for a few lines-of-sights were
the disks are seen edge-on.  The overall extinction along all
lines-of-sight reaches a peak during the merger-triggered starburst
and drops to very low values after star formation has ceased.

Finally, in computing the observer-frame apparent magnitudes, we
redshift the attenuated SED and convolve it with the same set of
filtercurves that we have deep observations for in the Chandra Deep Field
South (CDFS; Wuyts et al. 2008): $B_{435}$, $V_{606}$, $i_{775}$,
$z_{850}$, $J$, $H$, $K_s$, $[3.6 {\rm \mu m}]$, $[4.5 {\rm \mu m}]$,
$[5.8 {\rm \mu m}]$, and $[8.0 {\rm \mu m}]$.  Here, we apply the
depression factors $D_A(z)$ and $D_B(z)$ given by Madau (1995) for the
Lyman forest attenuation of the continuum between Ly$\alpha$ and
Ly$\beta$ and between Ly$\beta$ and the Lyman limit respectively.  The
flux blueward of the Lyman limit ($\lambda_L = 912$\AA) was set to 0,
as is done by the HYPERZ code (v1.1, Bolzonella et al. 2000) that we
use for SED modeling.

In practice, it is computationally more convenient to interpolate the
apparent magnitudes in a given passband for each of the stellar
particles on a precompiled grid of BC03 apparent magnitudes at the
redshift of interest.  The internal dust attenuation is then applied
using the value of the Calzetti et al. (2000) attenuation law at the
effective wavelength for that passband.  We tested that this method,
as opposed to attenuating the full resolution BC03 spectrum and then
convolving with the filtercurve, leads to photometric differences of
at most a few percent.

We note that we never attempt to separate the light into the
contribution from the two progenitors.  Instead, we always study the
total photometry, as if the merging system were unresolved.

In future studies, we will employ full-fledged radiative transfer
codes such as SUNRISE (Jonsson 2006) to translate the simulation
snapshot information to observables.  SUNRISE calculates scattering,
absorption, and reemission of light passing through the cold and/or
hot phase of the ISM, with the optional inclusion of a subgrid model
that accounts for the obscuring effects of birth clouds in which young
star-forming regions are embedded (Groves et al. 2008).  Reemission by
dust only influences the spectral shape longward of the wavelength
bands used in our SED modeling.  Preliminary analysis of SUNRISE SEDs
indicates that the other effects might lead to a stronger extinction
during the star-forming phases, while leaving the spheroid photometry
unchanged with respect to the photometry computed in this paper.  It
will be interesting, in the light of the findings presented here, to
investigate the shape of the effective attenuation law obtained with
such a more sophisticated code, and its dependence on the parameters
of the radiative transfer.  For the sake of this paper, our
line-of-sight photometry has the advantage of being transparent and
simple.

\subsection {The color evolution of merging galaxies}
\label{stamps.sec}

\begin {figure*} [htbp]
\centering
\plotone{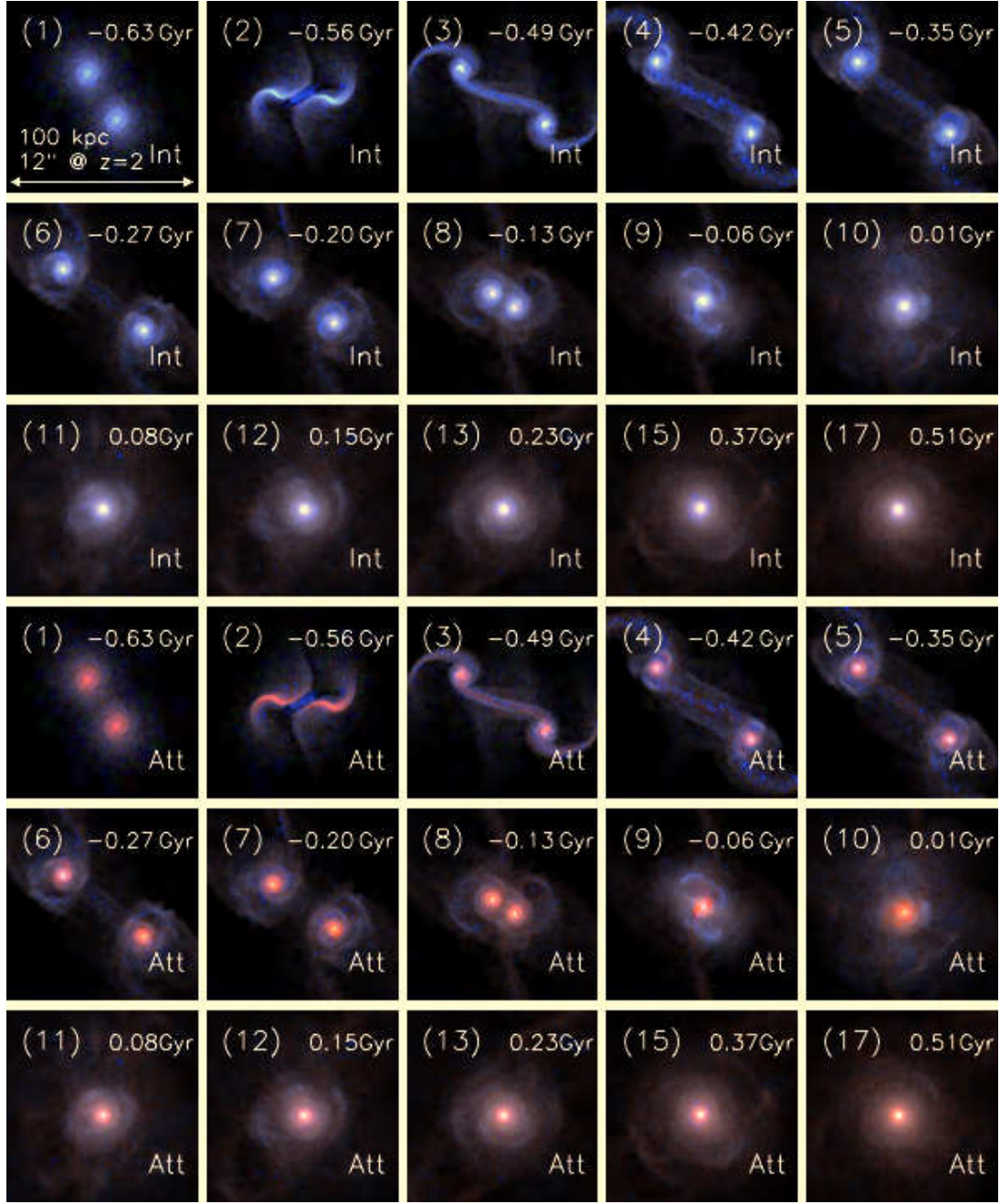} 
\caption{Three-color
 postage stamps (rest-frame $U$ ({\it blue}), $V$ ({\it green}), $J$
 ({\it red})) of a typical merger simulation.  The top three rows
 illustrate the intrinsic evolution, ignoring extinction.  The bottom
 three rows show the evolution when taking into account attenuation by
 dust.  Significant changes in color are observed over time, and between
 intrinsic and attenuated images.
\label {stamps.fig}
}
\end {figure*}

With the synthetic photometry at hand, we are now able to follow the
color evolution of simulated galaxies throughout the merger event.  We
illustrate the spatially resolved color evolution for a face-on view
of a typical simulation in Fig.\ \ref{stamps.fig}.  This particular
simulation has an initial gas fraction of 80\% and final stellar mass
of $1.2 \times 10^{11}\ M_{\sun}$.  The 3-color images are composed of
rest-frame $U$ ({\it blue}), $V$ ({\it green}), and $J$ ({\it red}).
On each postage stamp, we indicate the snapshot number, time since the
merger, and whether the intrinsic (Int) or attenuated (Att) colors are
plotted.  The interval between consecutive snapshots is 70 Myr.  The
color distribution is clearly not homogeneous throughout the merger
evolution.  Instead, color gradients are present, with the nuclear
regions being intrinsically bluer due to new star formation, but
effectively redder than the surrounding material due to the presence
of dust.  We return to this point in an analysis of observed versus
simulated galaxy colors by Wuyts et al. (2009b).  In this paper, we
restrict ourselves to the study of integrated colors.

\begin {figure*} [htbp]
\centering
\epsscale{0.8}\plotone{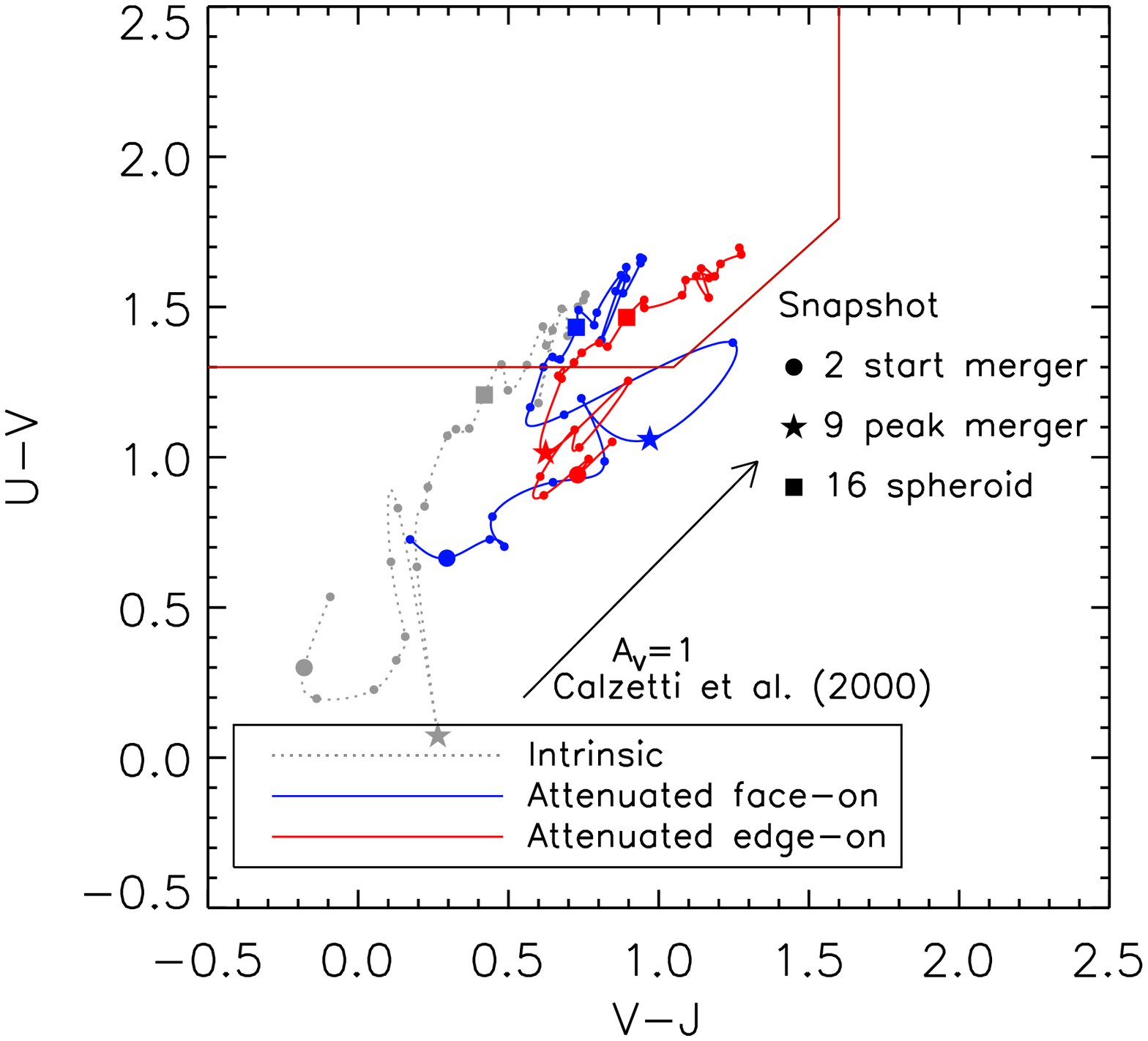} 
\epsscale{1}
\caption{Evolutionary tracks in a rest-frame $U-V$ versus $V-J$
color-color diagram for the typical merger simulation shown in Fig.\
\ref{stamps.fig}.  We illustrate the intrinsic color evolution as well
as the evolution of attenuated colors with time as we view the merging
pair face- and edge-on.  Snapshots 70 Myr apart are marked with small
dots.  Key phases corresponding to the numbers indicated in Fig.\
\ref{stamps.fig} and Fig.\ \ref{prefextinct.fig} are indicated with
large symbols.  Time since (or before) then actual moment of merging
is indicated in the top right corner of each panel.  Colors can vary
significantly at a given time (by $\sim 0.5$ mag) as we view the
merging system from different lines of sight.  The starburst occuring
during the merger is very dusty, leading to redder colors at that
stage.
\label{coltracks.fig}
}
\end {figure*}
The corresponding evolutionary tracks in integrated rest-frame $U-V$
versus $V-J$ color-color space are presented in Fig.\
\ref{coltracks.fig}.  Labb\'{e} et al. (2005) first introduced the 
observed-frame equivalent of this diagram to illustrate the wide range
of galaxy types at high redshift ranging from blue, relatively
unobscured star-forming systems to dusty starbursts to quiescent red
galaxies.  The color criterion proposed by Labb\'{e} et al. (in
preparation) to select galaxies of quiescent nature is plotted as the
red wedge in Fig.\ \ref{coltracks.fig}.

Considering the intrinsic (unattenuated) colors, the evolutionary
track has a direct relation to the mean light-weighted age (and
depends to a lesser degree on the gradual increase in stellar
metallicities).  It bends down to bluer $U-V$ as new star formation is
triggered during first passage of the progenitors, and again during
the shorter-lived nuclear starburst phase.  Eventually, the color
track leads to the region in color-color space where also quiescent
observed galaxies reside.  In reality, the path taken in color-color
space will not merely depend on the star formation history.  Taking
into account attenuation by dust, the integrated colors get redder.
This is particularly true during the phases of active star formation.
We note that the precise shape of the color track depends not only on
initial conditions (gas fraction, mass, disk orientations, ...) but
also on viewing angle.  In fact, as illustrated in Fig.\
\ref{coltracks.fig}, color differences between different viewing angles at 
a given time can be of similar size as color changes between different
times for a given viewing angle.  At later times, the role of dust is
minimal and the remnant converges to colors typical for quiescent
galaxies (i.e., within the red wedge).

\subsection {The colors and SEDs of simulated and observed galaxies}
\label{sim_color.sec}
\begin {figure} [htbp]
\centering
\plotone{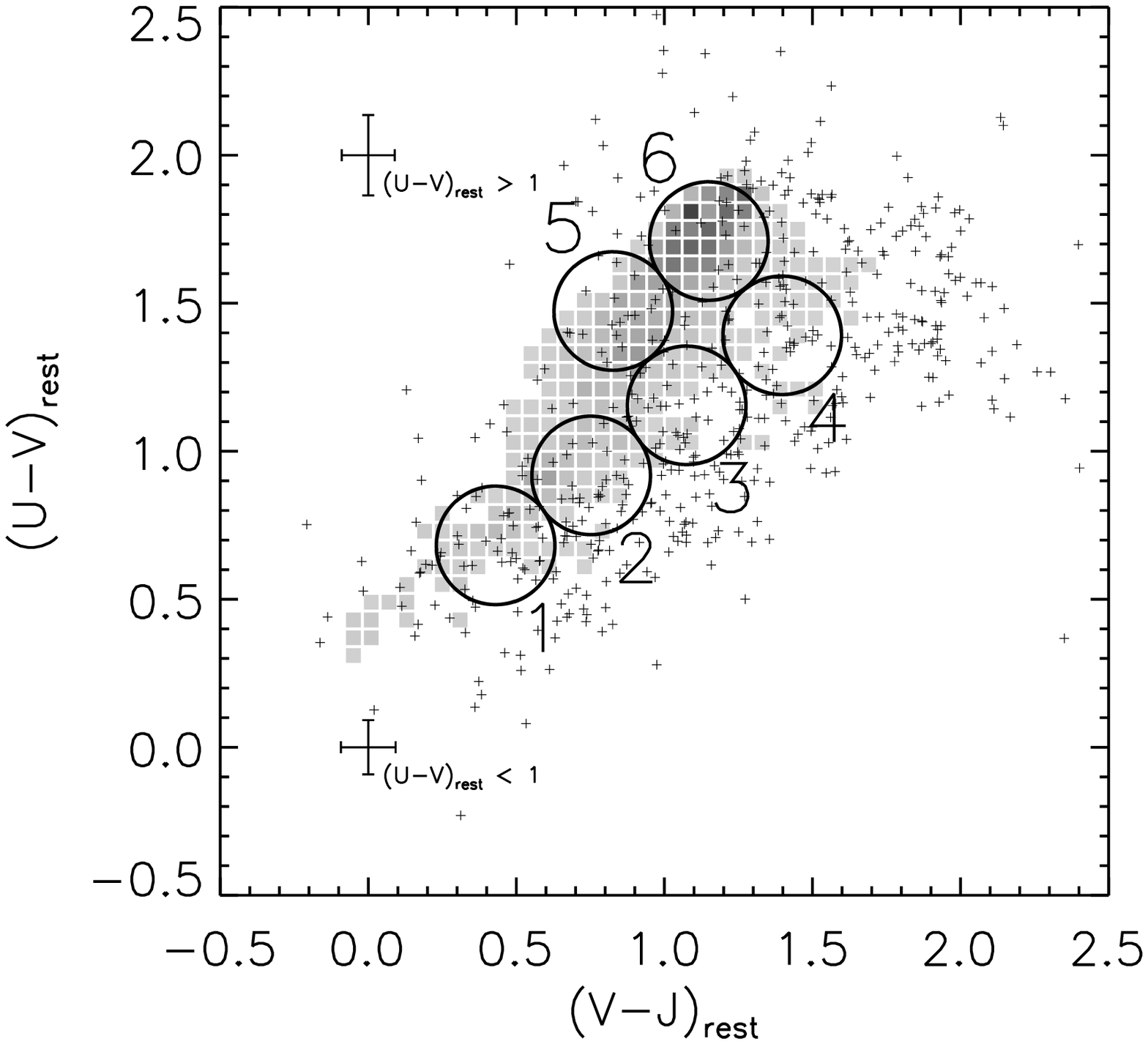} 
\caption{Rest-frame $U-V$ versus $V-J$ color-color diagram showing the
binned color distribution of the simulations seen under different
viewing angles and at different epochs.  Overplotted ({\it plus
symbols}) are the rest-frame colors of observed galaxies with $M_* >
1.4 \times 10^{10}\ M_{\sun}$ at $1.5<z<3$ in the HDFS, \1054, and the
CDFS.  Observed galaxies with matching colors are found for all
simulated galaxies.  The reddest observed sources in $U-V$ and $V-J$
are not reproduced by the considered set of simulations.  Rest-frame
SEDs for sources in regions 1-6 are displayed in Fig.\ \ref{SED.fig}.
\label {colcol.fig}
}
\end {figure}
\begin {figure*} [htbp]
\centering
\plotone{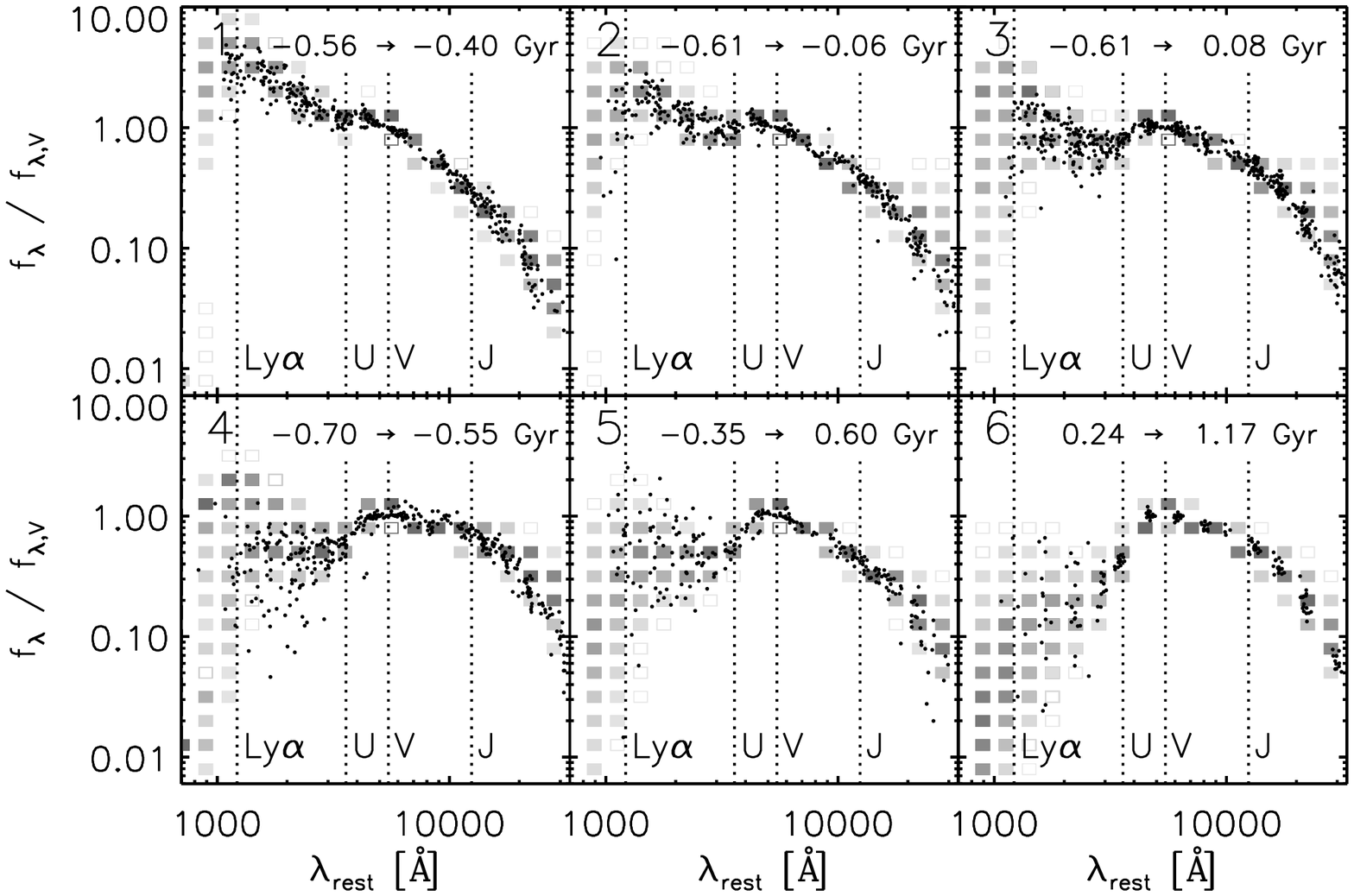} 
\caption{Rest-frame SEDs of simulated galaxies in regions 1-6 of Fig.\
\ref{colcol.fig}.  A darker intensity of the binned representation
indicates a larger density of simulated galaxies with that flux level.
In each panel, we give the central 68\% interval of the distribution
of times before or since the merger for the simulation snapshots with
photometry in the respective region.  Overplotted ({\it black dots})
are the rest-frame broad-band SEDs of observed $1.5<z<3$ galaxies with
$M_* > 1.4 \times 10^{10}\ M_{\sun}$ in the HDFS, \1054, and the CDFS.
A general agreement between observed and simulated spectral shapes is
observed, also outside the $U$-to-$J$ range where the correspondence
was not imposed by selection.
\label {SED.fig}
}
\end {figure*}

Prior to analyzing the performance of our SED modeling procedure, it
is important to confirm that the simulated galaxies have spectral
shapes resembling those of real high-redshift galaxies in observed
deep fields, thus validating their role as test objects.  To this end,
we indicate the binned color distribution of simulated galaxies,
viewed from different angles and during different phases of their
evolution, in a rest-frame $U-V$ versus $V-J$ color-color diagram
(Fig.\ \ref{colcol.fig}).  Plus symbols show the location of observed
galaxies in the HDFS (Labb\'{e} et al. 2003), \1054 (F\"{o}rster
Schreiber et al. 2006), and the CDFS (Wuyts et al. 2008) selected by
their photometric redshift (or spectroscopic when available) to lie in
the same redshift range ($1.5<z<3.0$).  We also applied a stellar mass
cut at $M_* > 1.4 \times 10^{10}\ M_{\sun}$ for the observed sample;
the lowest initial stellar mass for the considered set of simulations.
Here, we do not attempt to statistically compare the two samples.  The
abundances of different types of galaxies as predicted from the
simulations will be addressed by Wuyts et al. (2009b).  For our
current purpose of analyzing the effects of star formation history,
dust, metallicity and AGN on SED modeling, it is sufficient to note
that there is a large overlap between the color-color space spanned by
the simulated and observed galaxies.  Every simulated galaxy
considered in this paper has counterparts in the real universe with
similar rest-frame optical and optical-to-NIR colors.  However, the
observed distribution extends to colors that are redder by a few 0.1
mag, both in $U-V$ and in $V-J$, than the simulations considered here.
Given the one-sided nature of the different color spread, it is
unlikely that this can be attributed to photometric uncertainties
alone.  Therefore, we caution that our results may not necessarily be
extrapolated to the reddest galaxies present in observed samples.
Wuyts et al. (2009b) discuss several possible origins of the color
discrepancy, particularly in $V-J$, ranging from the method to compute
model photometry to differences between the evolutionary history of
high-redshift galaxies in the real universe and in our merger
simulations.  There we analyze, for example, the presence of color
gradients, and the dependence of the color distribution on the adopted
attenuation law and stellar population synthesis code.  Other models
than BC03 that include a more significant contribution of the
thermally pulsing asymptotic giant branch (particularly in the
near-IR) result in a color distribution that is shifted towards redder
$V-J$ colors (Maraston 2005; Maraston et al. 2006).

To ascertain that observed and simulated galaxies with similar $U-V$ and
$V-J$ colors have similar SEDs over the whole spectral range, Fig.\
\ref{SED.fig} presents the rest-frame SEDs of objects in region 1-6 of
Fig.\ \ref{colcol.fig}.  Again, the binned distribution represents the
simulations, with darker grayscales indicating a larger number of objects.
Overplotted with black dots is the broad-band photometry of our
observed sample within the same region of color-color space, placed at
the respective rest-frame wavelength.  The SEDs are normalized to the
rest-frame $V$-band.  By selection, the observed and simulated
photometry matches well at rest-frame $U$ and $J$.  In between the $UVJ$
filters, and outside the U-to-J range, no correspondence was imposed.
The fact that the UV spectral shape and the NIR tail of the observed
and simulated SEDs show a general agreement, is encouraging.  We
conclude that the simulated photometry can be adopted as a realistic
input to our SED modeling procedure.  The results of our analysis will
be applicable to observed galaxies with similar colors.

\section {SED modeling: methodology}
\label {SED_modeling.sec}
%
We characterize physical parameters such as stellar mass, stellar age,
 dust attenuation, and SFR by matching the observed-frame broad-band
photometry to synthetic templates from the stellar population
synthesis code by BC03.  We use the HYPERZ stellar population fitting
code, version 1.1 (Bolzonella et al. 2000) and fit the SED twice:
first fixing the redshift to the true value (for which we computed the
simulated photometry), next adopting a photometric redshift estimate
obtained from the EAZY version 1.0 photometric redshift code (Brammer
et al. in preparation).  In each case, the full $B_{435}$-to-8 $\mu$m SED,
sampled with identical passbands as available for the GOODS-CDFS
($B_{435}$, $V_{606}$, $i_{775}$, $z_{850}$, $J$, $H$, $K_s$, [3.6
$\mu$m], [4.5 $\mu$m], [5.8 $\mu$m], [8.0 $\mu$m]), was fed to
HYPERZ.  Random photometric uncertainties were assigned as to mimic
real observations in the CDFS, and fluxes in each band were perturbed
accordingly.  Precisely, for each of the 43200 SEDs corresponding to a
simulated galaxy observed during a certain phase of its evolution,
placed at a certain redshift, and observed along a certain
line-of-sight, we compute 5 realizations of the SED by introducing a
gaussian perturbation in all bands with the amplitude derived from the
depth of GOODS-CDFS observations in the respective bands.  A minimum
error of 0.08 mag was adopted for all bands, preventing small errors
from dominating the fit.

As in Wuyts et al. (2007), we selected the least $\chi^2$ solution out
of three possible star formation histories: a single stellar
population (SSP) without dust, a constant star formation (CSF) history
with dust ($A_V$ varying from 0 to 4 in steps of 0.2), and an
exponentially declining star formation history with a fixed {\it
e}-folding timescale of 300 Myr ($\tau_{300}$) and identical range of
$A_V$ values.  Ages were constrained to be larger than 50 Myr, to
prevent improbably young ages, and smaller than the age of the
universe at the observed redshift.  We used a Calzetti et al. (2000)
attenuation law, and assumed solar metallicity and a Salpeter (1955)
IMF with lower and upper mass cut-offs $0.1M_{\sun}$ and
$100M_{\sun}$.  In recent literature, several authors have used an IMF
with fewer low mass stars, such as presented by Kroupa (2001).  We
note that the choice of IMF is not the focus of this study, and our
results remain valid as long as we use the same IMF to compute the
synthetic photometry as to model the resulting SEDs.  Likewise, we use
for consistency the same stellar population synthesis code (BC03) to
compute and to fit the synthetic SEDs.

When refering to the age derived from SED modeling, we mean the
mass-weighted age obtained by integrating over the different ages of
SSPs that build up the best-fit SFH, weighted with their mass fraction
taking into account mass loss over time.  This measure aims to
quantify the age of the bulk of the stars.  For an SSP, it equals the
time passed since the single burst.  For a CSF history, it is
essentially half the time passed since the onset of star formation.
The $\tau_{300}$ SFH represents an intermediate case.


\section {Results from SED modeling at fixed redshift}
\label {results_fixz.sec}

First, in \S\ref{overall.sec}, we present the overall performance of
the standard SED modeling applied to the 'full' photometry, taking
into account the effects of both attenuation, metallicity, and AGN
contribution as realistically as possible.  Second, we discuss the
impact from different aspects influencing the colors and luminosities
of galaxies separately.  In order to isolate effects from star
formation history (\S\ref{fixz_SFH.sec}), dust attenuation
(\S\ref{fixz_Av.sec}), metallicity variations (\S\ref{fixz_Z.sec}),
and AGN contribution (\S\ref{fixz_AGN.sec}), we computed the
photometry for each snapshot with and without attenuation, with and
without AGN contribution, and using solar metallicity, or the
metallicity as computed by the simulation for each stellar particle.
To each of these sets of SEDs, we applied the modeling described in
\S\ref{SED_modeling.sec}.  This approach enables us to reconstruct the
analysis step by step, adding one aspect at a time.  Finally, we
discuss what our analysis means for galaxies with different rest-frame
colors (\S\ref{lessons.sec}).

\subsection {Overall performance}
\label {overall.sec}

\begin {figure*} [htbp]
\centering
\plottwo{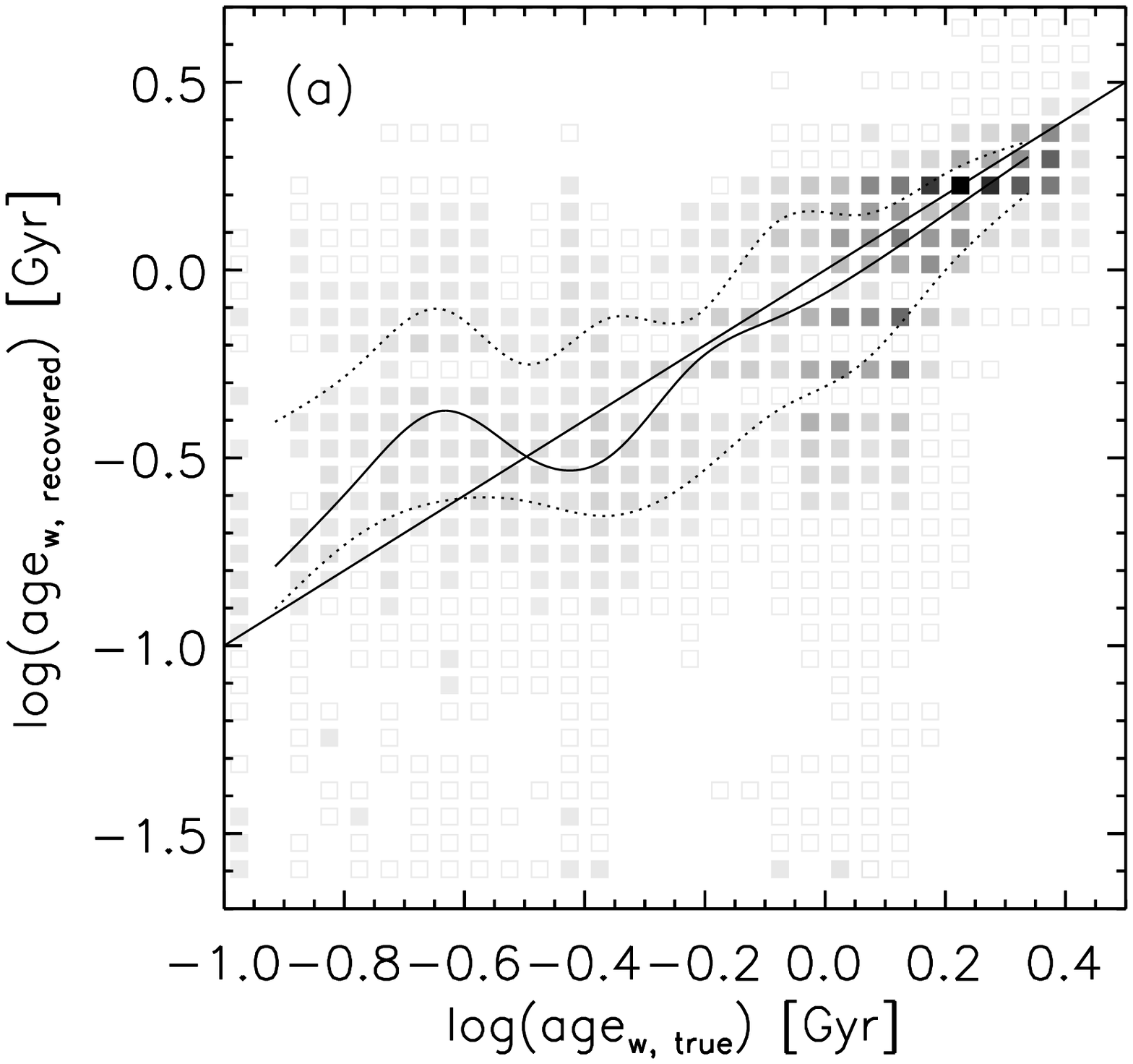}{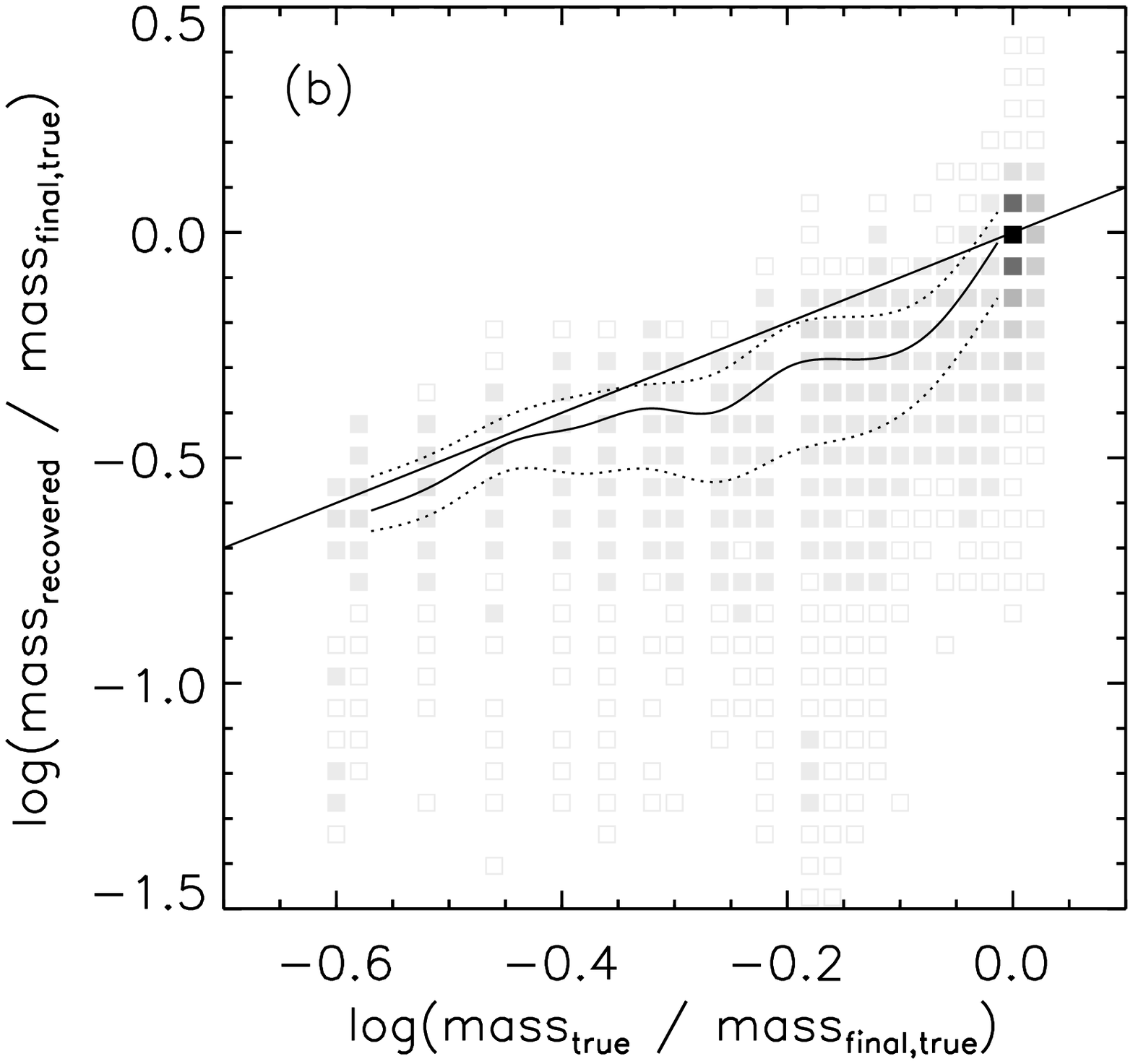}
\plottwo{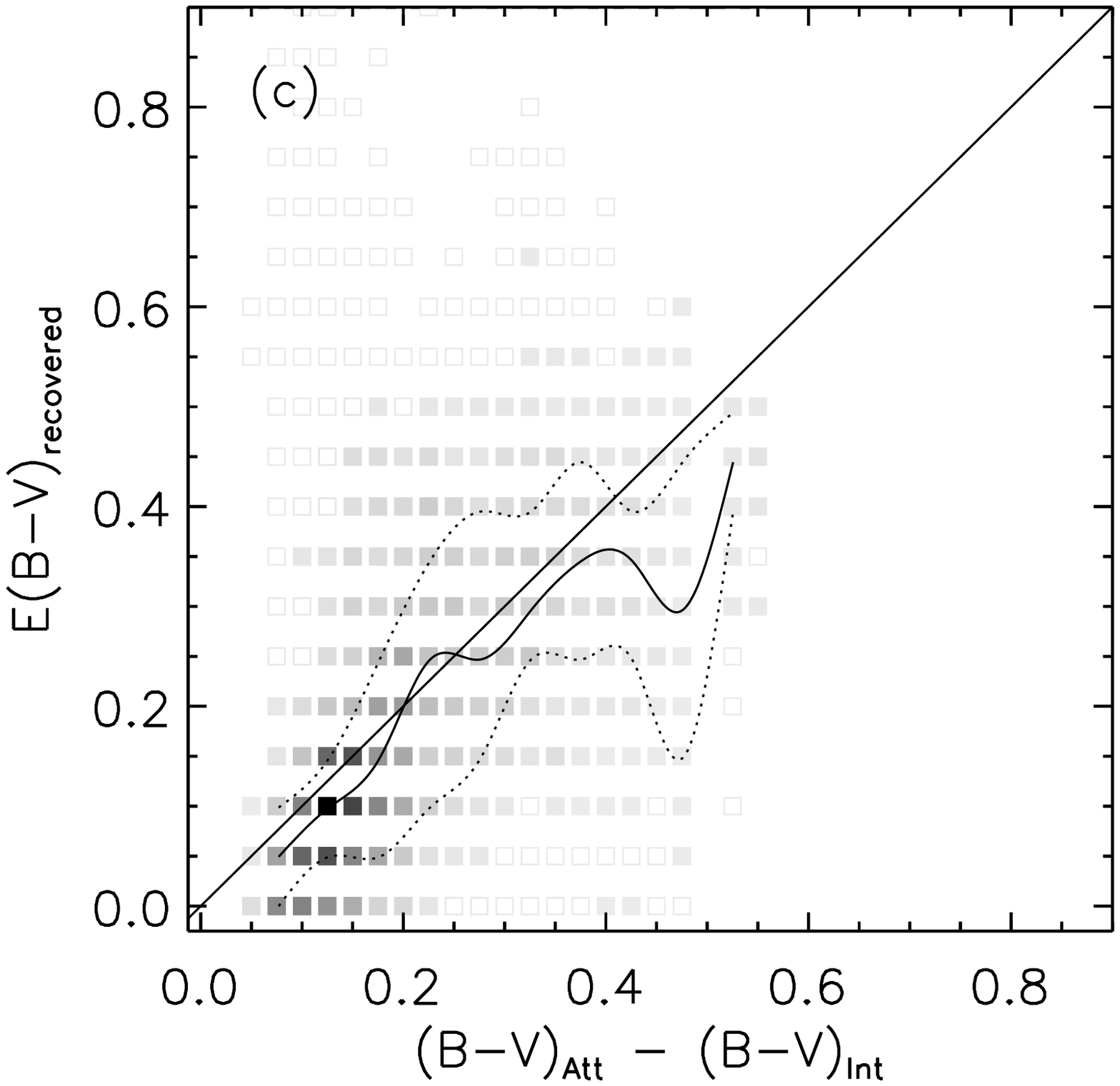}{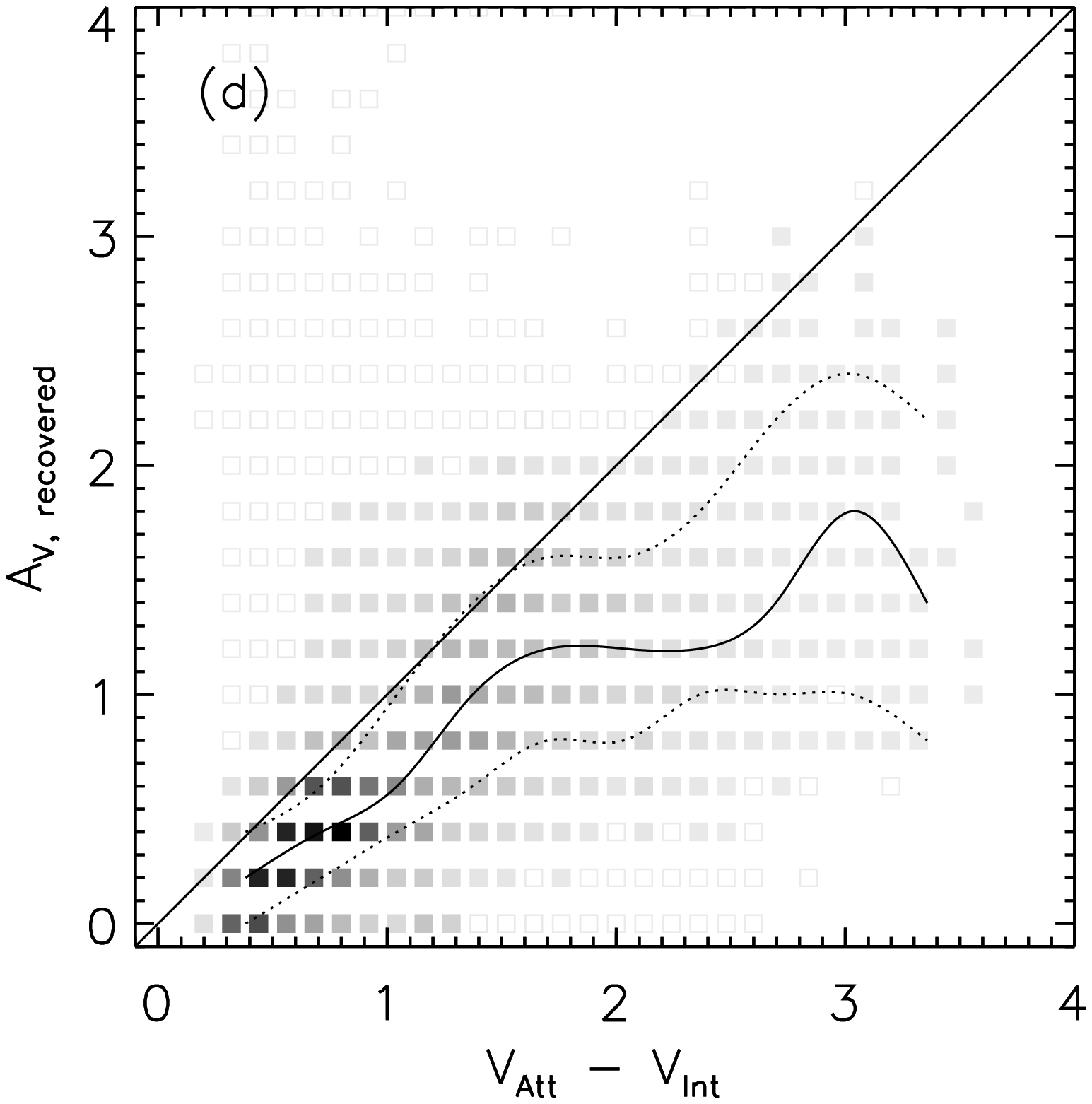}
\epsscale{0.5}\plotone{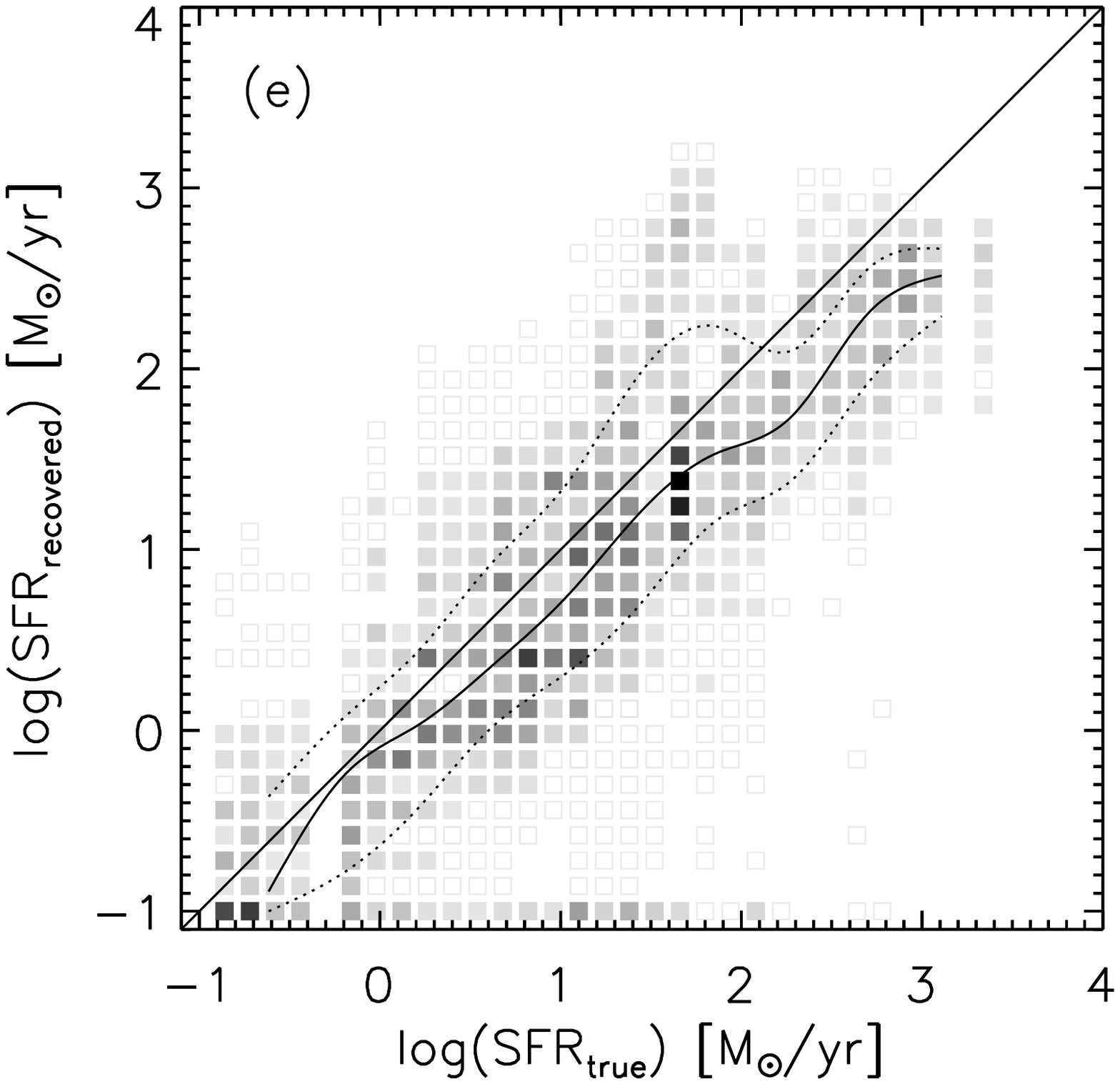}
\epsscale{1}
\caption{Overall performance of the SED modeling.  Recovered versus true (a) mass-weighted stellar age, (b) ratio of current to final
stellar mass, (c) effective reddening (i.e., attenuated minus
intrinsic $B-V$ color), (d) effective visual extinction (i.e.,
attenuated minus intrinsic V-band magnitude), and (e) star formation
rate.  The SED modeling was performed on the total (stellar+AGN)
attenuated photometry corresponding to simulations with a range of
masses, starting with 40\% and 80\% gas fractions, and seen at
different timesteps and under different viewing angles.  The solid
curves indicate the median and dotted curves comprise the central 68\%
of the distribution.  The total visual extinction $A_V$ is the least
constrained of the studied parameters.  In particular for heavily
extincted galaxies the $A_V$ is greatly underestimated.
\label {overall.fig}
}
\end {figure*}
\begin {figure*} [htbp]
\centering
\plottwo{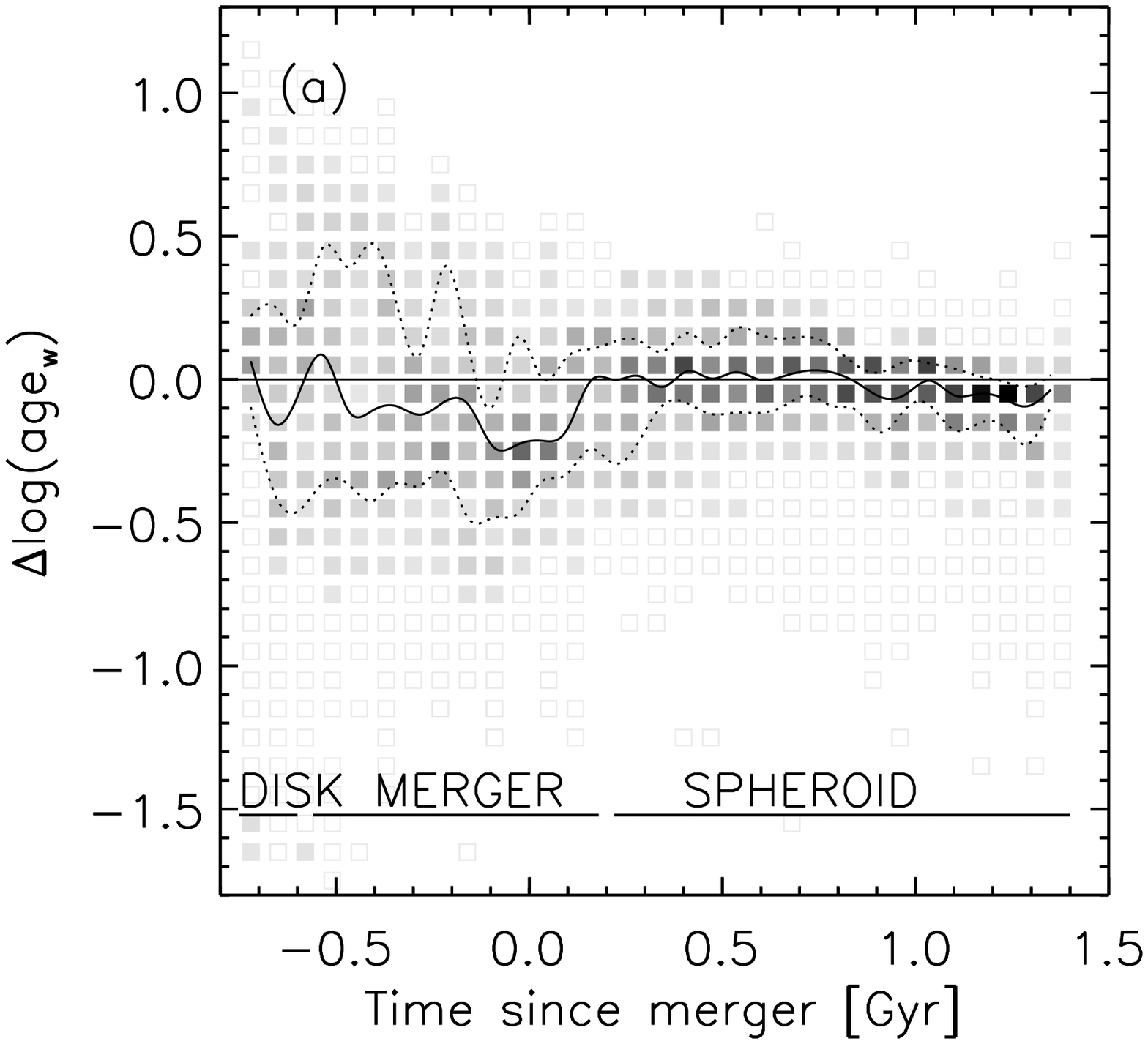}{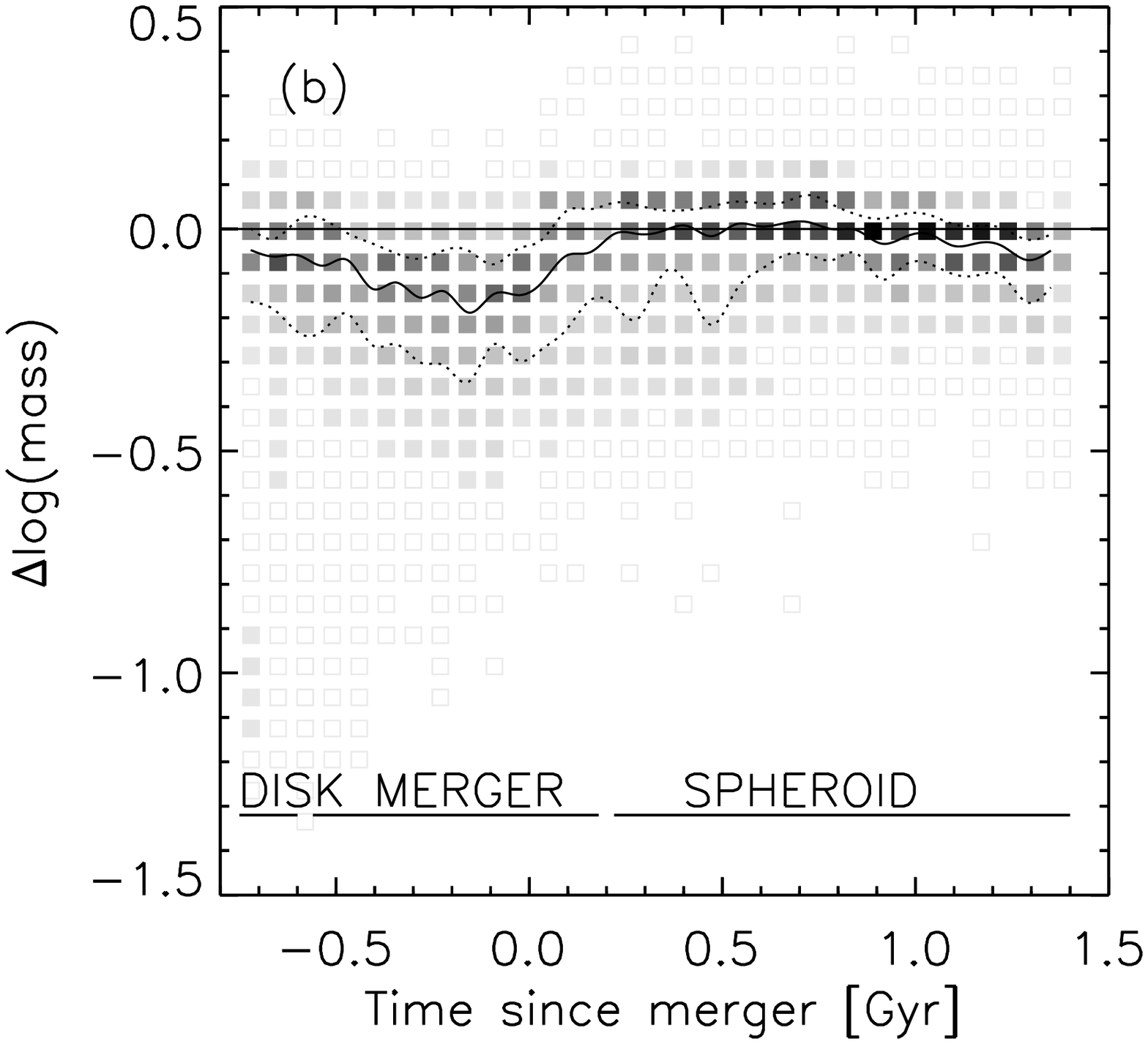}
\plottwo{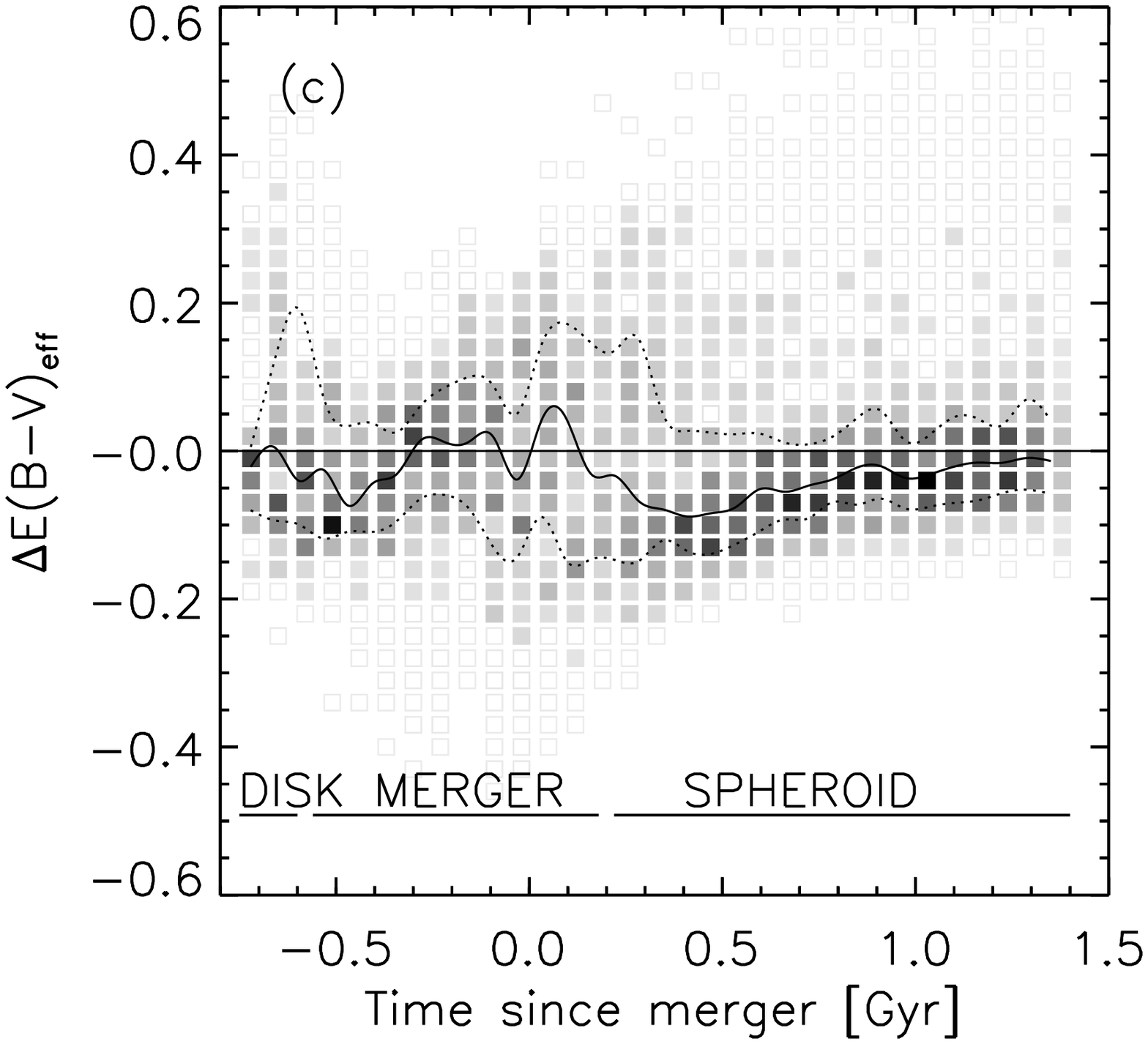}{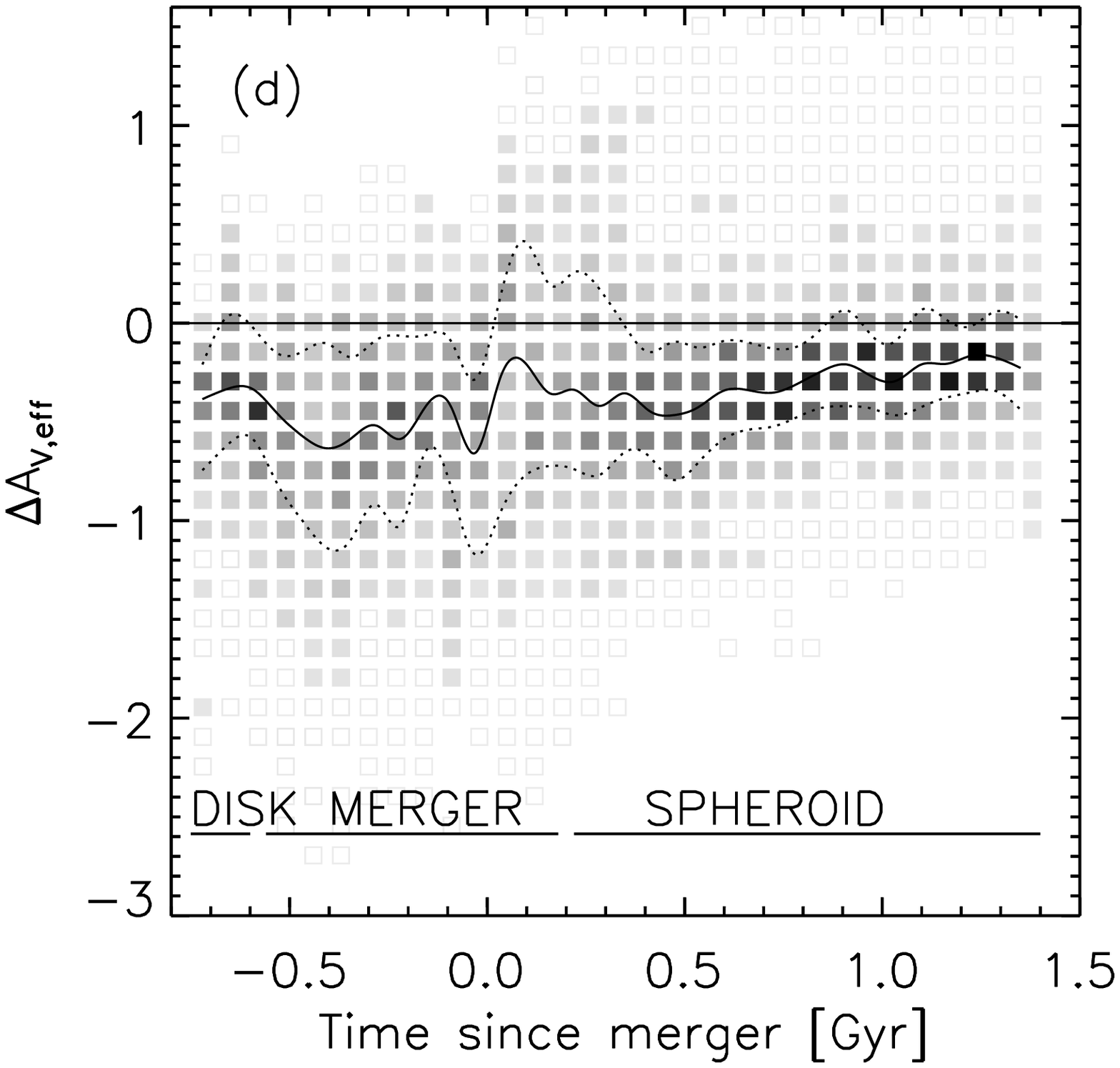}
\epsscale{0.5}\plotone{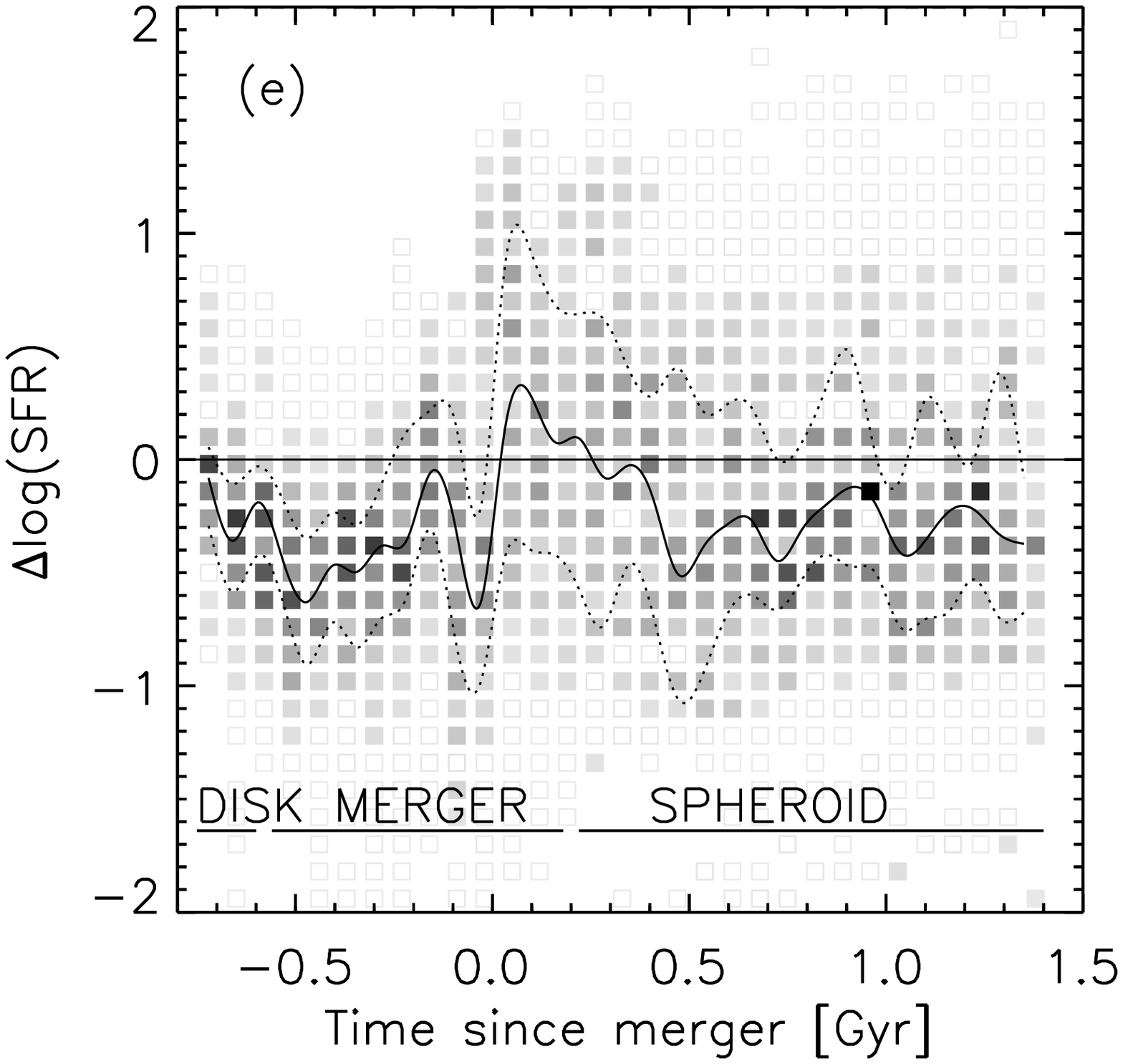}
\epsscale{1}
\caption{Overall performance of the SED modeling.  The difference
between estimated and true (a) mass-weighted age, (b) stellar mass,
(c) effective reddening, (d) effective visual extinction, and (e) star
formation rate as a function of time since the merger.  The SED
modeling was performed on the total (stellar+AGN) attenuated
photometry.  The solid curves indicate the median and dotted lines
comprise the central 68\% of the distribution.  The properties of
merger remnants are well reproduced.  The results for star-forming
galaxies, especially for those in the phase of merging, show
underestimates in both age, extinction, star formation rate, and mass.
\label {overalltime.fig}
}
\end {figure*}

The combined effects of mismatch between true and template SFH,
attenuation by dust, metallicity variations and AGN activity on our
ability to characterize the stellar population of a galaxy are
summarized in Fig.\ \ref{overall.fig}.  Here, we plot the best
estimate from SED modeling versus the true\footnote[1]{Hereafter, we use
'true' when referring to the value of a stellar population property as
computed using information on all SPH particles in the simulation.
Quantities derived by fitting templates to integrated SEDs are
referred to as 'estimates'.  While 'true' mass and mass-weighted age
can be read directly from the simulation, we note that parameters as
'true' reddening and visual extinction depend on the adopted stellar
population synthesis code and details of the radiative transfer, which
themselves may have biases.} value of the considered stellar
population property.  We bin the distribution of points for different
initial conditions, timesteps and lines-of-sight.  Darker intensities
represent a higher density in the bin.  The solid curve represents the
median of the distribution and the dotted curves mark the central 68\%
interval.

The youngest ages in Fig.\ \ref{overall.fig}(a) correspond to the
early stages of simulations with an initial gas fraction of 80\%.  The
stars present at the start of these simulations were assigned young
ages and low metallicities (see \S\ref{sim_main.sec}).  In
\S\ref{fixz_Z.sec}, we describe how modeling sub-solar metallicity
populations with solar metallicity templates can lead to age
overestimates.  Since the lower gas fraction runs start with higher
stellar ages and metallicities, they do not show this trend.  Fig.\
\ref{overall.fig}(b) compares the recovered and true stellar mass,
normalized to the final stellar mass of the simulation.  At low $M /
M_{\rm final}$ ratios, i.e., at the start of the high gas fraction
simulations, the mass estimates agree well with the true values.
Here, the reason is that biases in age and extinction average out.  On
the one hand, the overestimated age leads to an overestimate of the
intrinsic mass-to-light ratio ($\left [ \frac{M}{L_{\rm Int}} \right
]_{\rm estimate} > \left [ \frac{M}{L_{\rm Int}} \right ]_{\rm
true}$).  On the other hand, the amount of extinction tends to be
underestimated ($\left [ \frac{L_{\rm Int}}{L_{\rm Att}} \right ]
_{\rm estimate} < \left [ \frac{L_{\rm Int}}{L_{\rm Att}} \right ]
_{\rm true}$).  As a result, the mass estimate for this phase is
relatively robust:
\begin {eqnarray}
M_{\rm estimate} & = & L_{\rm Att} \left [ \frac{L_{\rm Int}}{L_{\rm Att}} \right ] _{\rm estimate} \left [\frac{M}{L_{\rm Int}} \right ]_{\rm estimate} \\
                 & \approx & L_{\rm Att} \left [ \frac{L_{\rm Int}}{L_{\rm Att}} \right ] _{\rm true} \left [\frac{M}{L_{\rm Int}} \right ]_{\rm true} \\
                 & = & M_{\rm true}
\end {eqnarray}
where $L_{\rm Int}$ is the intrinsic and $L_{\rm Att}$ is the
attenuated (i.e., observed) luminosity.  The largest systematic mass
underestimates occur at intermediate $M / M_{\rm final}$, during the
merger-triggered star-forming phases of the simulation.  This can be
explained in terms of a mismatch between the true SFH and the SFH of
the best-fitting template (see \S\ref{fixz_SFH.sec}), in combination
with an underestimate of the extinction (see \S\ref{fixz_Av.sec}).
Briefly, such a SFH mismatch leads to age underestimates, hence $\frac
{M}{L_{\rm Int}}$ underestimates, and, added to an insufficient dust
correction, to mass underestimates.  Finally, the correspondence is
best at $\log (M/M_{\rm final}) \sim 0$, where the merger remnants
reside.

The reddening (Fig.\ \ref{overall.fig}(c)) is overall well
reproduced.  Only at the highest reddening levels, the agreement
deteriorates.  The latter correspond to the times when and viewing
angles under which the effect of increased extinction toward young
stars is maximal, as will be explained in \S\ref{fixz_Av.sec}.  As
opposed to the reddening, however, the extinction (Fig.\
\ref{overall.fig}(d)) shows large
systematic underestimates.  Using
\begin {equation}
R_V = \frac{A_V}{E(B-V)}=4.05
\label {Rv.eq}
\end {equation}
to translate the selective absorption $E(B-V)$ into a total
visual absorption $A_V$ results in an underestimate over the whole
range of $A_V$ values, particularly during the highly obscured phases.
An observer is limited by the light that he/she receives.  In
\S\ref{fixz_Av.sec}, we discuss in depth how the sum of emitting
sources that are each attenuated according to Calzetti et al. (2000)
does not follow that same reddening law.

Finally, since the reliability of SFR indicators from X-rays over
optical and IR to radio wavelengths has been topic of much recent
debate (e.g., Reddy et al. 2006; Papovich et al. 2006; 2007; Daddi et
al. 2007a), it is interesting to investigate its recovery by modeling
the optical-to-8 $\mu$m SEDs.  This is presented in Fig.\
\ref{overall.fig}(e).  At low and intermediate SFRs ($SFR_{\rm true} <
100\ M_{\sun}/{\rm yr}$), we find a satisfying agreement, with a
median offset that increases slightly towards higher SFRs.  For more
than 84\% of the galaxies with $SFR_{\rm true} > 100\ M_{\sun}/{\rm
yr}$, we underestimate the SFR, typically by factor of 3.  These
systems often suffer strong dust obscuration, and therefore this bias
in SFR directly relates to the large underestimates of extinction at
the high $A_V$ end in Fig.\ \ref{overall.fig}(d).

In order to quantify the performance of the SED modeling in estimating
stellar mass-weighted ages, we define $\Delta log(age_w)$ as
$log(age_{w,{\rm recovered}}) - log(age_{w,{\rm true}})$.  Similar
definitions are used to quantify the offset in mass, reddening and
extinction, always indicating an underestimate with a negative value
of $\Delta = parameter_{recovered} - parameter_{true}$.  Fig.\
\ref{overalltime.fig} presents the performance of the SED modeling on
the full photometry (including dust, metallicity variations, and AGN),
expressed by the $\Delta$ values, as a function of time since the
merger.  We adopt the same binned plot style as in Fig.\
\ref{overall.fig}, with darker intensities meaning a higher density in
the bin.  Open boxes contain less than 1\% of the total number of SEDs
at that timestep.

We quantify the performance of the SED modeling separately for
galaxies in the 'disk', 'merger', and 'spheroid' regime by computing
the median and the central 68\% interval of the distribution of
$\Delta$ values for all simulation snapshots (under a range of viewing
angles) in that phase.  We find $\Delta \log age_{w,{\rm disk}} =
+0.03^{+0.19}_{-0.42}$, $\Delta \log age_{w,{\rm merger}} =
-0.12^{+0.40}_{-0.26}$, and $\Delta \log age_{w,{\rm spheroid}} =
-0.03^{+0.12}_{-0.14}$.  The underestimate and scatter is largest for
the phases of merger-triggered star formation.  These are the
statistics for low and high gas fraction runs combined.  Considering
simulations with an initial gas fraction of 40\% separately, the age
underestimates during the star-forming phases increase: $\Delta \log
age_{w, {\rm disk, 40\%}} = -0.13^{+0.25}_{-0.34}$ and $\Delta \log
age_{w, {\rm merger, 40\%}} = -0.30^{+0.08}_{-0.11}$.  The $f_{\rm
gas} = 80\%$ runs instead show age overestimates until $\sim 200$ Myr
before the merger, mainly due to their lower initial metallicity (see
\S\ref{fixz_Z.sec}).  During the spheroid phase, ages are recovered
similarly well for the low and high gas fraction runs.  The recovery
of other stellar population properties does not depend strongly on gas
fraction unless stated otherwise.

Fig.\ \ref{overalltime.fig}(b) shows the same scenario as described
for Fig.\ \ref{overall.fig}(b).  Namely, mass estimates are robust
once the merger remnant phase is reached ($\Delta \log M_{\rm
spheroid} = -0.02^{+0.06}_{-0.11}$), but more significant mass
underestimates occur for star-forming disks ($\Delta \log M_{\rm disk}
= -0.06^{+0.06}_{-0.14}$), and especially during phases of
merger-enhanced star formation ($\Delta \log M_{\rm merger} =
-0.13^{+0.10}_{-0.14}$).

Averaged over all SEDs for a given time with respect to the merger
(representing different viewing angles, masses, and gas fractions),
our reddening estimates are well behaved: $\Delta E(B-V)_{\rm disk} =
-0.02^{+0.13}_{-0.07}$, $\Delta E(B-V)_{\rm merger} =
-0.02^{+0.08}_{-0.08}$, and $\Delta E(B-V)_{\rm spheroid} =
-0.03^{+0.11}_{-0.07}$.  However, as noted already in Fig.\
\ref{overall.fig}(d), the bias in estimated visual extinction is often
much larger, particularly during the star-forming (disk and merger)
phases: $\Delta A_{V,{\rm disk}} = -0.35^{+0.29}_{-0.34}$, $\Delta
A_{V,{\rm merger}} = -0.54^{+0.40}_{-0.46}$, and $\Delta A_{V,{\rm
spheroid}} = -0.29^{+0.32}_{-0.30}$.  Part of the scatter quoted for
$\Delta E(B-V)$ and $\Delta A_V$ results from differences between the
low and high gas fraction runs.  The three key differences, with
reference to where their impact is detailed, are the following.
First, the high gas fraction runs start out with lower metallicities
(\S\ref{fixz_Z.sec}).  Second, the low gas fraction runs start with
older stars.  The larger age range makes the effect of age-dependent
extinction more pronounced (\S\ref{fixz_prefextinc.sec}).  Finally,
the high gas fraction runs feed relatively more gas to the SMBH,
resulting in a larger AGN contribution to the integrated light during
final coalescence (\S\ref{fixz_AGN.sec}).  Since the obscuration
during the merger phase tends to be to the newly formed stars mainly,
the underestimated extinction leads to an underestimate of the SFR.
Consequently, the evolution of $\Delta \log SFR$ with time is as a
mirrored version of the star formation history, showing the largest
underestimates during first passage and final coalescence.  We find
$\Delta \log SFR_{\rm disk} = -0.22^{+0.23}_{-0.28}$, $\Delta \log
SFR_{\rm merger} = -0.44^{+0.32}_{-0.31}$, and $\Delta \log SFR_{\rm
spheroid} = -0.23^{+0.62}_{-0.47}$.  Note that the offset for the
spheroid phase merely means that we are unable to distinguish low (a
few 0.1 M$_{\sun}/$yr) from very low ($\sim$0.1 M$_{\sun}/$yr)
star-forming systems.  We are able to robustly classify them as
quiescent.

In the above statistics, each $\Delta$ value corresponding to a
particular snapshot was given equal weight.  Alternatively, we can
quantify the performance of our SED modeling in a global sense by
summing the SFRs of all snapshots and comparing this to the sum of all
recovered SFRs.  Such an approach gives more weight to phases with
higher SFR, as is also the case for observational studies of the
cosmic SFR density.  Considering the entire timespan for all
simulations, we find that our SED modeling underpredicts the total
mass of stars being formed per unit time by a factor 1.8.  A similar
exercise shows that the total assembled stellar mass is underestimated
by only 14\%.  When focussing on the merger regime alone, the
underestimates increase to a factor 2 for the summed SFR and 30\% for
the summed stellar mass.

\begin {figure}
\centering
\plotone{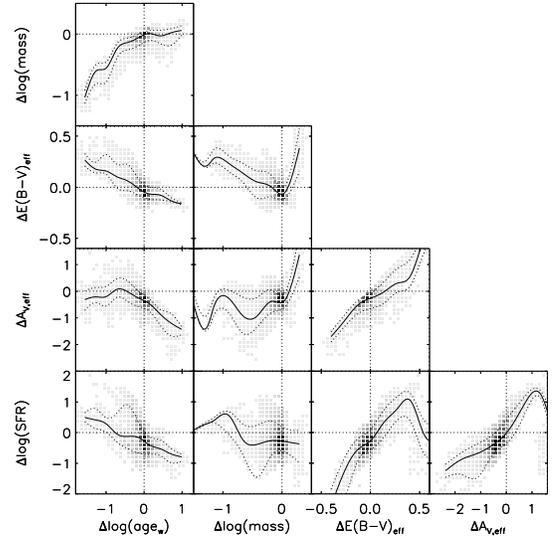} 
\caption{
Correlations between biases in the estimated age, mass, reddening,
extinction, and SFR obtained from broad-band SED modeling.  Style as
in Fig.\ \ref{overalltime.fig}.  Positive values of $\Delta$ mean
overestimates, negative values refer to underestimates with respect to
the true age, mass, $E(B-V)_{\rm eff}$, $A_{V, {\rm eff}}$, and SFR in
the simulations.  Strong correlations exist between $\Delta$ values
for different stellar population properties.  E.g.,
over/underestimates in reddening $E(B-V)_{\rm eff}$ lead to offsets of
the same sign in extinction $A_{V, {\rm eff}}$ and SFR.
\label {correl.fig}
}
\end {figure}

We present an overview of correlations between the $\Delta$ values of
all considered properties in Fig.\ \ref{correl.fig}.  It is clear that
biases in estimates of stellar population properties are often strongly
correlated.  For example, when the age is severely underestimated, the
same will be true for stellar mass.  Likewise, a strong positive
correlation exists between biases in the estimated reddening,
extinction, and star formation rate.  In the following sections, we
shed more light on how the estimates of age, mass, reddening,
extinction, and SFR are connected.

In the light of the upcoming Spitzer warm mission, we note that the
interval containing the central 68\% of $\Delta \log M$ values becomes
twice as broad when omitting the IRAC bands.  This test demonstrates
the value of extending SEDs to longer wavelengths.  Similar
improvements by including IRAC photometry are found for the other
stellar population parameters considered in this paper.

We performed our analysis on synthetic photometry of galaxies placed
at redshifts $z=1.5$ to $z=2.9$.  We note that the results do not
change if you change the redshift of the merger.  Furthermore, the
trends described in this section are all systematic and cannot be
attributed to signal-to-noise effects (e.g. more extincted galaxies at
the highest redshifts being fainter and therefore less well
recovered).  We tested this in two ways: first by omitting the
perturbation of the synthetic fluxes, second by applying a
conservative cut in the observed $K_s$-band magnitude: $K_{s,{\rm
obs}}< 23.6$, corresponding to $S/N_{K_s} > 10$.  In both cases, the
same trends described in this Section are still present.

\subsection {Impact of mismatch between true and template SFH}
\label {fixz_SFH.sec}
The biases in estimates of stellar population properties
(\S\ref{overall.sec}) result from a complex interplay between SFH,
dust distribution, metallicity variations, and AGN contribution.  We
now separate these effects in order to obtain a better understanding
of the origin of the biases.  In this section, we repeat the SED
modeling on the intrinsic stellar photometry (i.e., no dust extinction
or AGN contribution), with the metallicity of the stars fixed to solar
(i.e., no metallicity variations).  Each stellar particle is still
treated as a SSP with an age given by the GADGET-2 output (i.e., the
SFH is identical to that on which the 'full' photometry in
\S\ref{overall.sec} was based), but we extract the photometry from
solar metallicity SSP models, even if the stellar particles were
recorded to have lower/higher metallicities in the GADGET-2
simulation.  Since we use solar metallicity templates to compute the
integrated light of the simulations as well as to fit the resulting
synthetic SEDs, this procedure removes effects due to mismatch in
metallicity and isolates the role of the SFH.  This illustrates the
consequences of a possible mismatch between the true SFH in the
simulation and the allowed template SFHs in our SED modeling.  Our aim
is to constrain the properties of the bulk of the stars (e.g.,
mass-weighted age), whereas our input SEDs are obviously of a
light-weighted nature.  Since massive O and B stars make young stellar
populations brighter than older stellar populations, giving them more
weight in the integrated SED, the light-weighted stellar age will be
younger than the mass-weighted stellar age.  This is always the case,
but provided we have a template representing the correct SFH, it is
possible to account for this effect and still find the correct age of
the bulk of the stars.  Our three allowed SFHs are an SSP, where all
stars formed in a single burst, a model with $SFR \propto e^{-t/\tau}$
with $\tau$ = 300 Myr, and a constant star formation history.  These
are standard choices in analyses of distant galaxies.  However, they
do not encompass a star formation history where the rate of star
formation was lower in the past than it is now, as is the case during
first passage and during the actual merger-triggered starburst (see
Fig.\ \ref{sim.fig}).  In general, fitting a template SFH that has
$\left[\frac{dSFR}{dt}\right]_{template} <
\left[\frac{dSFR}{dt}\right]_{true}$, the older population will be
lost to some degree under the glare of newly formed stars, leading to
an underestimate of the age.

\begin {figure} [htbp]
\centering
\plotone{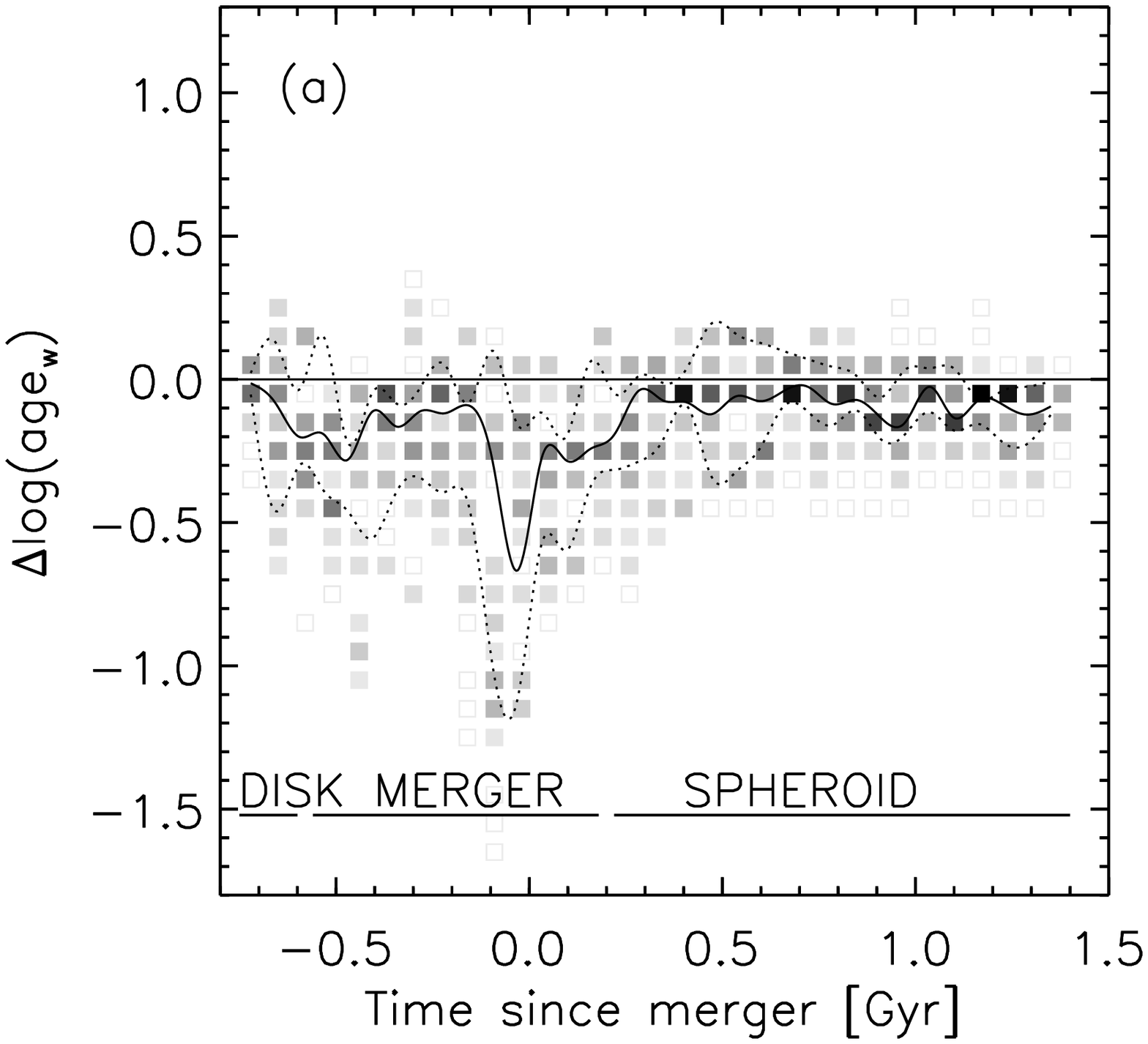} 
\plotone{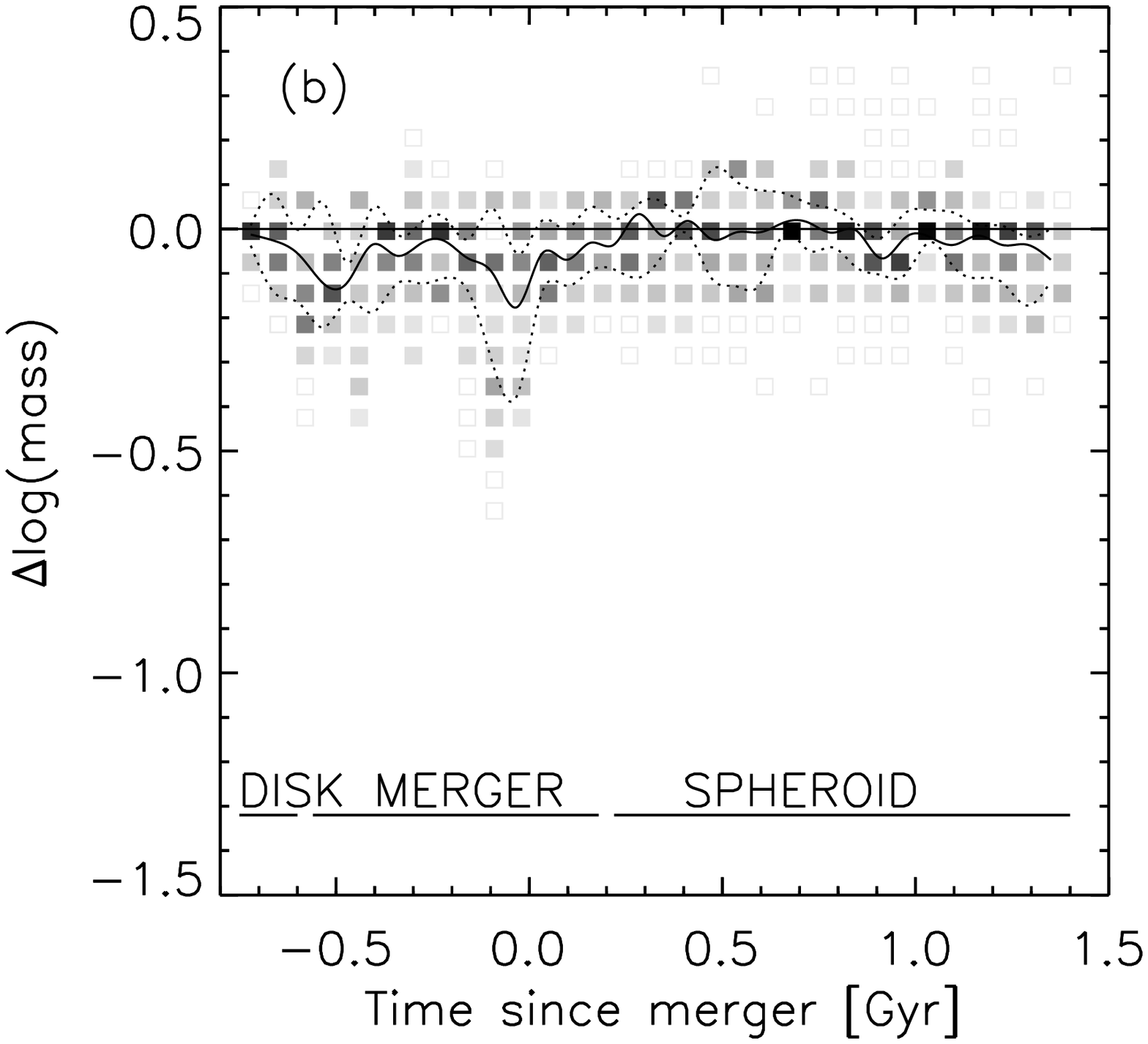} 
\plotone{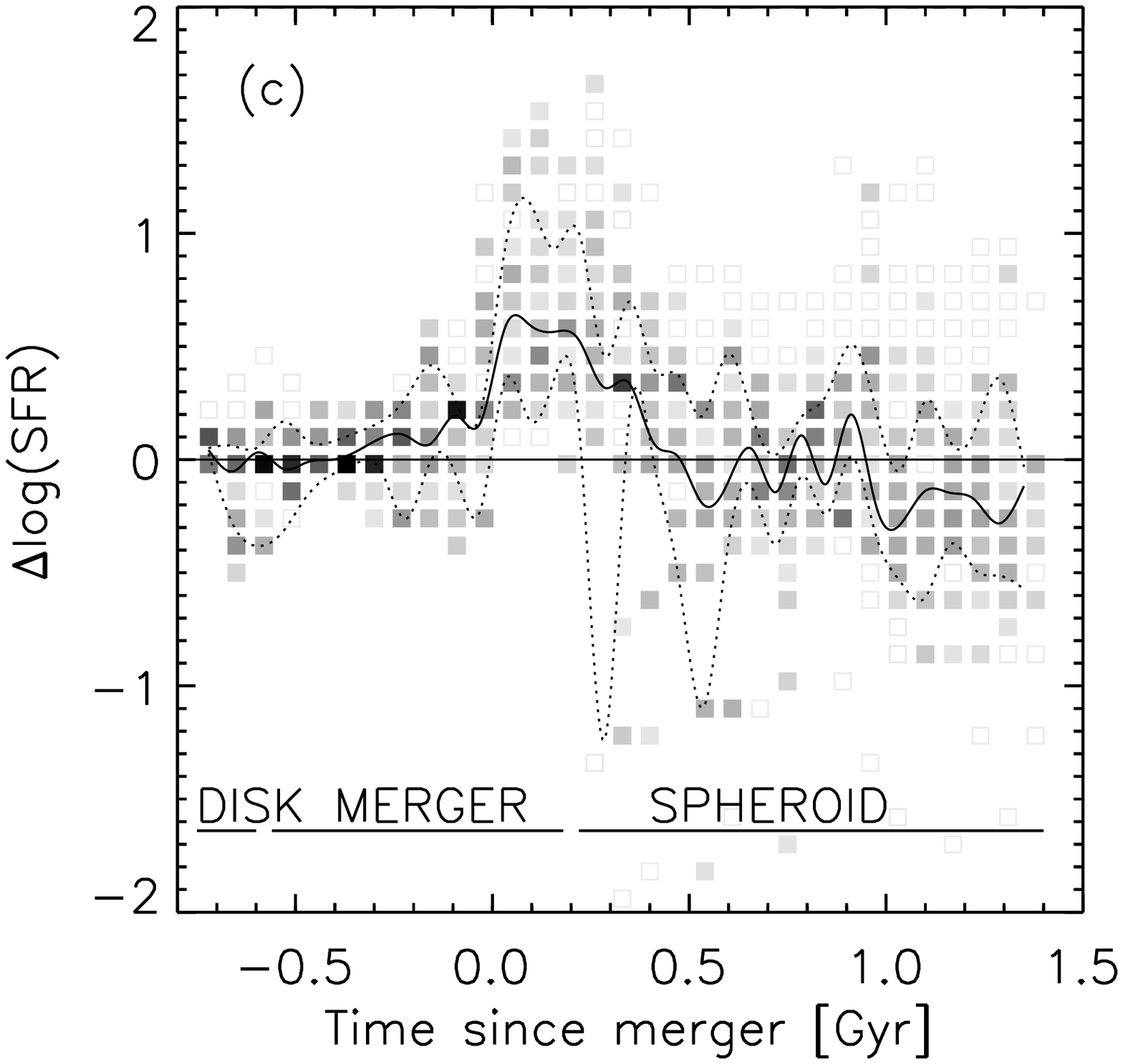} 
\caption{Impact of star formation history.  The difference between
estimated and true (a) mass-weighted age, (b) mass, and (c) SFR as a
function of time for all simulations, with the SED modeling performed
on the intrinsic (i.e., unattenuated) stellar photometry with all
stars set to solar metallicity.  Style as in Fig.\
\ref{overalltime.fig}.  Deviations from 0 (negative indicating an
underestimate) are due to mismatch between the actual star formation
history and the histories allowed in our SED modeling
(SSP/CSF+dust/$\tau_{300}$+dust).  Maximal underestimates of age and
mass are reached during the merger itself.  A secondary minimum is
reached during first passage of the progenitors, $\sim 0.4$ Gyr
before.  The SFR, however, is well recovered during the episodes of
merger-enhanced star formation, but not in the first few 100 Myr after.
\label {impactSFH.fig}
}
\end {figure}

This is illustrated in Fig.\ \ref{impactSFH.fig}(a), where the
difference between recovered and true mass-weighted stellar age is
plotted as a function of time since the merger between the SMBHs.
During the first snapshot, when the star formation history matches (by
construction) our CSF template, we find no systematic offset and a low
scatter, purely resulting from photometric uncertainties.  Soon after,
we start to underestimate the age, with minima coinciding with the
moment of first passage (500 Myr before the actual merger) and that of
the actual merger-triggered starburst.  It is precisely at these
moments that the real SFH deviates most from the allowed template SFHs
(during the merger, the $\tau_{300}$ and CSF templates are most
frequently preferred by the fitting routine).  During the starburst
phase itself, the median offset of true mass-weighted age versus
recovered age exceeds 0.6 dex, with a large scatter due to differences
in the SFH for different initial conditions.  For example, the ratio
of SFR at first passage over SFR during the central starburst
increases with gas fraction.  After all activity has quieted down, the
derived ages lie within 0.1 dex of their true value.

Since one tends to count the young light only, mass will be
underestimated as well during first passage and the final nuclear
starburst (see Fig.\ \ref{impactSFH.fig}(b)).  For the same reason,
models allowing for a secondary burst of star formation on top of an
older stellar population were found to reveal larger total stellar
masses, in particular for blue objects (Papovich et al. 2006; Erb et
al. 2006b; Wuyts et al. 2007).  As for the age measurement, the
derived masses for the merger remnants lie within 0.1 dex of their
true value.

Since no knowledge about underlying old stellar populations is needed
to constrain the ongoing star formation rate, SED modeling of the
intrinsic light reproduces the SFR well during the episodes of
enhanced star formation (Fig.\ \ref{impactSFH.fig}(c)), when age and
mass determinations are least reliable.  However, the effect of
template mismatch between true and best-fit SFH comes into play later.
In the simulations, the SFR drops rapidly after reaching its peak
during final coalescence.  The preferred SFH of the best-fit model
most frequently has a slower decrease, over an e-folding time of 300
Myr.  By accounting for the bulk of recent star formation, the model
consequently overpredicts the ongoing SFR during the first few 100 Myr
after the merger, by $\sim 0.6$ dex.  Once more time has passed, the
SED modeling correctly identifies the remnant as a low SFR galaxy.

We check that the systematic offsets and scatter in Fig.\
\ref{impactSFH.fig} are indeed due to mismatches between the true and
template SFHs by performing the following test.  For each simulation,
we change the distribution of ages of stellar particles to reflect the
assumed SFHs that are used in the SED fitting (see
\S\ref{SED_modeling.sec}).  We then compute the synthetic photometry
in the redshifted passbands as described in \S\ref{sim_phot.sec},
still ignoring dust reddening and assuming solar metallicity.
Modeling the resulting idealized synthetic SEDs, we recover the
mass-weighted age, mass, and SFR essentially without systematic
offsets ($\lesssim 1 \%$), and with little scatter (a fifth of the
scatter obtained when the true SFH from the simulations was adopted).
The scatter in $\Delta \log M$ decreases by only 40\% when omitting
the perturbation of the synthetic fluxes, and for $\Delta \log age_w$
and $\Delta \log SFR$ the change is even smaller.  This indicates that
the discrete sampling of the SED by the observed passbands and the
discreteness of the age steps allowed in the fit contribute by a
similar amount as photometric uncertainties to the very small $\Delta$
values.  This test demonstrates that, for a survey as GOODS, mismatch
between true and template SFH is a much larger limitation to constrain
stellar population properties of $1.5<z<3$ galaxies than, e.g.,
photometric uncertainties.

\subsection {Impact of attenuation}
\label {fixz_Av.sec}

\subsubsection {Non-uniform attenuation uncorrelated to properties of emitters}
As illustrated in \S\ref{simulations.sec}, dust attenuation has a
large effect on colors, SEDs, and fluxes in all star-forming phases.
This is the case for both isolated spiral galaxies (see also Rocha \&
Jonsson 2008) and interacting systems (Jonsson et al. 2006).  As
described in \S\ref{SED_modeling.sec}, we use the approach of a
uniform foreground screen to account for the attenuation by dust in
our SED modeling.  Fig.\ \ref{sim.fig}(f), illustrating the range of
effective visual extinction values (attenuated minus intrinsic
$V$-band magnitude) for a random simulation depending on the viewing
angle, proves that such a representation is not valid.  Here we
address the impact that a non-uniform distribution of the dust will
have when modeled by a uniform foreground screen.

\begin {figure} [t]
\centering
\plotone{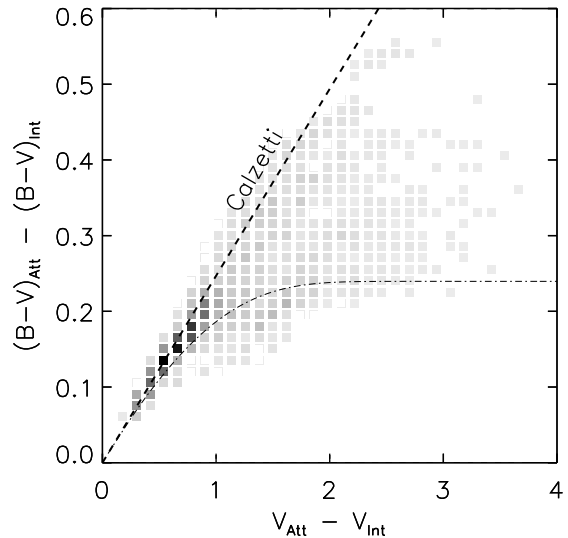} 
\caption{The effective reddening (attenuated minus intrinsic $B-V$
color) versus total absorption in the $V$-band for all timesteps,
viewing angles and initial conditions.  The intensity of the binned
distribution indicates the number of simulations in the respective
part of the diagram.  A ratio of total to selective absorption
$R_V=4.05$ as by Calzetti et al. (2000) is plotted with the thick
dashed line.  The dot-dashed curve indicates a series of toy models where the
distribution of $A_V$ values is uniform between 0 and a maximum value,
and all emitting sources are identical.  Stellar particles
individually have $R_V=4.05$, but in the case of a non-uniform dust
distribution the sum of all stellar particles has an effective
$R_V>4.05$.
\label {EBVAv.fig}
}
\end {figure}

First, we consider a situation where the optical depth is not the same
to all stellar particles, but the variations are uncorrelated with the
intrinsic properties of the stellar particles.  Such a scenario is by
construction the case at the start of the simulation.  For each
stellar particle individually the ratio of total to selective
absorption, $R_V = \frac{A_V}{E(B-V)}=4.05$, was taken from Calzetti
et al. (2000).  Since less extincted regions are also less reddened
and have a larger weight in the integrated SED, the effective
extinction $A_{V,{\rm eff}} \equiv V_{\rm Att} - V_{\rm Int}$ and
effective reddening $E(B-V)_{\rm eff} \equiv (B-V)_{\rm Att} -
(B-V)_{\rm Int}$ of the galaxy as a whole will not be related by the
same factor 4.05 as for the individual particles.  Instead, the
overall reddening for a given $A_V$ will be smaller than predicted by
Calzetti (i.e., the extinction is grayer).  This is illustrated in
Fig.\ \ref{EBVAv.fig} where the dashed line represents the $A_V=4.05
\times E(B-V)$ scaling by Calzetti et al. (2000) and the dot-dashed
curve represents a series of toy models for different $A_{V,{\rm
max}}$ with a uniform distribution of $A_V$ values between 0 and
$A_{V,{\rm max}}$ to stellar particles that all emit at identical
intrinsic luminosities.  The effective visual extinction for such a
model consisting of N stellar particles emitting with identical
luminosities $L_i$ is calculated as follows:

\begin{eqnarray}
\displaystyle A_{V,{\rm eff}} & = & V_{\rm Att} - V_{\rm Int} \\
                              & = & -2.5 \log \left[ \frac{\sum\limits_{i=1}^N L_i\ 10^{-0.4A_{V,i}}}{\sum\limits_{i=1}^N L_i} \right] \\
                              & = & -2.5 \log \left[ \frac{\int\limits_{0}^{A_{V,{\rm max}}} 10^{-0.4A_V} dA_V}{\int\limits_{0}^{A_{V,{\rm max}}} dA_V} \right] \\
                              & = & -2.5 \log \left[ \frac{10^{-0.4 A_{V, {\rm max}}} - 1} {-0.4 A_{V, {\rm max}} ln(10)}\right]
\label{Aveff.eq}
\end{eqnarray}

With $A_{B,{\rm eff}}$ calculated in the same way as $A_{V,{\rm eff}}$
in Eq.\ \ref{Aveff.eq}, and using the relation between $A_B$ and $A_V$
to individual stellar particles from Eq.\ \ref{Rv.eq}, the toy model
gives an effective reddening of:

\begin{equation}
E(B-V)_{\rm eff} = 2.5 \log \left[ \frac{\left(1+\frac{1}{4.05}\right) \left[10^{-0.4 A_{V, {\rm max}}} - 1\right]} {10^{-0.4 \left( 1 + \frac{1}{4.05}\right) A_{V, {\rm max}}} - 1}\right].
\label{EBVeff.eq}
\end{equation}

It is clear that the relation between total and selective absorption
for the series of toy models always lies below the Calzetti relation,
and more so for the toy models with the largest $A_{V,{\rm eff}}$
(i.e., largest $A_{V,{\rm max}}$).  The reason is that stellar
particles with little extinction are relatively dominant in the
integrated observed SED.  The binned distribution in Fig.\
\ref{EBVAv.fig}, presenting total versus selective absorption for all
considered simulation snapshots and viewing angles, is roughly
encompassed by the Calzetti relation (Eq.\ \ref{Rv.eq}) on the one hand
and our series of toy models (Eq.\ \ref{Aveff.eq} and Eq.\
\ref{EBVeff.eq}) on the other hand.

\begin {figure*} [htbp]
\centering
\plotone{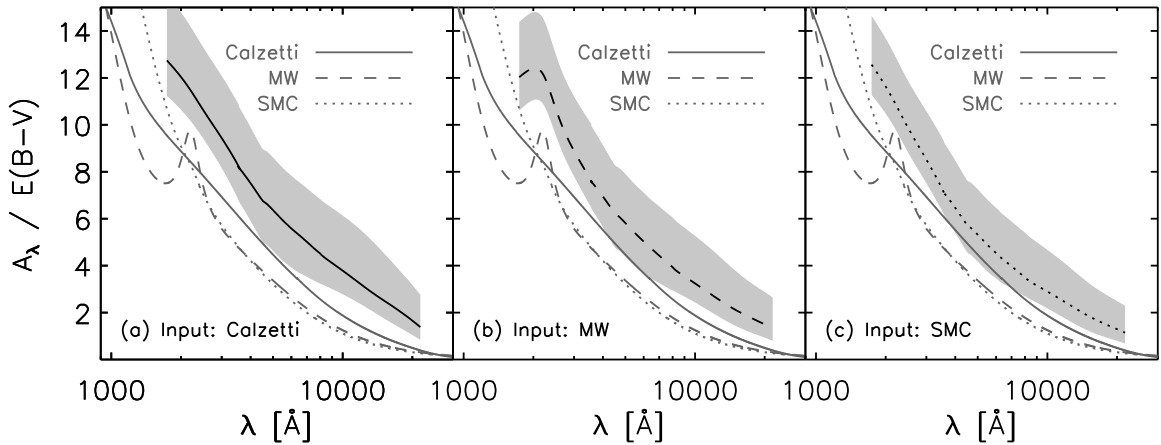} 
\caption{Effective extinction curves of simulated galaxies with
$A_{V,{\rm eff}} > 1$ for different input attenuation laws: (a) the Calzetti
et al. (2000) law, MW-like reddening from Pei (1992), and (c) SMC-like
reddening from Pei (1992).  The black curve indicates the median over
all snapshots and viewing angles with $A_{V,{\rm eff}}>1$.  The light gray
polygon indicates the central 68\% interval.  The Calzetti, MW, and
SMC attenuation laws are plotted in gray.  In all cases, the effective
extinction of simulated galaxies with large $A_{V,{\rm eff}}$ is grayer than
the Calzetti et al. (2000) law that is used in standard SED modeling.
The offset is smallest when each stellar particle is attenuated
according to the SMC-like law.
\label {atten_curve.fig}
}
\end {figure*}

Since the Calzetti et al. (2000) attenuation law was derived
empirically for galaxies as a whole, it is arguably not the
appropriate law to apply to the individual stellar particles, i.e.,
the smallest stellar populations that our simulation can resolve,
typically $10^5 - 10^6\ M_{\sun}$.  We investigated the changes in
photometry when adopting a MW and SMC-like reddening curve by Pei
(1992), which were derived in a more bottom-up fashion from the
physics of interstellar dust grains.  Again, we scaled the optical
depth with the metallicity along the line of sight.  For the SMC
reddening curve, the resulting colors become redder by up to 0.05,
0.1, and 0.2 mag in rest-frame $B-V$, $U-V$, and $V-J$ respectively.
The MW-like attenuation law is also less gray than Calzetti, thus
producing slightly redder colors, though less so than for the SMC law.
The effective extinction curve, expressed as
$\frac{A_{\lambda}}{E(B-V)}$ as a function of wavelength, of snapshots
and viewing angles with large optical depths ($A_{V,{\rm eff}}>1$) is
presented for different input attenuation laws in Fig.\
\ref{atten_curve.fig}.  In each case, we find the typical effective
attenuation law to lie above (i.e., show less reddening for a given
extinction than) the Calzetti et al. (2000) law.  Even in the K-band,
such an offset is still notable, and of similar size as would be
expected for our toy model with $A_{V,{\rm eff}} = 2$ (corresponding
to $A_{V,{\rm max}} = 7$, i.e., many particles are heavily extincted).

\begin {figure} [htbp]
\centering
\plotone{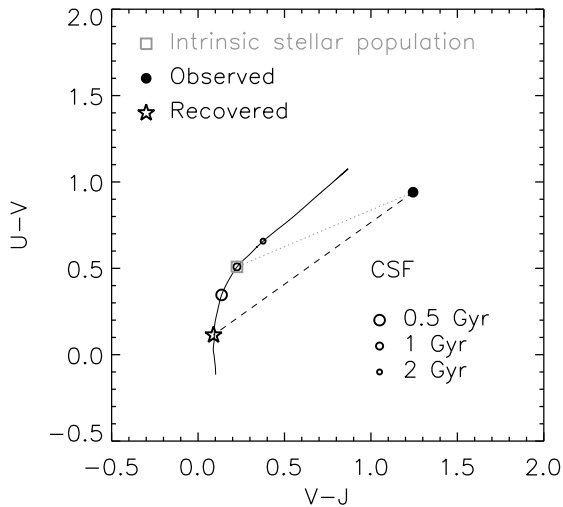} 
\caption{Rest-frame $U-V$ versus $V-J$ color-color diagram illustrating
the effect of a non-uniform distribution of $A_V$ values that is
uncorrelated with the intrinsic properties of the emitting sources.
The black curve indicates a CSF population with age between 50 Myr and
2 Gyr.  Suppose an intrinsic population of age 1 Gyr ({\it box
symbol}) is reddened by such a dust distribution to the location in
color-color space of the filled circle.  Under the assumption of a
uniform foreground screen of dust, the observed colors will then be
traced back along the Calzetti et al. (2000) reddening vector ({\it
dashed black line}), resulting in an artificially young age ({\it star
symbol}).
\label {toyAv.fig}
}
\end {figure}

Not only does non-uniform extinction change the reddening
($\frac{dA_{\lambda}}{d\lambda}$) at a given $A_V$, it also affects
the dependence of the reddening on wavelength
($\frac{d^2A_{\lambda}}{d\lambda^2}$).  For extinction that is
uncorrelated to the properties of the emitting sources, this gives the
effective dust vector in the $U-V$ versus $V-J$ color-color diagram a
shallower slope.  I.e., for a given reddening in $V-J$, the reddening
in $U-V$ is smaller than predicted by the Calzetti et al. (2000) law.
For example, for the simple toy models described in Eq.\
\ref{Aveff.eq} and Eq.\ \ref{EBVeff.eq}, the slope $\frac{A_{U,{\rm
eff}} - A_{V,{\rm eff}}} {A_{V,{\rm eff}} - A_{J,{\rm eff}}}$ of the
effective reddening vector decreases asymptotically with increasing
$A_{V,{\rm max}}$ to half the slope of the Calzetti reddening vector.
The consequence of a different $\frac{d^2A_{\lambda}}{d\lambda^2}$
than Calzetti is clarified in Fig.\ \ref{toyAv.fig}.  The solid curve
represents the evolutionary track of a stellar population following a
CSF history.  The track starts 50 Myr after the onset of star
formation and ends 2 Gyr later.  Suppose different parts of a galaxy
all contain a 1 Gyr old CSF population whose intrinsic location in
color-color space is marked by the open gray box.  A distribution of
dust as described by the above mentioned toy model will redden the
galaxy along the dotted line.  Interpreting the observed colors ({\it
filled circle}) as a CSF population attenuated by a foreground screen
according to the Calzetti et al. (2000) law, will lead to a best-fit
age ({\it star symbol}) that is too young and reddening that is too
large.

\begin {figure*} [htbp]
\centering
\plottwo{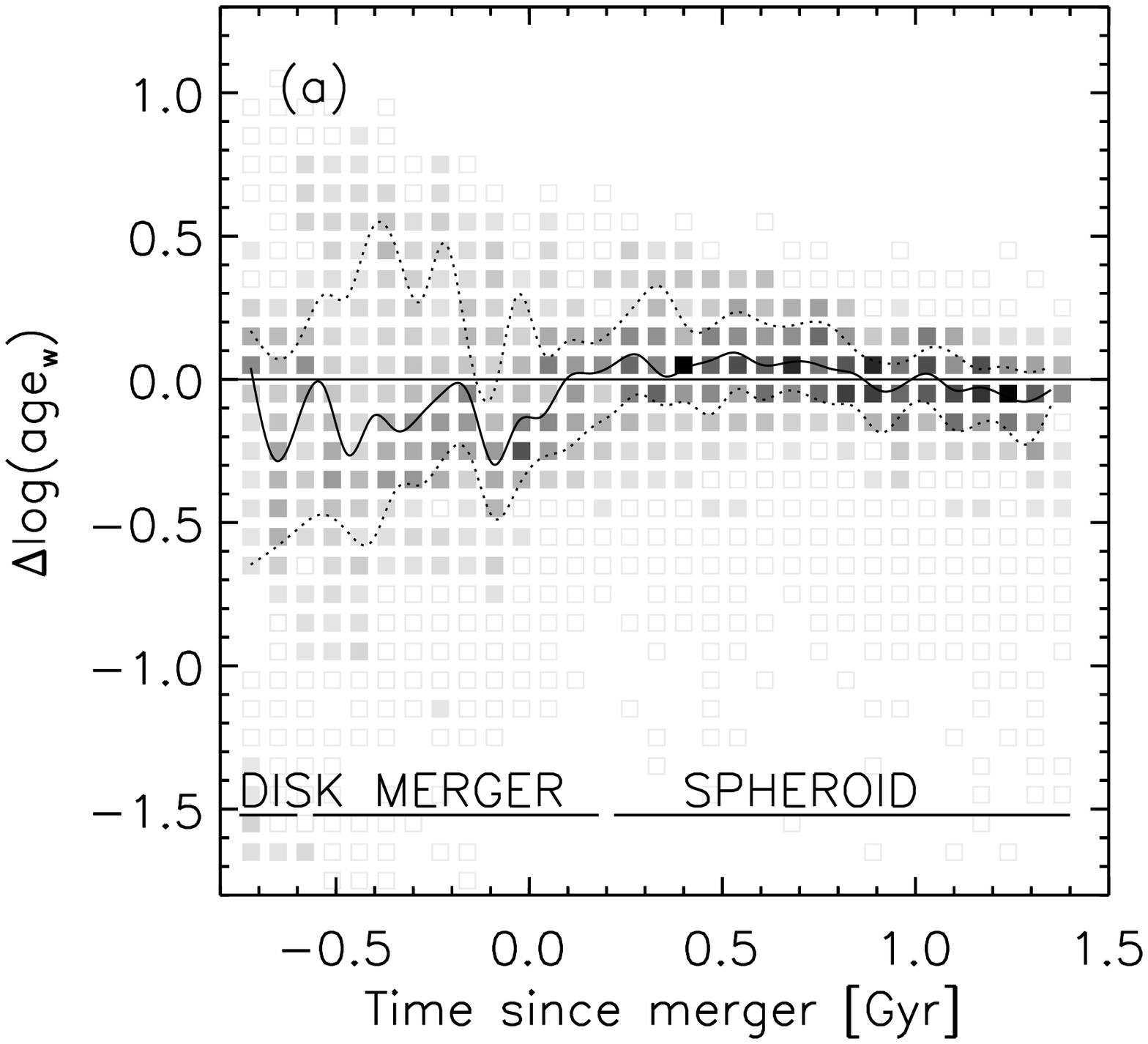}{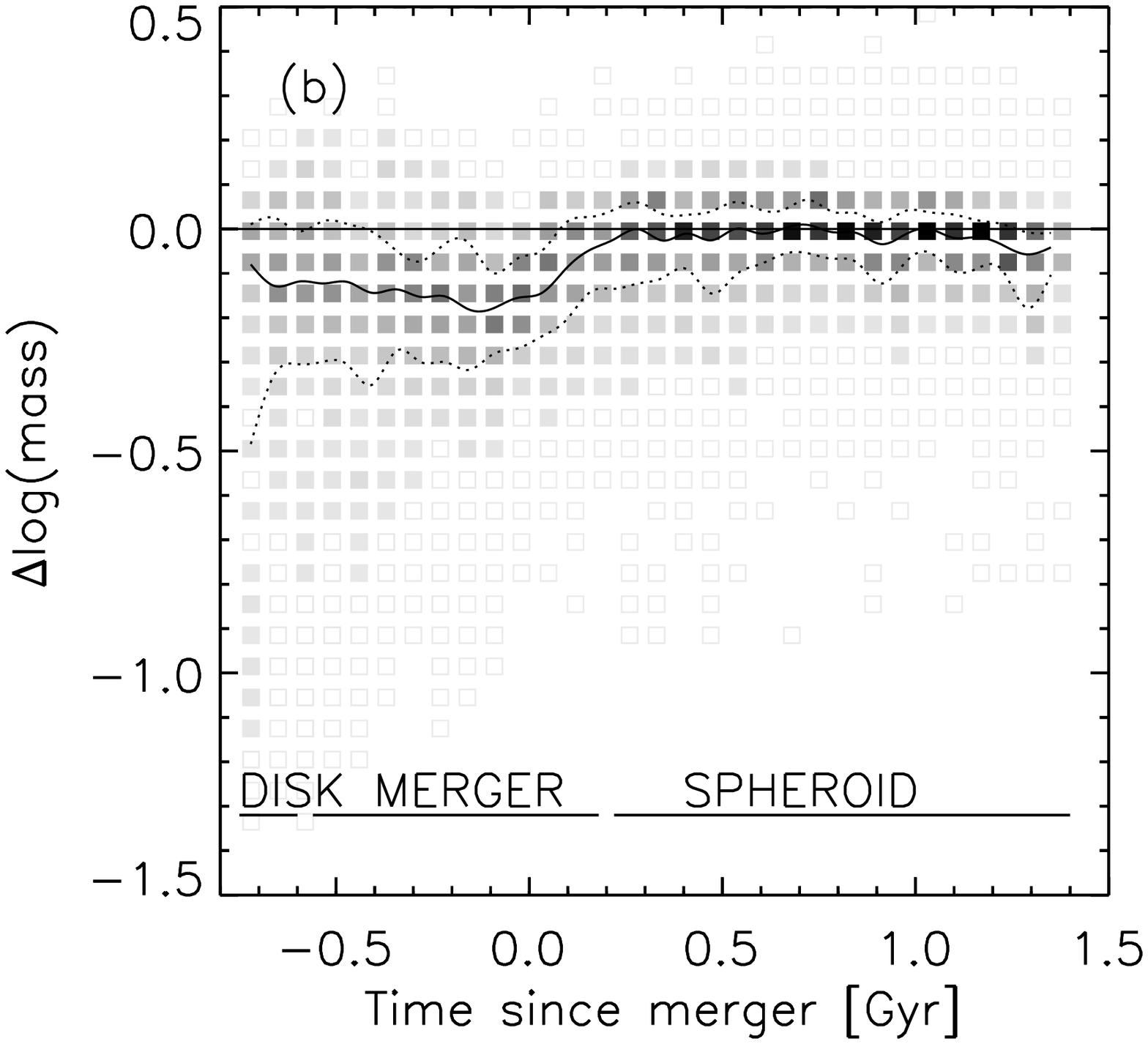}
\plottwo{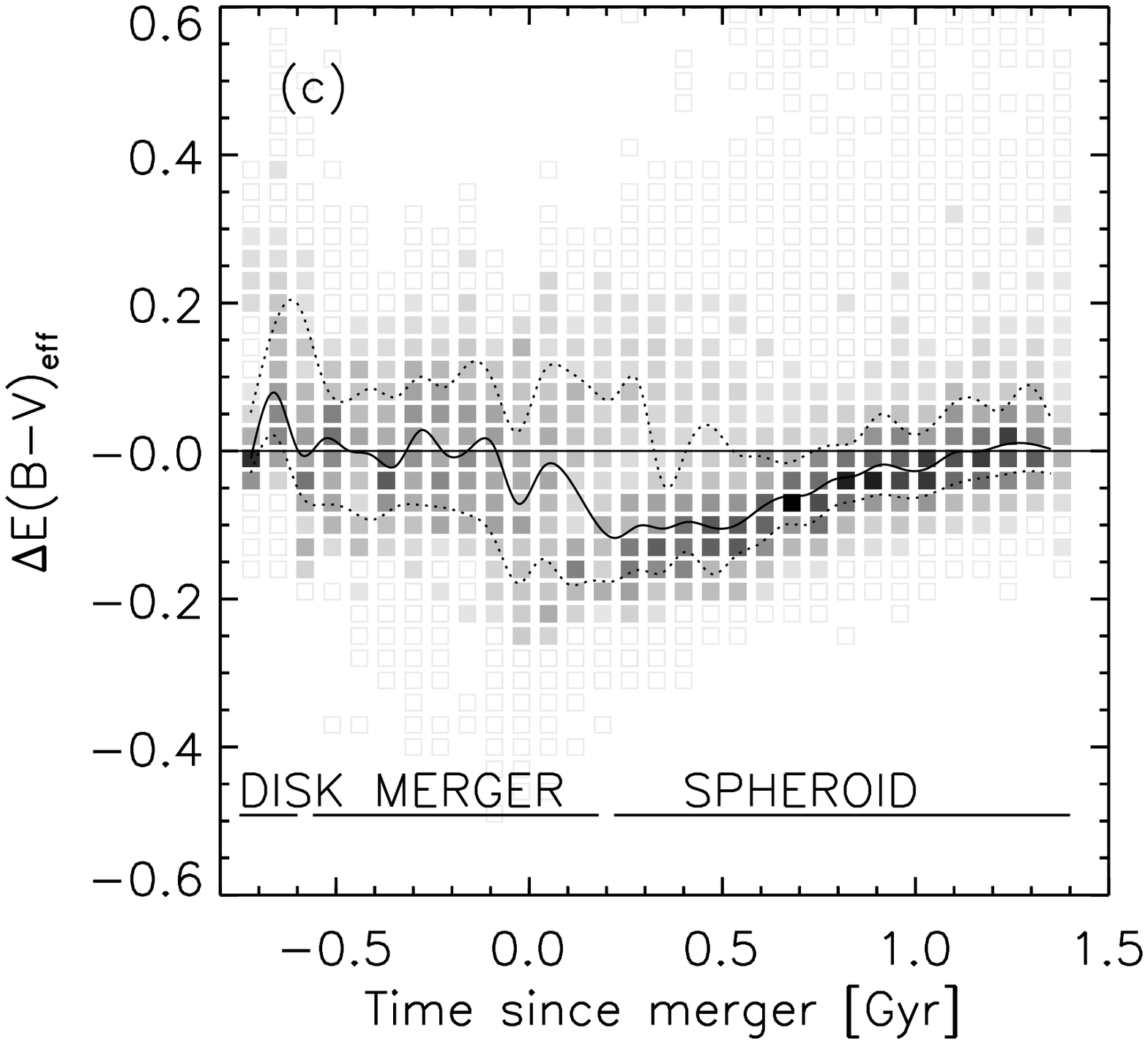}{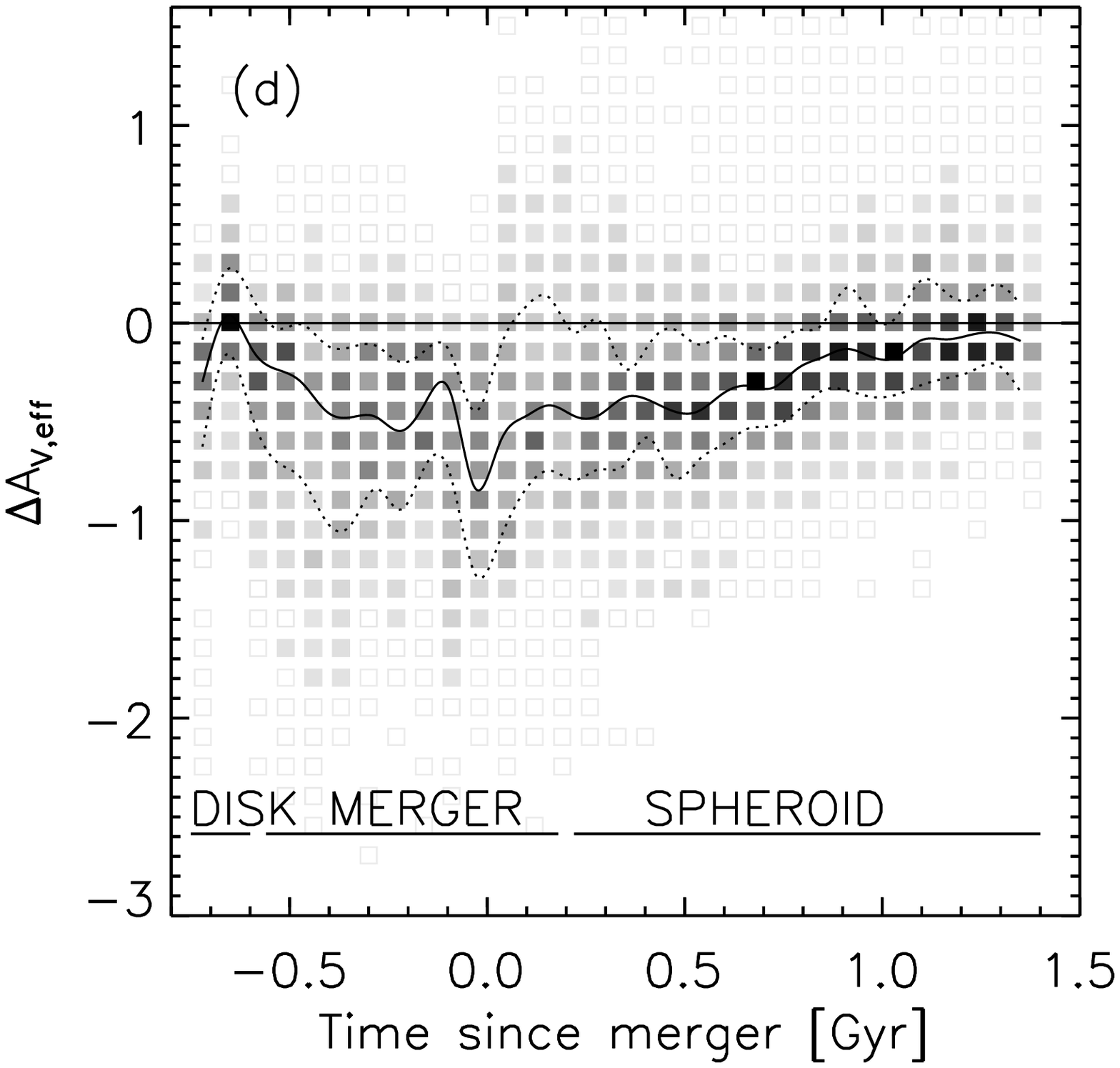}
\epsscale{0.5}\plotone{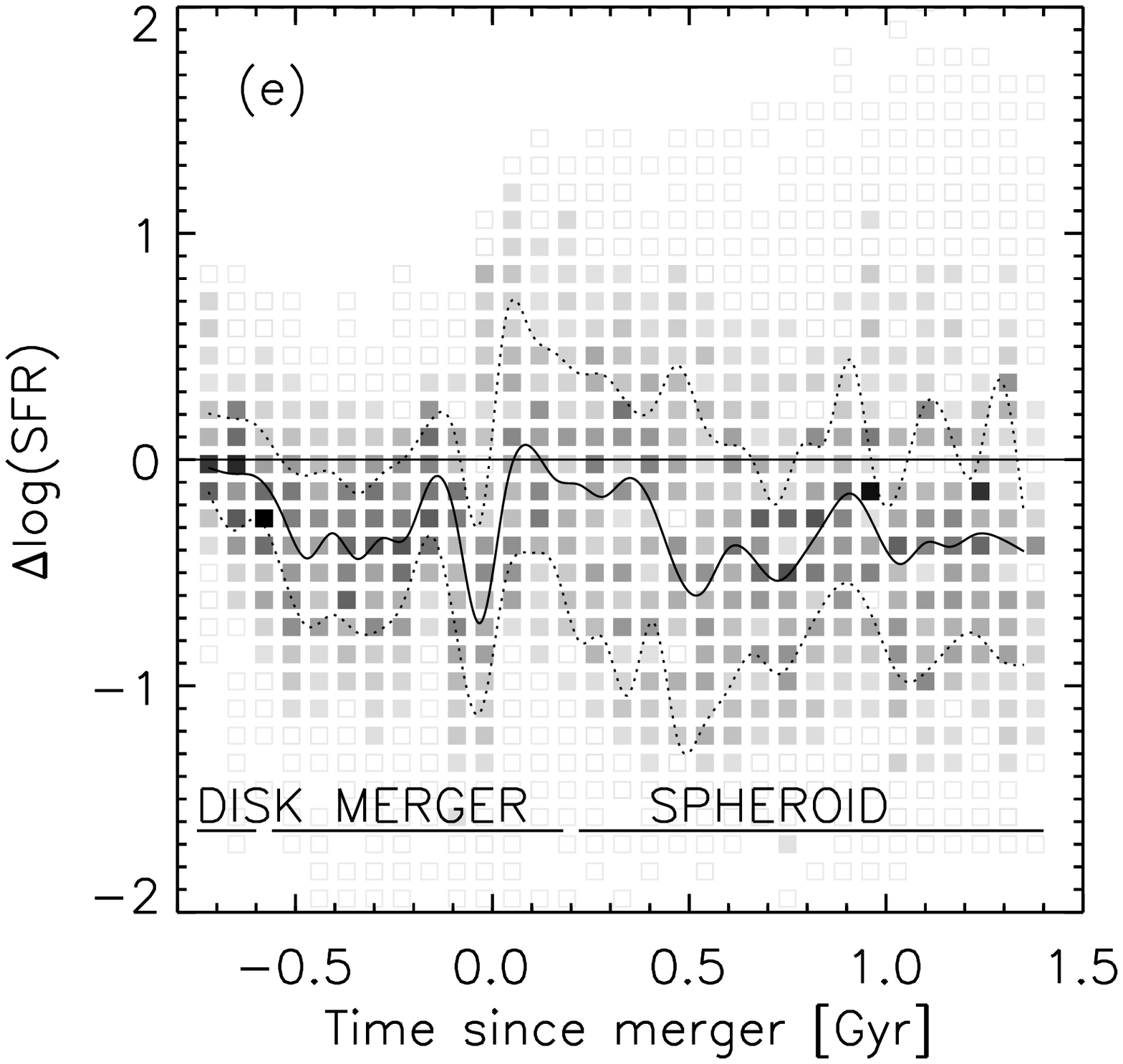}
\epsscale{1}
\caption{\small Effect of extinction.  The difference between estimated and true (a) mass-weighted age, (b) stellar mass, (c) effective reddening, (d) effective visual extinction, and (e) star formation rate as a
function of time since the merger.  The SED modeling was performed on
the attenuated stellar photometry with all stars set to solar
metallicity.  Style as in Fig.\ \ref{overalltime.fig}.  Ages are still
underestimated for the first 0.8 Gyr of the evolution, but to a lesser
degree than estimates based on the intrinsic light.  Added to the
underestimated $A_V$, this leads to a characterization of the stellar
mass that is too low by 0.1 - 0.15 dex.  The recovery of SFRs shows
minima around first passage ($\sim -0.4$ Gyr) and final coalescence
(-0.1 to 0 Gyr).
\label {Avtimesince.fig}
}
\end {figure*}

Since in our simulations the ages of the stellar particles (that are
each treated as SSPs) present at the start of the simulation were
drawn randomly from a uniform distribution, the system has a CSF
history in the earliest snapshots without a correlation between the
optical depth and intrinsic luminosities and colors of the stellar particles.
Therefore, it comes as no surprise that, when looking at the
attenuated stellar photometry in Fig.\ \ref{Avtimesince.fig} (for now
all stars still set to solar metallicity), the central 68\% interval
in $\Delta \log age_w$ reaches to more negative values (to -0.5 dex)
during the earliest phases than was the case for the unattenuated
photometry (Fig.\ \ref{impactSFH.fig}).  The estimated reddening is
slightly larger than the true value, but nevertheless the use of Eq.\
\ref{Rv.eq} still causes an underestimated $A_V$, as can be understood
from Fig.\ \ref{EBVAv.fig}.  The systematic underestimate in age and
$A_V$ combined cause the evaluation of the stellar mass during the
first snapshots, when template mismatch due to the SFH is still
negligible, to be too small by $\sim 0.12$ dex.

After a few 100 Myr after the beginning of the simulation however,
Fig.\ \ref{Avtimesince.fig} reveals an improved recovery of the
mass-weighted stellar age compared to that obtained by SED modeling of
the intrinsic light (Fig.\ \ref{impactSFH.fig}).  Clearly, the
assumption of a non-uniform dust distribution that is uncorrelated
with the intrinsic properties of the emitting sources breaks down.

\begin {figure} [htbp]
\centering
\plotone{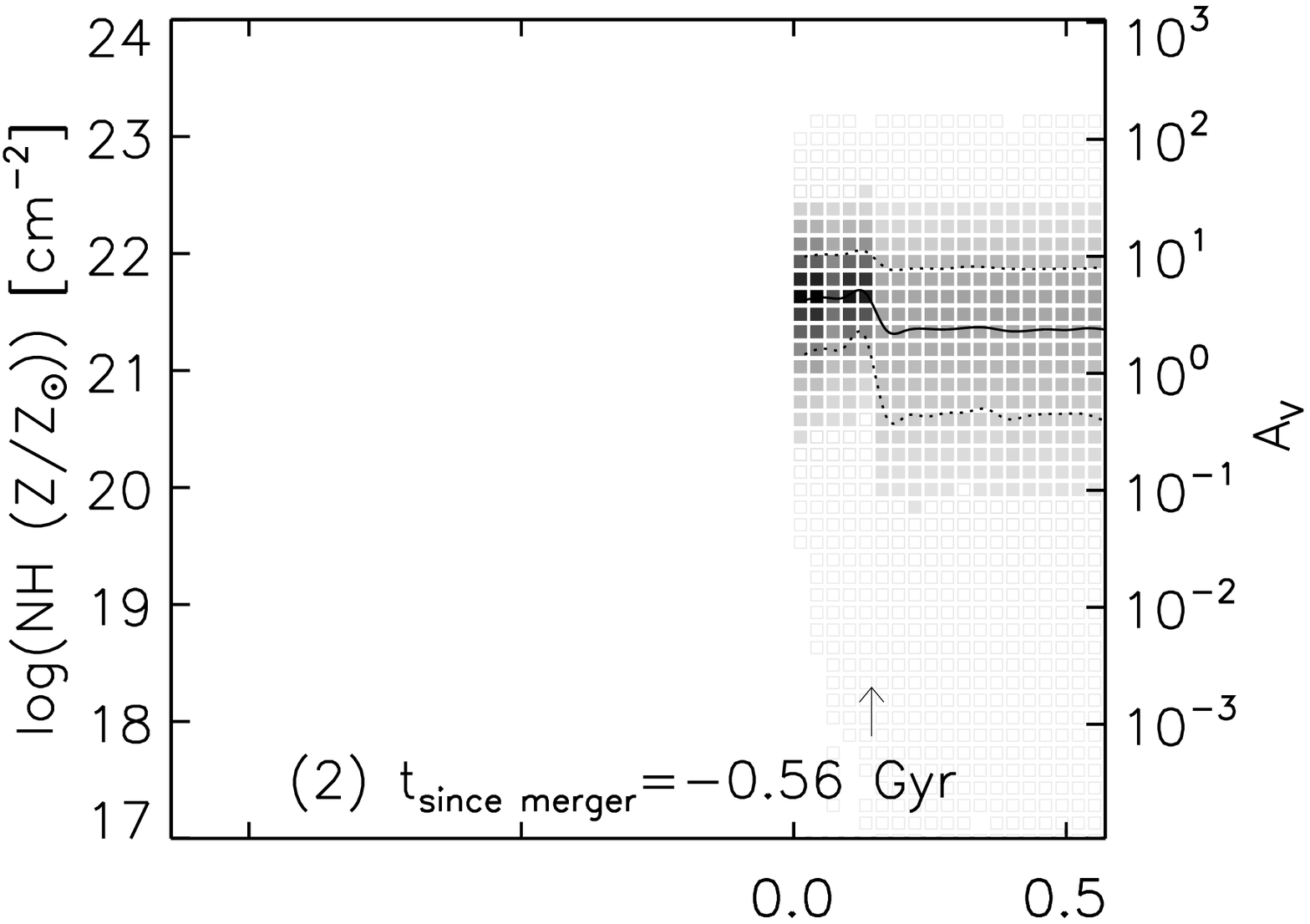} 
\vspace{0.5cm}
\plotone{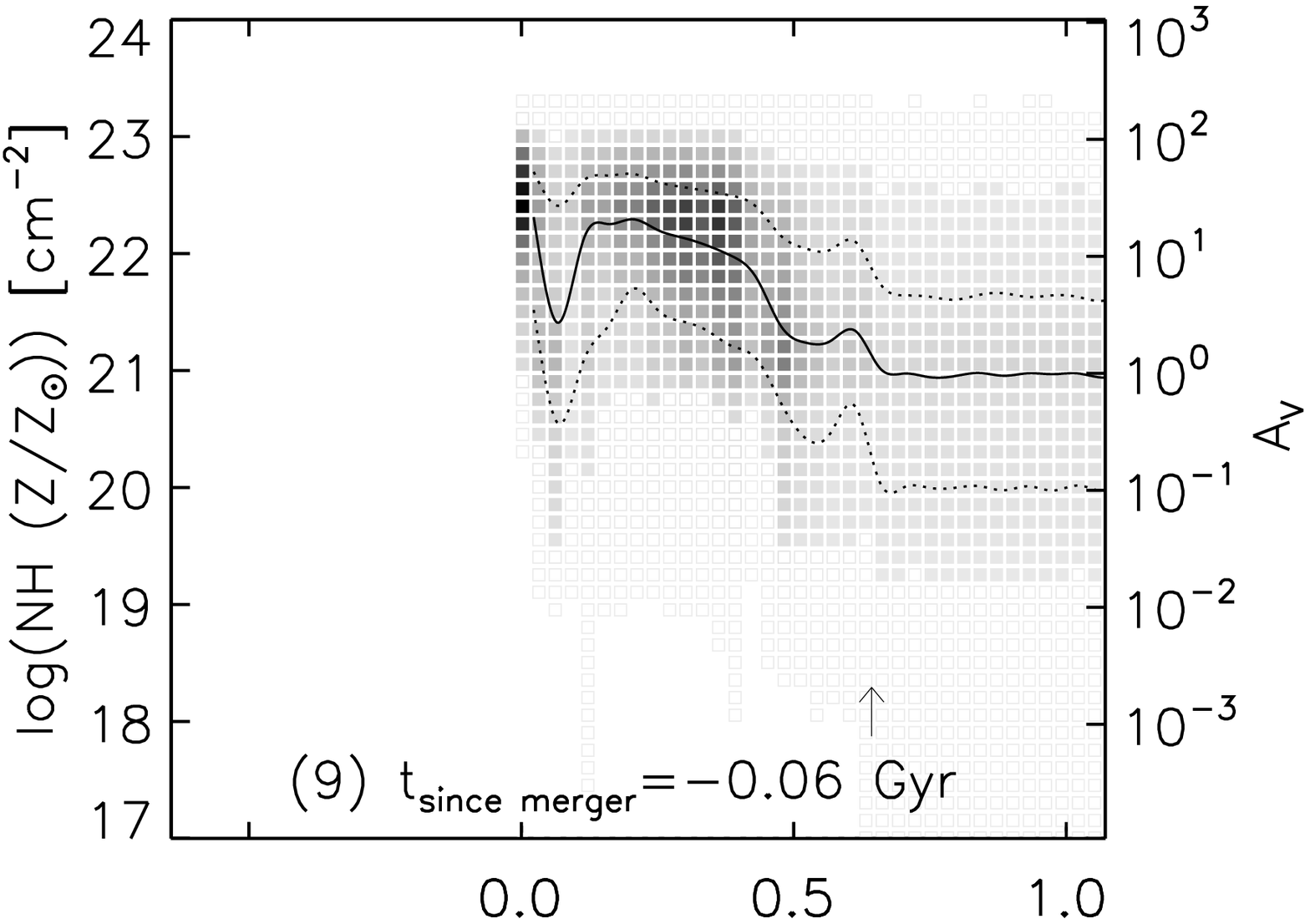} 
\vspace{0.5cm}
\plotone{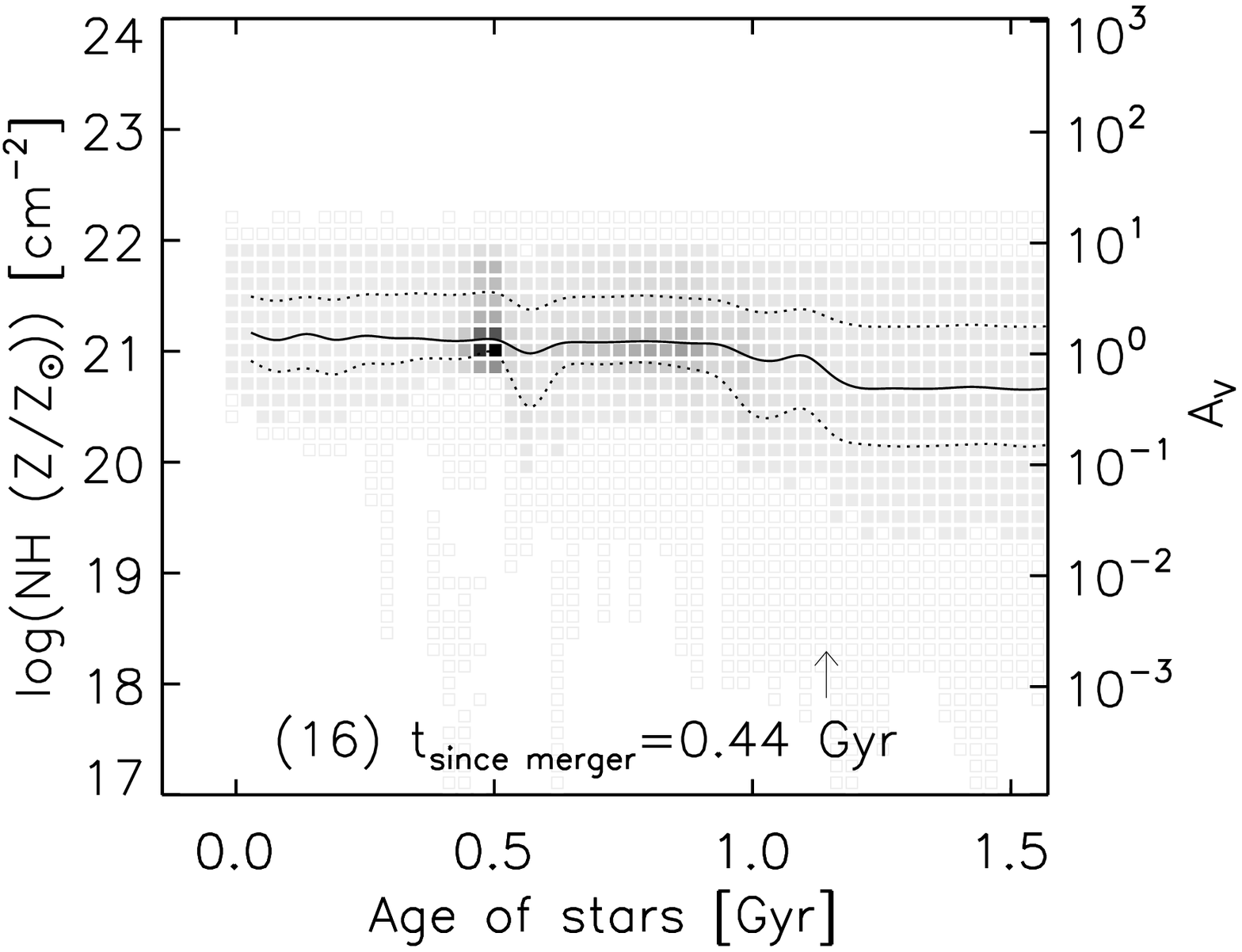} 
\caption{Distribution of hydrogen column densities, linearly scaled
with the metallicity of the gas along different lines-of-sight to the
stellar particles versus the age of the respective stellar particle.
The relation between column density and stellar age is plotted for 3
snapshots: before (2), during (9), and after (16) the merger (see
Fig.\ \ref{sim.fig}).  The solid and dotted curves indicate the median
and 68\% interval of the distribution respectively.  Darker intensity
means a larger number of stars is present with that age.  All stellar
ages rightward of the arrow correspond to initial stars and were set
by hand.  The optical depth, which is proportional to the
metallicity-scaled gas density, is larger toward newly formed stars
during the merger-triggered starburst.  The signature of this
age-dependent extinction weakens during more quiescent episodes of
star formation.
\label {prefextinct.fig}
}
\end {figure}

\subsubsection {Preferential extinction toward young stars}
\label{fixz_prefextinc.sec}
Fig.\ \ref{prefextinct.fig} demonstrates the occurence of preferential
extinction toward young star forming regions in one of our
simulations.  The three panels indicate the binned distribution of the
metallicity-scaled hydrogen column density measured along various
lines-of-sight versus the age of the stellar particle to which the
column density was computed for the 3 snapshots marked in Fig.\
\ref{sim.fig}(a).  The vertical arrow indicates the start of the
simulation.  All stellar ages older than this value (cut off for
illustrational purposes) were set by hand as explained in
\S\ref{sim_main.sec}.  As we already pointed out in Fig.\ \ref{sim.fig}(f), 
the typical column densities are higher during the merger (panel b)
than before (a) or after (c).  Moreover, Fig.\ \ref{prefextinct.fig}
shows that the ratio of column densities toward ongoing star formation
over column densities toward older populations reaches a maximum
during the merger (b).  Using sticky particle simulations of dusty
starburst mergers, Bekki \& Shioya (2001) found a similar
age-dependent extinction, confirming that this is a generic feature of
merging systems and not determined by the method used to model
dissipative processes.  Poggianti \& Wu (2000) inferred age-dependent
extinction during a starburst to explain the nature of so-called e(a)
galaxies: galaxies with [OII] in emission and strong Balmer absorption
lines, frequently associated with merger morphologies.

From a physical perspective, it is expected that during the merging
process hydrodynamical and gravitational forces channel gas and dust
to the central regions where it triggers a starburst.  Once started,
supernovae going of on a few $10^7$ yr timescale further increase the
dust content of the regions where newly formed stars reside.  The fact
that the distribution of younger (and thus intrinsically bluer)
stellar populations does not trace that of the older populations of
stars and that it is intimately correlated with the dust distribution
leads to an overestimate in age.  In Fig.\ \ref{toyAv.fig}, a
distribution of $A_V$ values that was uncorrelated to the intrinsic
properties of the emitters caused an effective reddening along a
shallower slope in the $U-V$ versus $V-J$ diagram than for the
Calzetti curve.  In case of a distribution of $A_V$ values that is
correlated with the intrinsic colors of the emitters (larger $A_V$ to
intrinsically bluer stellar populations), the slope of the effective
reddening vector in the $U-V$ versus $V-J$ diagram is steeper than for
the Calzetti et al. (2000) law.  Since an observer will mistakenly
model the galaxy with an intrinsically redder template, the reddening
by dust $E(B-V)$ will be underestimated.  Although a given total
absorption corresponds to a stronger reddening in the presence of
age-dependent extinction compared to uncorrelated non-uniform
extinction, the Calzetti et al. (2000) relation between $E(B-V)$ and
$A_V$ given by Eq.\
\ref{Rv.eq} can still be considered as an upper limit.  Therefore, the
total absorption will be underestimated.  This is illustrated in Fig.\
\ref{EBVAv.fig} where we plotted the binned distribution of true $E(B-V)$
versus true $A_V$ for all of our simulation snapshots, viewed under a
range of viewing angles.  Finally, the derived stellar mass owes its
more robust character to the compensating effects of systematic
offsets in age and absorption.

\begin {figure} [t]
\centering
\plotone{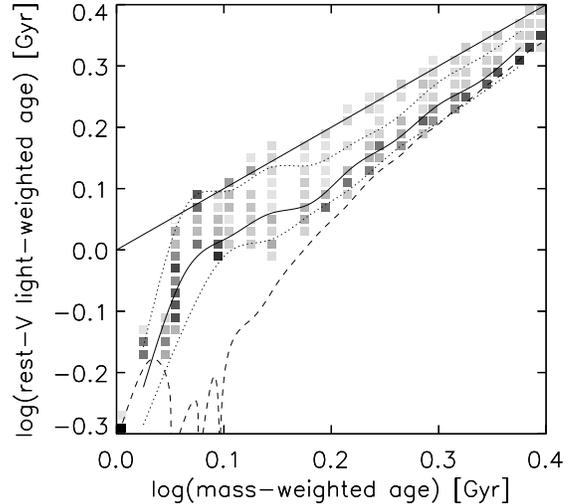} 
\caption{Rest-frame attenuated and intrinsic $V$-band light-weighted age versus mass-weighted age
for an initially 40\% gas fraction simulation.  The boxes mark the
mean age weighted with the attenuated $V$-band light.  Darker
intensities indicate a larger number of viewing angles.  The solid and
dotted curves mark the median and central 68\% interval respectively.
The dashed curve indicates the mean age weighted with the intrinsic
$V$-band light (no attenuation).  The attenuated light-weighted age is
a better approximation of the mass-weighted age than the intrinsic
light-weighted age, increasingly so for younger stellar populations.
Larger optical depths to young than to old stars are responsible for
this effect.
\label {lightweighted.fig}
}
\end {figure}

The effect of the larger extinction toward young stars will in
practice be superposed on the effect of mismatch between template and
true SFH, that prevents us from fully accounting for the difference
between light- and mass-weighted stellar age (see
\S\ref{fixz_SFH.sec}).  Fig.\ \ref{lightweighted.fig} illustrates how
an increased extinction toward young stars reduces the difference
between the light-weighted and mass-weighted measure of age.  We
conclude that the SED modeling on galaxies with solar metallicity
stars and dust distributed in between still underestimates the age
during the merger-enhanced star-forming phases, but adding dust has
improved our best guess to an overall median offset of -0.04 dex
(compare Fig.\ \ref{Avtimesince.fig}(a) to Fig.\
\ref{impactSFH.fig}(a)).  Similar conclusions were drawn by Bell \& de
Jong (2001) who examine the reddening and dimming effects of dust and
its impact on estimating stellar mass-to-light ratios.  At later
times, the merger remnant is largely devoid of gas and dust, making
accurate determinations of age and mass possible, as described already
in \S\ref{fixz_SFH.sec}.

Finally, the evolution of $\Delta \log SFR$ over time (Fig.\
\ref{Avtimesince.fig}(e)) differs significantly from that obtained
from modeling the intrinsic light (Fig.\ \ref{impactSFH.fig}(c)).
With the SED modeling now performed on attenuated photometry, we find
underestimates in the SFR of -0.4 dex and -0.7 dex during first
passage and final coalescence respectively.  The agreement between
recovered and true SFR improves during the first few 100 Myr after the
starburst.  Later, during the spheroid phase, large values of $\Delta
\log SFR$ merely reflect our inability to distinguish between low
(a few $0.1\ M_{\sun}/yr$) and very low ($\sim 0.1\ M_{\sun}/yr$) SFRs.
We therefore focus on the underestimates during the star-forming
phases.  In the presence of dust, there is a component of the
merger-triggered starburst that is completely enshrouded.  This allows
young stars to be hidden without revealing their presence by an
increased reddening of the integrated SED.  Since these heavily
obscured star-forming regions are not accounted for by the best-fit
model, the SFR during these phases is significantly underestimated.

\clearpage

\subsection {Impact of stellar metallicity}
\label {fixz_Z.sec}

\begin {figure} [t]
\centering
\plotone{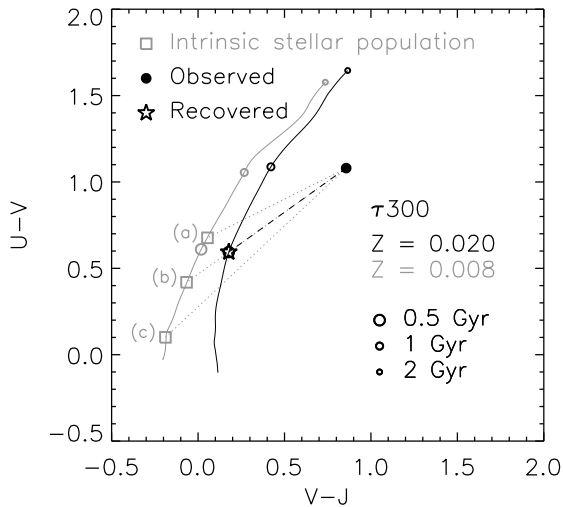} 
\caption{Rest-frame $U-V$ versus $V-J$ color-color diagram illustrating
the effect of fitting solar metallicity templates to stellar
populations of sub-solar metallicity.  The black and gray curves
represent evolutionary tracks for an exponentially declining star
formation history with {\it e}-folding time of 300 Myr for solar ($Z=0.02$)
and sub-solar ($Z=0.008$) metallicity respectively, each starting at
50 Myr.  A stellar population with intrinsic colors indicated by the
gray boxes will be reddened to the location in color-space marked by
the filled circle in the case of (a) non-uniform age-independent
extinction, (b) extinction by a uniform foreground screen, and (c)
age-dependent extinction.  In all three cases, the assumption of solar
metallicity and Calzetti attenuation will lead to the conclusion that
the stellar population formed its first stars 0.5 Gyr ago.  This is an
underestimate (a) or overestimate (b, c) respectively.  The reddening
is always underestimated.
\label {metal.fig}
}
\end {figure}
In \S\ref{fixz_SFH.sec} and \S\ref{fixz_Av.sec}, we tested our SED
modeling on synthetic photometry that was computed assuming a solar
metallicity for all emitting sources.  In reality, stars with a range
of metallicities will be present, reflecting the level of enrichment
in the gas at the epoch of their formation.  Before we repeat our
analysis now setting the stellar metallicities to their appropriate
value calculated by the GADGET-2 code, we anticipate the effect using
the diagnostic $U-V$ versus $V-J$ color-color diagram in Fig.\
\ref{metal.fig}.

The tracks represent exponentially declining SFHs for metallicities of
Z=0.008 ({\it gray}) and 0.02 (solar, in {\it black}).  Both
evolutionary tracks are drawn from 50 Myr to 2 Gyr after the onset of
star formation.  The classic age-metallicity degeneracy states that
the optical broad-band colors of a young stellar population are nearly
indistinguishable from that of an older, more metal-poor population
(O'Connell 1986).  For the $\tau_{300}$ star formation history drawn
here, this effect becomes only notable at later times: 2 Gyr after the
onset of star formation the sub-solar metallicity track has the same
$U-V$ color as a solar metallicity population that started forming
stars 1.8 Gyr ago.  On the one hand, the addition of dust will
complicate the age-metallicity degeneracy.  On the other hand, the
addition of rest-frame NIR photometry helps to separate the
evolutionary tracks for different metallicities (see Fig.\
\ref{metal.fig}).  A galaxy whose attenuated light has colors marked
by the filled circle may correspond with one of the intrinsic colors
indicated by the gray boxes depending on the kind of extinction: (a)
for non-uniform age-independent extinction, (b) for a foreground
screen of dust, and (c) for age-dependent extinction.  In case (a),
the assumption of Calzetti attenuation and solar metallicity in our
SED modeling leads to a recovered evolutionary stage that is too
young, marked with the star symbol on the solar metallicity track.  In
case (b) and (c), the same recovered evolutionary stage is too old.
In all cases, the determination of the reddening will be too low, as
will consequently be the case for the $A_V$ and the stellar mass, and
increasingly so for lower metallicities.  Obviously, the effects
described will again be superposed on the previously discussed effects
of star formation history and dust.  It is also noteworthy that taking
into account the enrichment by heavy elements reduces the effect of
age-dependent extinction.  Young stellar populations are still
intrinsically bluer than old populations, but to a lesser degree since
they have formed at later times from gas that was more enriched.

In our recovery analysis of stellar population properties, we find
that at metallicities of a quarter solar and below, the age is
overestimated by 0 to 0.5 dex (central 68\% interval of $\Delta \log
age_w$).  However, the underestimate in reddening and therefore
extinction for these low-metallicity galaxies is such that the mass
estimate (which is dependent on both age and $A_V$) stays within $\pm
0.1$ dex of its true value for 68\% of the cases.  We find no
dependence of $\Delta \log SFR$ on metallicity.

\subsection {Impact of AGN contribution}
\label {fixz_AGN.sec}

Since the merger simulations described in this paper take into account
the role of supermassive black holes on its environment (see e.g. Di
Matteo et al. 2005; Springel et al. 2005b), we now include
its contribution to the integrated galaxy SED.  We note that only
during a timespan of the order of a Salpeter time, a few $10^7$ to
$10^8$ year, the AGN emission amounts to a significant fraction of the
stellar emission.  Admittedly, the peak of AGN activity can be
missed by the time sampling of our snapshots.  Nevertheless, the
current dataset provides a useful insight on its impact on the SED
modeling.

\begin {figure} [t]
\centering
\plotone{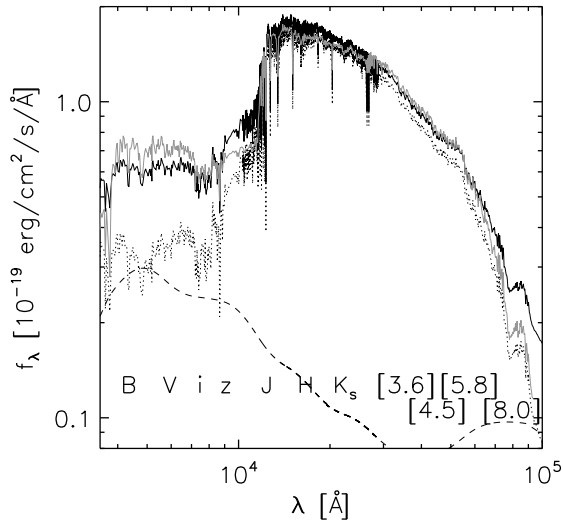} 
\caption{Attenuated spectrum of a simulated merger placed at $z=2.1$
during the peak of AGN activity.  The total attenuated light ({\it black
solid curve}) is decomposed into a contribution from stars ({\it dotted curve})
and AGN (dashed line).  An observer who samples the total attenuated
light with an identical set of broad-band filters as available for
GOODS-CDFS and models the SED using stellar population synthesis only,
will find as best-fitting model the spectrum in gray.  Its age is too
young by $\Delta \log age_w = -0.4$.  The reddening $E(B-V)$ and
absorption $A_V$ are overestimated by 0.1 and 0.4 mag respectively.
The opposite sign of offsets in age and $A_V$ leads to a mass recovery
that is only 0.05 dex below its true value.
\label {bhspec.fig}
}
\end {figure}
We illustrate the typical behavior in Fig.\ \ref{bhspec.fig} showing
the photometry computed at the time of merging when the accretion onto
the supermassive black hole is maximal.  Here, the solid black curve
represents the light received by an observer.  We break down the
attenuated SED in a stellar ({\it dotted curve}) and AGN contribution ({\it dashed
curve}).  Finally, the best-fit model (in this case an exponentially
declining star formation history that started 0.8 Gyr ago) is plotted
in gray.  Although resulting in a low $\chi_{\rm reduced}^2 \sim 1$, the
SED modeling is mislead by a degeneracy between the stellar+AGN light
and the stellar light of a younger population obscured by large
columns of dust.  The addition of AGN light, when exceeding 10\% of
the total emission, adds another -0.1 to -0.15 dex to $\Delta
\log age_w$, +0.05 to +0.1 mag to $\Delta E(B-V)$, and +0.3 to +0.5
mag to $\Delta A_V$.  Since all observed light is treated as having a
stellar origin, the star formation rates during the phase of and under
viewing angles with significant AGN contribution to the observed SED
are overestimated by an order of magnitude.  These cases typically
show a larger $\chi_{\rm reduced}^2$ (70\% have $\chi_{\rm reduced}^2
> 5$).  As expected, simulations with an initial gas fraction of 80\%
have relatively more stars formed during final coalescence and more
gas fed to the supermassive black hole than lower gas fraction runs.
Hence the bias due to a contribution of AGN light is more prominent
for simulations with a higher gas fraction.

\subsection {Lessons for SED modeling}
\label{lessons.sec}

\begin {figure*} [htbp]
\centering
\plotone{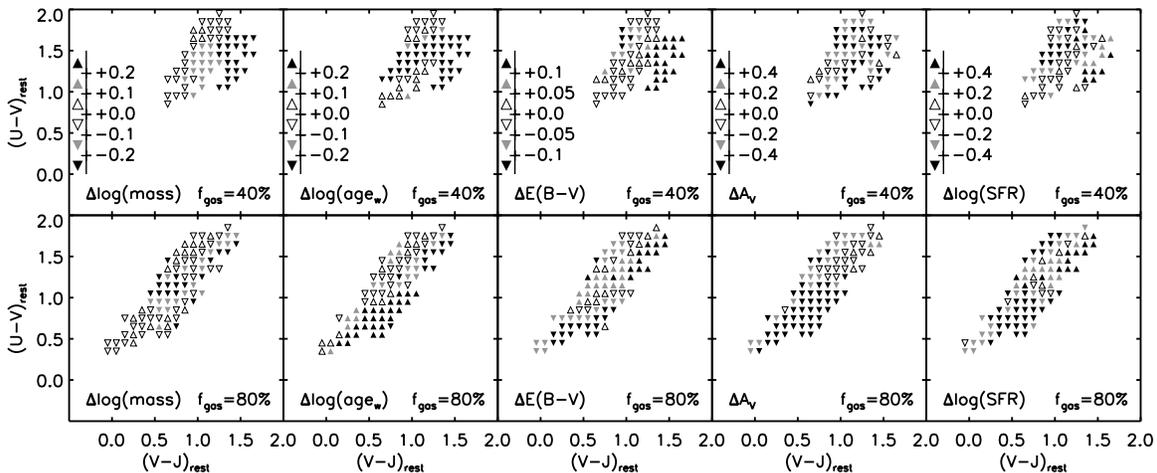} 
\caption{Median quality of recovered stellar population properties for
simulations with a gas fraction of 40\% and 80\% as a function of
rest-frame optical and optical-to-NIR color.  Downward
triangles indicate underestimates with respect to the true value.
Upward triangles mark overpredictions.  Galaxies in regions with white
triangles are characterized most accurately.  Simulations with
$f_{\rm gas}=80\%$ reach bluer colors in $(U-V)_{\rm rest}$ than those with
$f_{\rm gas}=40\%$ since their initial stars were set to younger ages (see
\S\ref{sim_main.sec}).  At red optical colors [$(U-V)_{\rm rest} > 1$],
galaxies with relatively blue $(V-J)_{\rm rest}$ colors are better
recovered than those at the red $(V-J)_{\rm rest}$ end.  The former are
older, less obscured systems, while the latter have a young and dusty
nature.  At blue optical colors [$(U-V)_{\rm rest}<1$], large systematics
in the determination of age and $A_V$ occur.  Their opposite signs
cancel out in the derivation of stellar mass.
\label {lesson.fig}
}
\end {figure*}

How can the modeling of real high-redshift galaxies benefit from our
analysis of merger simulations?  In order to answer this question, we
need to translate the mechanisms described above in terms of the
physical properties of stars and dust into their combined effect as a
function of observables.

An observable that correlates crudely with time since the merger, the
parameter as a function of which we quantified the quality of SED
modeling, is specific star formation rate ($sSFR = SFR / M$).
Generally, merger remnants are characterized by a low $sSFR \lesssim
10^{-9.5}$ yr$^{-1}$, whereas the sSFR is large ($sSFR \gtrsim
10^{8.5}$ yr$^{-1}$) during the merger-triggered starburst.  However,
very gas-rich isolated galaxies (the 'disk' phase of simulations with
$f_{\rm gas, init} = 80\%$) may reach similarly high sSFRs.

Since the results of our SED modeling showed no trend with redshift
for which the synthetic photometry was computed, we can also describe
the quality of recovering stellar population properties in terms of
rest-frame colors.  Although not true observables, their computation
as described by Rudnick et al. (2003) suffers from only a minor
template dependence compared to parameters such as mass, age, dust
content, and SFR.  In Fig.\ \ref{lesson.fig}, we present the
performance of our SED modeling as a function of location in the
rest-frame $U-V$ versus $V-J$ color-color diagram.  Simulations with
initial gas fractions of 40\% and 80\% are shown separately.  The SED
modeling was based on the full (stellar+AGN) photometry in both cases.
Downward triangles indicate a median value of $\Delta =
parameter_{recovered} - parameter_{true}$ that is negative for
simulations with the respective colors.  Upward triangles represent an
overprediction of the true value.  Lighter symbols are used for a
better correspondence between the true and modeled parameter value.
The different initial stellar ages and lower metallicities for $f_{\rm
gas}=80\%$ runs explain why they extend to bluer $U-V$ colors.  At
blue optical colors ($U-V < 1$), the attenuation is seriously
underestimated.  An overestimate of the age by a similar order of
magnitude (at $U-V < 0.65$ mostly due to the lower metallicity and at
$0.65 < U-V < 1$ due to age-dependent extinction) leads to a
relatively robust estimate of the mass.

At red optical colors ($U-V > 1$), sources with relatively blue $V-J$
colors are generally better modeled than those with the reddest $V-J$
colors in our sample.  In the $1 < U-V < 1.8$ color regime, the
effects of dust attenuation play an important role.  Here we find
objects that are heavily extincted during the merger-triggered
starburst, but also disk galaxies seen edge-on during the earliest
phases of the simulation.  All sources with $V-J > 1.5$ in our sample
belong the latter category.  Since we did not impose an age gradient
for the initial stars, the dust distribution in thoses cases is
non-uniform but uncorrelated with intrinsic color of the stellar
particles.  This explains why they have the strongest overestimate in
$E(B-V)$ (see \S\ref{fixz_Z.sec}).  The upper part of the color-color distribution
($U-V > 1.8$) is where each galaxy ends up by aging without
further inflow of gas.  In such a system, the $A_V$ values are modest
to low and the lack of a template exactly matching the SFH is less
problematic as the epoch of major star formation lies further back in
time.


\section {Results from SED modeling with free redshift}
\label{results_freez.sec}

In practice, complete spectroscopic surveys of mass-limited samples of
high-redshift galaxies are rare.  Consequently, we often are not able
to fix the redshift in the SED modeling procedure to its exact true
value.  Over the past few years, several codes have improved on
estimating the redshift based on the broad-band SED, by experimenting
with different template sets and fitting algorithms.  In this section,
we will use the new photometric redshift code EAZY (Brammer et al. in
preparation) to establish the quality of photometric redshift
($z_{\rm phot}$) estimates derived from our synthetic photometry and
analyze the impact of $z_{\rm phot}$ uncertainties on the derived stellar
population properties.

\subsection {The photometric redshift code EAZY}
\label{freez_eazy.sec}
Here, we summarize the main characteristics of the EAZY photometric
redshift code.  A full description of the algorithm and template sets
will be presented by Brammer et al. (in preparation).  We test our
ability to recover the redshifts using a template set based on a set
of $\sim 3000$ P\'{E}GASE 2.0 (Fioc \& Rocca-Volmerange 1997)
templates.  The large number of templates was reduced to 5 principal
components necessary to span the colors of galaxies in the
semi-analytic model by De Lucia \& Blaizot (2007).  In addition to
these 5 templates, a dusty template with an age of 50 Myr and
$A_V=2.75$ accounts for the existence of dustier (and thus redder)
galaxies than present in the semi-analytic model.

The code fits a non-negative superposition of templates to the
$B$-to-$8\ \mu$m photometry, using a template error function that
effectively downweights the rest-frame UV and NIR portion of the
templates in the fit.  The value of the redshift marginalized over the
total redshift probability distribution is adopted as best $z_{\rm
phot}$ estimate.

EAZY allows for the use of a magnitude prior function, constructed
from observed or simulated number counts as a function of apparent
magnitude and redshift.  However, since we shift each simulation over
the entire redshift range $1.5 \leq z \leq 2.9$, our methodology does
not allow to test this feature.


\subsection {Recovering redshifts and stellar population properties from broad-band photometry}
\label{freez_full.sec}

\subsubsection {Recovering redshifts}
\label{zquality.sec}

\begin {figure} [t]
\centering
\plotone{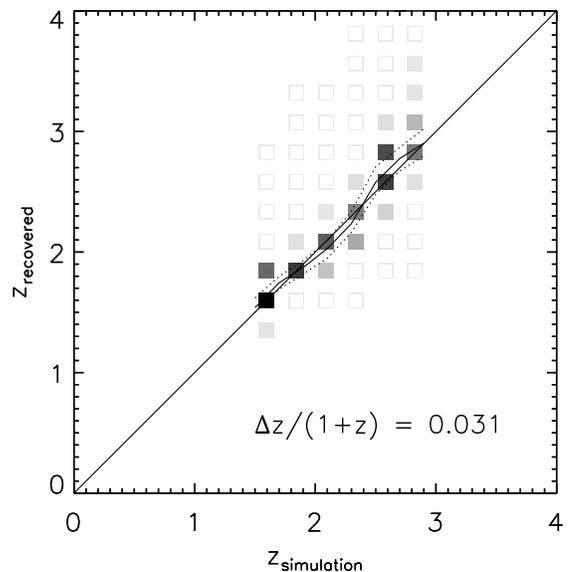} 
\caption{Comparison of photometric redshifts by EAZY versus
true redshift.  The correspondence is good ($\sigma_{\rm NMAD}(\Delta
z / (1+z)) = 0.031$), without significant systematic offsets.
\label {redshift.fig}
}
\end {figure}

In Fig.\ \ref{redshift.fig}, we compare the photometric redshifts
obtained with EAZY with the input redshifts for which the synthetic
broad-band SEDs were computed.  The measure commonly used to quantify
the photometric redshift quality is $\Delta z / (1+z)$.  Its
normalized median absolute deviation is 0.031.  We find no dependence
of $\Delta z / (1+z)$ on redshift, time since (or before) the merger,
or mass-weighted age.  A slight trend with $A_V$ was found in the
sense that the median $\Delta z / (1+z)$, which varies between -0.01
and 0.01 for $0<A_V<2.5$, increases to a few 0.01 for simulated
galaxies with $A_V > 2.5$.  However, this systematic offset is still
smaller than the scatter in $\Delta z / (1+z)$ for these highly
obscured galaxies.  The performance of EAZY is very good and
competitive with that of other codes presented in the literature.  We
note that the high $z_{\rm phot}$ quality owes greatly to the dense
sampling of the rest-frame UV-to-NIR SED by our synthetic photometry
(filter set and depths were adopted from the GOODS-South survey).
Leaving out 1 of the 11 filters, especially when located near a
spectral break at the considered redshift, may boost the uncertainty
in $z_{\rm phot}$.  E.g., running EAZY on the synthetic SEDs without
$H$-band results in a scatter over the $1.5 \leq z \leq 2.9$ range
that is 35\% larger: $\sigma_{\rm NMAD}(\Delta z / (1+z)) = 0.042$.
At $z \sim 2.7$, where the Balmer/4000\AA\ break has shifted in
between $J$ and $H$, we find a median offset and scatter in $\Delta z
/ (1+z)$ of 0.050 ($z_{\rm phot}$ being larger than $z_{\rm spec}$)
and 0.071 respectively when omitting the H-band.

This exercise offers a valuable complementary test to the empirical
comparison with spectroscopic samples of high-redshift galaxies.  The
latter are direct measurements and therefore insensitive to our
knowledge of stellar tracks and population synthesis.  On the other
hand, spectroscopic samples of high-redshift galaxies often suffer
from selection biases, especially against galaxies lacking emission
lines.

\subsubsection {Impact of $z_{\rm phot}$ uncertainties}
\label{zph_impact.sec}

Having quantified the quality of photometric redshifts, we now repeat
the SED modeling fixing the redshift to its best-fit value.  The same
mechanisms as discussed in
\S\ref{results_fixz.sec} are still influencing the recovery of
physical parameters.  Given the partially random nature of the $z_{\rm
phot}$ uncertainties, it comes as no surprise that the central 68\%
interval broadens with respect to the SED modeling at fixed redshift.
Averaged over time, the broadening is limited to 1\%, 7\%, 5\%, 9\%,
and 1\% in $\Delta \log age_w $, $\Delta \log M$, $\Delta E(B-V)_{\rm
eff}$, $\Delta A_{V,{\rm eff}}$, and $\Delta \log SFR$ respectively,
owing to the small scatter in $z_{\rm true}$ versus $z_{\rm phot}$.

Any systematic part of the $z_{\rm phot}$ uncertainty translates into
additional systematic offsets in the stellar population properties.
Qualitatively, when a source is mistakenly placed at higher redshift,
a larger mass estimate and lower dust reddening are required to match
the observed SED.  Although we established that systematic offsets in
$\Delta z / (1+z)$ are small (\S\ref{zquality.sec}), such an effect is
noticeable for the most obscured ($A_V > 2.5$) galaxies.  The slight
overprediction of their redshifts translates into an additional offset
in the estimated mass of $\log M_{\rm fixz} - \log M_{\rm photz} \sim
-0.05$ dex, in the estimated reddening of $E(B-V)_{\rm fixz} -
E(B-V)_{\rm photz} \sim 0.018$ mag, and in the estimated extinction of
$A_{V, {\rm fixz}} - A_{V, {\rm photz}} \sim 0.11$ mag.  We note that,
for an excellent photometric redshift accuracy as presented here,
these effects are up to an order of magnitude smaller than the
systematic offsets from the true stellar population properties as
introduced by our poor knowledge of the star formation history, dust
content and distribution, and metallicity.  Furthermore, the impact of
$z_{\rm phot}$ uncertainties discussed here is specific for the
particular template set used.  Other template sets may introduce
different redshift biases and therefore propagate into different
biases when characterizing stellar populations.


\section {Summary}
\label{summary.sec}

\begin{deluxetable*}{llllll}
\tablecolumns{2}
\tablewidth{0pc}
\tablecaption{Qualitative summary of systematic trends in recovering stellar population properties. \label{overview.tab}
}
\tablehead{
\colhead{Effect} & \colhead{on age} & \colhead{on $E(B-V)$} & \colhead{on $A_V$} & \colhead{on mass} & \colhead{on SFR}
}
\startdata
SFH mismatch (higher SFR than in the past) & - & + & - & - & + \\
Non-uniform extinction (uncorrelated with intrinsic colors of emitters) & - & + & - & - & - \\
Age-dependent extinction (more extinction toward young stars)           & + & - & - & -\tablenotemark{a} & - \\
$Z < Z_{\sun}$ & + & - & - & +\tablenotemark{b} & 0 \\
AGN & - & + & + & +\tablenotemark{c} & +
\enddata
\tablenotetext{a}{The overestimated age and underestimated $A_V$
compete in the determination of stellar mass.  Since in practice we
find this effect to be most outspoken in phases where the SFH mismatch
is largest, the stellar mass will effectively be underestimated.}
\tablenotetext{b}{Here too, the systematic offsets in age and $A_V$
have an opposite sign.  From comparison of mass estimates based on the
full attenuated photometry and its equivalent with all stars fixed to
$Z_{\sun}$, we find the effective $\Delta \log M$ is positive.}
\tablenotetext{c}{Similar to (b), we compared the results from SED
modeling with/out AGN contribution and find that in the median the addition of AGN
light increases the mass estimate slightly.}
\end{deluxetable*}

We analyzed the performance of a simple SED modeling procedure applied
to synthetic optical-to-NIR broad-band SEDs of merger simulations
placed at redshifts z=1.5, 1.7, ..., 2.9.  The synthetic SEDs span a
large range of colors and shapes, and are significantly affected by
dust attenuation.  For all simulated galaxies, we find counterparts in
the real high-redshift universe with similar SED types.  First, we
modeled the SEDs assuming the redshift was known.  The masses, ages,
$E(B-V)$, and $A_V$ of simulated ellipticals are very well reproduced,
with an average value of $\Delta \log M = -0.02^{+0.06}_{-0.11}$,
$\Delta \log age_w = -0.03^{+0.12}_{-0.14}$, $\Delta E(B-V) =
-0.03^{+0.11}_{-0.07}$, and $\Delta A_V = -0.29^{+0.32}_{-0.30}$.
Here the errors indicate the central 68\% interval of the distribution
of $\Delta parameter$ values of all the simulations (different masses,
gas fractions, viewing angles) in the spheroid regime.  The simulated
ellipticals are correctly classified as having low SFRs ($\Delta \log
SFR = -0.23^{+0.62}_{-0.47}$).  In earlier, actively star-forming,
phases, the scatter in recovered stellar population properties with
respect to the true value increases, and larger systematic
underestimates of age, mass, extinction, and SFR occur.  This is
particularly the case for the simulation snapshots of phases with
merger-triggered star formation, where we find the following offsets
and scatter (averaged over the merger regime indicated in Figure\
\ref{sim.fig}): $\Delta \log M = -0.13^{+0.10}_{-0.14}$, $\Delta \log
age_w = -0.12^{+0.40}_{-0.26}$, $\Delta E(B-V) =
-0.02^{+0.08}_{-0.08}$, $\Delta A_V = -0.54^{+0.40}_{-0.46}$, and
$\Delta \log SFR = -0.44^{+0.32}_{-0.31}$.  The SED modeling performs
better on regular star-forming disks than on galaxies during the
merging event.  Compared to spheroids however, the results of the SED
modeling on disks show a larger scatter and larger systematic
underestimates: $\Delta \log M = -0.06^{+0.06}_{-0.14}$, $\Delta \log
age_w = +0.03^{+0.19}_{-0.42}$, $\Delta E(B-V) =
-0.02^{+0.13}_{-0.07}$, $\Delta A_V = -0.35^{+0.29}_{-0.34}$, and
$\Delta \log SFR = -0.22^{+0.23}_{-0.34}$.

By adding the effects of dust attenuation, metallicity variations and
AGN step by step to the basic intrinsic photometry, we were able to
disentangle the different mechanisms at play and their impact on the
estimation of the galaxy's stellar population properties.  The
qualitative impact on the SED modeling results by different aspects of
the galaxy content is summarized in Table\
\ref{overview.tab}.

A mismatch between the real SFH and the allowed template SFHs leads to
an inability to account for the difference between light-weighted and
mass-weighted properties such as stellar age and mass.  If the optical
depth toward intrinsically bluer emitting sources is larger than to
intrinsically redder stellar populations, the nett effect of the age
and mass underestimate due to mismatch between true and template SFH
is less severe.  We find proof of such an increased extinction toward
younger stars during the merger-triggered starburst.  Applying the
Calzetti et al. (2000) reddening law toward each stellar particle, we
find that the overall reddening for a given $A_V$ is less than
predicted by the Calzetti et al. (2000) law, particularly when the
optical depth is uncorrelated with the intrinsic colors of the sources
it is hiding.  In the latter case, the effective dust vector has a shallower
slope in the $U-V$ versus $V-J$ color-color diagram than the Calzetti et
al. (2000) vector.  In the case of larger optical depths toward young
(blue) stellar populations, there is relatively more reddening in $U-V$
for a given reddening in $V-J$.  Applying a MW or SMC-like attenuation
law to the individual stellar particles in the simulation increases
the reddening, but the effective extinction is still grayer than the
Calzetti et al. (2000) law.

All other properties remaining the same, the effect of applying our
SED modeling to stellar populations with sub-solar metallicities is
that one would underpredict the reddening.  For the young ages where such
sub-solar metallicities could be expected, interpreting the light as
coming from a solar-metallicity population will lead to an
overestimate of the stellar age.

Finally, our SED modeling is based on purely stellar emission.  During
the brief period when the AGN contribution is significant, the
addition of its light will make the galaxy look younger and dustier,
with a larger SFR.

Although some of the biases described above could be anticipated from
simpler models (see, e.g., Fig.\ \ref{toyAv.fig} and Fig.\
\ref{metal.fig}), the simulations used in this paper prove very
valuable in illustrating and quantifying the effects at play.
Moreover, by following the formation of stars and distribution of gas
and dust on a physically motivated basis, they provide insights that
could not be obtained from simple toy models, such as the preferential
extinction of young stars during the merger and its impact on
recovering stellar population properties by SED modeling.

We next repeated our analysis adopting the best-fit photometric
redshift estimate.  Using the photometric redshift code EAZY (Brammer
et al. in preparation), we obtain a median normalized absolute
deviation $\sigma_{\rm NMAD}(\Delta z / (1+z)) = 0.031$.  The random
uncertainty in $z_{\rm phot}$ boosts the scatter in the quality
measures $\Delta \log M$, $\Delta \log age_w$, $\Delta E(B-V)$,
$\Delta A_V$, and $\Delta \log SFR$ by less than 10\%.  A slight
dependence of $\Delta z / (1+z)$ on extinction propagates into an
additional systematic difference in the estimated stellar mass, on the
10\% level for galaxies with $A_V > 2.5$.  Offsets in reddening and
visual extinction $A_V$ are anti-correlated with $\Delta z / (1+z)$.

S. W. and T. J. C. acknowledge support from the W. M. Keck Foundation.
B. E. R. gratefully acknowledges support from a Spitzer Fellowship
through a NASA grant administrated by the Spitzer Science Center.  We
thank Natascha F\"{o}rster Schreiber for useful discussions on stellar
populations.

\begin{references}
{\small
\reference{} Barnes, J. E.,\& Hernquist, L. 1996, ApJ, 471, 115
\reference{} Barnes, J. E.,\& Hernquist, L. 1991, ApJ, 370, L65
\reference{} Bell, E. F.,\& de Jong, R. S. 2001, ApJ, 550, 212
\reference{} Bolzonella, M., Miralles, J.-M.,\& Pell\'{o}, R. 2000, A\&A, 363, 476
\reference{} Brammer, G. B., van Dokkum, P. G.,\& Coppi, P. 2008, ApJ, 686, 1503
\reference{} Bruzual, G.,\& Charlot, S. 2003, MNRAS, 344, 1000 (BC03)
\reference{} Bullock, J. S., Kolatt, T. S., Sigad, Y., Somerville, R. S., Kravtsov, A. V., Klypin, A. A., Primack, J. R.,\& Dekel, A. 2001, MNRAS, 321, 559
\reference{} Calzetti, D., et al. 2000, ApJ, 533, 682
\reference{} Cox, T. J., Dutta, S. N., Di Matteo, T., Hernquist, L., Hopkins, P. F., Robertson, B.,\& Springel, V. 2006, ApJ, 650, 791
\reference{} Daddi, E., et al. 2007a, ApJ, 670, 156
\reference{} Daddi, E., et al. 2007b, ApJ, 670, 173
\reference{} Di Matteo, T., Springel, V.,\& Hernquist, L. 2005, Nature, 433, 604
\reference{} Driver, S. P., Windhorst, R. A.,\& Griffiths, R. E. 1995, ApJ, 453, 48
\reference{} Edmunds, M. G. 1990, MNRAS, 246, 678
\reference{} Erb, D. K., Shapley, A. E., Pettini, M., Steidel, C. C., Reddy, N. A.,\& Adelberger, K. L. 2006a, ApJ, 644, 813 
\reference{} Erb, D. K., Steidel, C. C., Shapley, A. E., Pettini, M., Reddy, N. A.,\& Adelberger, K. L. 2006b, ApJ, 646, 107
\reference{} Erb, D. K., Steidel, C. C., Shapley, A. E., Pettini, M., Reddy, N. A.,\& Adelberger, K. L. 2006c, ApJ, 647, 128
\reference{} Ferguson, H. C., Dickinson, M., Papovich, C. 2002, ApJ, 569, 65
\reference{} Fioc, M.,\& Rocca-Volmerange, B. 1997, A\&A, 326, 950
\reference{} F\"{o}rster Schreiber, N. M., et al. 2004, ApJ, 616, 40
\reference{} F\"{o}rster Schreiber, N. M., et al. 2006, AJ, 131, 1891
\reference{} Franx, M., Illingworth, G. D., Kelson, D. D., van Dokkum, P. G.,\& Tran, K. 1997, ApJ, 486, L75
\reference{} Genzel, R., et al. 1998, ApJ, 498, 579
\reference{} Glazebrook, K., Ellis, R. S., Santiago, B.,\& Griffiths, R. 1995, MNRAS, 275, L19
\reference{} Grazian, A., et al. 2006a, A\&A, 449, 951
\reference{} Groves, B., Dopita, M. A., Sutherland, R. S., Kewley, L. J., Fischera, J., Leitherer, C., Brandl, B.,\& van Breugel, W. 2008, ApJS, 176, 438
\reference{} Hernquist, L. 1989, Nature, 340, 687
\reference{} Holden, B. P., et al. 2005, ApJ, 620, 83
\reference{} Holmberg, E. 1941, ApJ, 94, 385
\reference{} Hopkins, P. F., Hernquist, L., Martini, P., Cox, T. J., Robertson, B., Di Matteo, T.,\& Springel, V. 2005a, ApJ, 625, L71 
\reference{} Hopkins, P. F., Hernquist, L., Cox, T. J., Di Matteo, T., Martini, P., Robertson, B.,\& Springel, V. 2005b, ApJ, 630, 705
\reference{} Hopkins, P. F., Richards, G. T.,\& Hernquist, L. 2007, ApJ, 654, 731
\reference{} Jonsson, P. 2006, MNRAS, 372, 2
\reference{} Jonsson, P., Cox, T. J., Primack, J. R.,\& Somerville, R. S. 2006, ApJ, 637, 255
\reference{} Kriek, M., et al. 2006, ApJ, 645, 44
\reference{} Kriek, M., et al. 2007, ApJ, 669, 776
\reference{} Kroupa, P. 2001, MNRAS, 322, 231
\reference{} Labb\'{e}, I., et al. 2003, AJ, 125, 1107
\reference{} Larson, R. B.,\& Tinsley, B. M. 1978, ApJ, 219, 46
\reference{} Maiolino, R., et al. 2008, A\&A, 488, 463
\reference{} Maraston, C. 2005, MNRAS, 362, 799
\reference{} Maraston, C., Daddi, E., Renzini, A., Cimatti, A., Dickinson, M., Papovich, C., Pasquali, A.,\& Pirzkal, N. 2006, ApJ, 652, 85
\reference{} Mihos, J. C.,\& Hernquist, L. 1994, ApJ, 431, L9
\reference{} Mihos, J. C.,\& Hernquist, L. 1996, ApJ, 464, 641
\reference{} Mo, H. J., Mao, S.,\& White, S. D. M. 1998, MNRAS, 295, 319
\reference{} Noguchi, M. 1988, A\&A, 203, 259
\reference{} O'Connell, R. W. 1986, in Stellar Populations, ed. C. Norman, A. Renzini,\& M. Tosi (Cambridge: Cambridge Univ. Press), 213
\reference{} Panuzzo, P., Granato, G. L., Buat, V., Inoue, A. K., Silva, L., Iglesias-P\'{a}ramo, J.,\& Bressan, A. 2007, MNRAS, 375, 640
\reference{} Papovich, C., Dickinson, M., \& Ferguson, H. C. 2001, ApJ, 559, 620
\reference{} Papovich, C., Dickinson, M., Giavalisco, M., Conselice, C. J.,\& Ferguson, H. C. 2005, ApJ, 631, 101
\reference{} Papovich, C., et al. 2006, ApJ, 640, 92
\reference{} Papovich, C., et al. 2007, ApJ, 668, 45
\reference{} Pei, Y. C. 1992, ApJ, 395, 130
\reference{} Pettini, M., Kellogg, M., Steidel, C. C., Dickinson, M., Adelberger, K. L.,\& Giavalisco, M. 1998, ApJ, 508, 539
\reference{} Pettini, M., et al. 2001, ApJ, 554, 981
\reference{} Reddy, N. A., Erb, D. K., Steidel, C. C., Shapley, A. E., Adelberger, K. L,\& Pettini, M. 2005, ApJ, 633, 748
\reference{} Reddy, N. A., Steidel, C. C., Fadda, D., Yan, L., Pettini, M., Shapley, A. E., Erb, D. K., \& Adelberger, K. L., ApJ, 644, 792
\reference{} Richards, G. T., et al. 2006, ApJS, 166, 470
\reference{} Rigopoulou, D., et al. 2006, ApJ, 648, 81
\reference{} Robertson, B., Cox, T. J., Hernquist, L., Franx, M., Hopkins, P. F., Martini, P.,\& Springel, V. 2006, ApJ, 641, 21
\reference{} Robertson, B., Li, Y., Cox, T. J., Hernquist, L.,\& Hopkins, P. F. 2007, ApJ, 667, 60
\reference{} Rocha, M., Jonsson, P., Primack, J. R.,\& Cox, T. J. 2008, MNRAS, 383, 1281
\reference{} Sanders, D. B., Soifer, B. T., Elias, J. H., Madore, B. F., Matthews, K., Neugebauer, G.,\& Scoville, N. Z. 1988a, ApJ, 325, 74
\reference{} Sanders, D. B., Soifer, B. T., Elias, J. H., Neugebauer, G.,\& Matthews, K. 1988b, ApJ, 328, L35
\reference{} Sanders, D. B.,\& Mirabel, I. F. 1996, ARA\&A, 34, 749
\reference{} Shapley, A. E., Steidel, C. C., Adelberger, K. L., Dickinson, M., Giavalisco, M.,\& Pettini, M. 2001, ApJ, 562, 95
\reference{} Shapley, A. E., Steidel, C. C., Pettini, M.,\& Adelberger, K. L. 2003, ApJ, 588, 65
\reference{} Shapley, A. E., Steidel, C. C., Erb, D. K., Reddy, N. A., Adelberger, K. L., Pettini, M., Barmby, P.,\& Jiasheng, H. 2005, ApJ, 626, 698
\reference{} Springel, V.,\& Hernquist, L. 2003, MNRAS, 339, 289
\reference{} Springel, V., Di Matteo, T.,\& Hernquist, L. 2005a, ApJ, 620, L79
\reference{} Springel, V., Di Matteo, T.,\& Hernquist, L. 2005b, MNRAS, 361, 776
\reference{} Toomre, A.,\& Toomre, J. 1972, ApJ, 178, 623
\reference{} Toomre, A. 1977, in Evolution of Galaxies and Stellar Populations, 401, Yale Univ. Obs: New Haven
\reference{} van Dokkum, P. G.,\& Stanford, S. A. 2003, ApJ, 585, 78
\reference{} van Dokkum, P. G., et al. 2004, ApJ, 611, 703
\reference{} Vitvitska, M., Klypin, A. A., Kravtsov, A. V., Wechsler, R. H., Primack, J. R.,\& Bullock, J. S. 2002, ApJ, 581, 799
\reference{} Wuyts, S., et al. 2007, ApJ, 655, 51
\reference{} Wuyts, S., et al. 2008, ApJ, 682, 985
\reference{} Wuyts, S., et al. 2009b, submitted to ApJ
\reference{} Zwicky, F. 1956, Ergebnisse der Exakten Naturwissenschaften, 29, 344

}
\end {references}






















\end {document}